\documentclass[fleqn,usenatbib]{mnras}
\usepackage[switch]{lineno}
\usepackage{newtxtext,newtxmath}

\usepackage[T1]{fontenc}

\DeclareRobustCommand{\VAN}[3]{#2}
\let\VANthebibliography\thebibliography
\def\thebibliography{\DeclareRobustCommand{\VAN}[3]{##3}\VANthebibliography}

\usepackage{graphicx}	
\usepackage{amsmath}	
\usepackage{longtable}
\usepackage{float}
\usepackage{fix-cm}

\newcommand{\sqcm}{cm$^{-2}$}  
\newcommand{\kms}{$\rm km~s^{-1}$} 
\newcommand{\lya}{Ly$\alpha$}
\newcommand{\lyb}{Ly$\beta$}
\newcommand{\lyg}{Ly$\gamma$}

\newcommand{\HI}{\mbox{H\,{\sc i}}}
\newcommand{\HII}{\mbox{H\,{\sc ii}}}
\newcommand{\Nhi}{$N(\HI)$}

\newcommand{\logNhi}{${\rm log}_{10}($\Nhi$/{\rm cm}^{-2})$}

\newcommand{\OII}{\mbox{O\,{\sc ii}}}
\newcommand{\OIII}{\mbox{O\,{\sc iii}}}
\newcommand{\OIV}{\mbox{O\,{\sc iv}}}

\newcommand{\OVI}{\mbox{O\,{\sc vi}}}

\newcommand{\CIV}{\mbox{C\,{\sc iv}}}

\newcommand{\CIII}{\mbox{C\,{\sc iii}}}

\newcommand{\MgII}{\mbox{Mg\,{\sc ii}}}

\newcommand{\logm}{$\log_{10}~(M_{\star}/\rm M_{\odot})$}
\newcommand{\Msun}{$\rm M_{\odot}$}

\newcommand{\loghi}{${\rm log}_{10}(N(\HI)/{\rm cm}^{-2})$}

\usepackage{cancel}
\usepackage[flushleft]{threeparttable}
\usepackage{ragged2e}
\usepackage{graphicx}	
\usepackage{amsmath}	
\usepackage{tabularx}
\usepackage{newtxtext,newtxmath,ulem}
\usepackage{caption}
\usepackage{multirow}
\usepackage{cases}
\usepackage{subcaption}
\usepackage{booktabs,caption}
\usepackage[T1]{fontenc}
\usepackage{lastpage}

\title[MUSEQuBES: Cool \HI\ gas around low-$z$ galaxies]{MUSEQuBES: The Column Density, Covering Fraction, Mass, and Environmental Dependence of Cool \HI\ Gas Around Low-Redshift Galaxies}

\author[S. Dutta et al.]{Sayak Dutta,$^{1}$\thanks{E-mail: sayak18@iucaa.in}
Sowgat Muzahid,$^{1}$
Joop Schaye,$^{2}$
Sean Johnson,$^{3}$
Nicolas F. Bouch\'e,$^{4}$
Ramona Augustin,$^{5}$
\newauthor 
Sebastiano Cantalupo, $^{6}$
Hsiao-Wen Chen,$^{7}$
and
Martin Wendt$^{8}$
\\~\\ 
$^{1}$Inter-University Centre for Astronomy \& Astrophysics, Post Bag 04, Pune, India 411007\\
$^{2}$Leiden Observatory, Niels Bohrweg 02, 2333 CA Leiden, Netherlands\\
$^{3}$Department of Astronomy, University of Michigan, 1085 S. University Ave, Ann Arbor, MI 48109, USA\\
$^{4}$Centre de Recherche Astrophysique de Lyon (CRAL) UMR5574, Univ Lyon1, Ens de Lyon, CNRS, 69230 Saint-Genis-Laval, France\\
$^{5}$Leibniz-Institut f\"ur Astrophysik Potsdam (AIP), An der Sternwarte 16, 14482 Potsdam, Germany\\
$^{6}$Department of Physics, University of Milan Bicocca, Piazza della Scienza 3, 20126, Milano, Italy\\
$^{7}$Department of Astronomy and Astrophysics, The University of Chicago, 5640 S. Ellis Avenue, Chicago, IL 60637, USA\\
$^{8}$Institut f\"ur Physik und Astronomie, Universit\"at Potsdam, Karl-Liebknecht-Str. 24/25, 14476 Potsdam, Germany
}

\date{Accepted XXX. Received YYY; in original form ZZZ}

\pubyear{\the\year{}}

\begin{document}
\label{firstpage}
\pagerange{\pageref{firstpage}--\pageref{LastPage}}
\maketitle

\begin{abstract}

\noindent 
We investigate cool \HI\ gas traced by Lyman series absorption around 256 galaxies at $z\approx0.48$ (median stellar mass, \logm\ $= 8.7$) using 15 background quasars (median impact parameter, $D = 140$ pkpc), as part of the MUSE Quasar-fields Blind Emitters Survey (MUSEQuBES). We find that the \HI\ column density ($N(\HI)$) profile around isolated star-forming galaxies spanning $\approx 3$ dex in $M_{\star}$ is well described by a power law with slope $\approx -3$ when expressed as a function of normalized impact parameter $D/R_{\rm vir}$. The \HI\ covering fraction ($\kappa$) within the virial radius for \loghi\ $= 14$ is significantly lower in high-mass passive galaxies than in isolated star-forming galaxies. The $\kappa$-profile of isolated star-forming galaxies suggests a characteristic size of the \HI-rich CGM of $\approx 1.5 R_{\rm vir}$ across the stellar mass range. The mean \HI\ mass in the outer CGM ($0.3$–$1~R_{\rm vir}$) increases with $M_{\star}$, ranging from $\approx 10^{5.0}$ to $10^{6.6}$~\Msun. The $b$-parameters of the strongest \HI\ components correlate and anti-correlate with specific star-formation rate (sSFR) and mass, respectively, with $>2\sigma$ significance. Broad \lya\ absorbers (BLAs) with $b>60$~\kms\ are predominantly associated with high-mass galaxies, likely tracing the warm-hot phase of the CGM. The velocity centroids of \HI\ components indicate that absorbers at $D<R_{\rm vir}$ are largely consistent with being gravitationally bound to their galaxies, independent of stellar mass. Finally, leveraging $\approx 3000$ galaxies from the wide-field Magellan follow-up of six MUSEQuBES fields, we find that non-isolated galaxies exhibit an \HI-rich environment extending roughly three times farther than in isolated counterparts.

\end{abstract}




\section{Introduction} \label{sec:intro}

The presence of a gaseous medium surrounding the luminous parts of galaxies is now well established. This extended gas reservoir between the stellar disk and intergalactic medium (IGM), also known as the circumgalactic medium (CGM), is believed to play a crucial role in the formation and evolution of galaxies \citep[see][]{Tumlinson_2017, Chen_2024}. In the standard cosmological framework of galaxy evolution, intergalactic gas accretes onto galactic halos and, unlike dark matter, is shock-heated, subsequently cools, and condenses to form the luminous disks of galaxies \citep[]{white}. A significant fraction of baryons may remain in this diffuse, gaseous phase, forming the CGM that envelops the central galaxy. The CGM is dynamically shaped by galactic feedback processes, which can enrich it with gas and metals expelled from galaxies or deplete it by driving them out into the IGM.

As the most abundant element in the universe, hydrogen serves as an ideal tracer of the CGM. At low redshifts, the neutral component of this predominantly ionized element effectively traces the cool ($\sim10^{4}$~K) phase of baryonic matter within galactic halos. However, detecting the cool, neutral gas in the CGM in emission is difficult due to its low density. Absorption line spectroscopy of bright background sources, such as quasars, has emerged as a powerful tool for investigating the diffuse CGM \citep[e.g.,][]{Bergeron_1986, Petitjean_1990,Chen_2005,Chen_2009,
Steidel_2010,Prochaska_2011}. 
\color{black}

The Space Telescope Imaging Spectrograph (STIS) and Cosmic Origin Spectrograph (COS) onboard the {\it{Hubble Space Telescope (HST)}} have revolutionized the CGM studies at low redshift. Over the past decade, multiple studies have systematically investigated the spatial distribution and kinematic properties of \HI\ absorbing gas surrounding low-$z$ galaxies \citep[e.g.,][]{Chen_2010, Thom_2012,Tumlinson_2013,Borthakur_2016,Keeney_2017}. \citet[]{Chen_2005, Chen_2009} showed that the differential clustering strength of strong \lya\ absorbers with various galaxy types at projected comoving distances $<250h^{-1}$ kpc indicates a low incidence of strong \lya\ absorption near absorption-line-dominated galaxies. \citet{Tumlinson_2013} demonstrated that \HI\ is ubiquitous around star-forming galaxies in their COS-Halos survey, with non-detections primarily associated with passive galaxies. However, when \HI\ was detected, they found no statistically significant difference in column densities ($N(\HI)$) between star-forming and passive galaxies \citep[see also][]{Thom_2012}. 
\color{black} 
Combining the COS-Halos and COS-GASS galaxy samples, \citet[]{Borthakur_2016} reported a strong correlation between sSFR and \lya\ rest-frame equivalent width (REW) when the radial dependence is taken into consideration. Recently, based on the stacked \lya\ REW, \citet[]{Dutta1_2024} found suppressed \HI\ absorption for high-mass and passive galaxies compared to star-forming galaxies within the virial radius. Such variation of cool circumgalactic gas with halo mass has been predicted in simulations to result from i) different accretion mechanisms, namely `hot' and `cold' mode accretion for high- and low-mass halos \citep[]{Birnboim_2003, keres, Fukugita_2006, Voort_2011}, and ii) feedback processes due to star-formation or AGN activity; shock-heating the cooler CGM clouds \citep[]{Voort_2011}, and/or possibly even expelling/unbinding the CGM \citep[]{Borthakur_2013}.

A comprehensive understanding of how accretion and feedback processes shape the CGM requires studying baryon distributions across a wide stellar mass range. However, the current literature remains predominantly focused on $\approx L_*$ galaxies. 
Although dwarf galaxies ($M_\star \lesssim 10^{9}$~\Msun), owing to their shallow potential wells, are ideal laboratories for studying the effects of feedback, most studies of the CGM around such low-mass systems have been limited to the nearby universe \citep[$z < 0.1$; e.g.,][]{Bordoloi_2014,Zheng_2024}.
For such low-$z$ galaxies, accurately determining circumgalactic $N(\HI)$-- and consequently the total \HI\ mass, $M(\HI)$-- is challenging with COS far-ultraviolet (FUV) spectra due to the lack of coverage of higher-order Lyman series lines. 
Accurately measuring $M(\HI)$ is crucial, as it enables an estimate of the total baryonic mass in the CGM when combined with reasonable ionization corrections.

Advances with multi-object spectrographs (MOS) have extended the CGM studies for dwarfs to $z\approx0.3$ \citep[e.g.,][]{Johnson_17, Wilde_2021, Wilde_2023}, but the advent of integral-field spectroscopy \citep[IFS, e.g., MUSE;][]{Bacon_2010} now enables systematic searches for low-mass galaxies with \logm~$\approx 7-9$ out to $z\approx1.0$ \citep[see e.g.,][] {Weng_2023,Dutta1_2024,Dutta2_2025,Dutta3_2025,Mishra_2024,Ramona_2024}. Recent deep IFS observations have successfully detected both ionized gas and metal lines in emission from the CGM, providing new constraints on its spatial distribution and physical properties \citep[e.g.,][]{Zabl_2021,Dutta_2023,Guo_2023,Nielsen_2024,Dutta_R_2024}
\color{black}

In addition to host galaxy properties, the local environment plays a critical role in shaping the spatial distribution and kinematics of circumgalactic gas. Both cosmological simulations and observations indicate that environmental processes-- such as tidal interactions, ram pressure stripping, and satellite accretion-- can significantly alter the properties of galaxies and their CGM \citep[e.g.,][]{
Bahe_2019, 
Putman_2021, 
Dutta_2021, 
Rohr_2023,
MishraS_2024} 
Recent studies at higher redshifts ($z \approx 3$–$4$) have revealed systematic differences between the CGM of galaxies in overdense environments and their isolated counterparts. In particular, galaxies in overdense regions exhibit higher covering fractions and column densities (or REWs) of both gas (\HI) and metal ions (\CIV), suggesting enhanced CGM enrichment and/or retention in group environments \citep[]{Muzahid_2021,Banerjee_2023,Galbiati_2024,Banerjee_2025}. Similarly, at lower redshifts ($z \lesssim 2$), group galaxies exhibit flatter \MgII\ rest-frame equivalent width profiles, along with significantly higher covering fractions extending out to several 100s of kpc, compared to isolated galaxies. \citep[]{Chen_2010_part2,Huang_2021,
Dutta_2021,Cherrey_2025}. \citet{Johnson_15} found no such difference for \HI\ for their sample of $z\approx0.2$ galaxies. However, a comprehensive investigation into how the environment influences the \HI\ distribution in and around low-redshift galaxies—based on sufficiently large sample sizes—is still largely lacking in the literature.
\color{black}

\begin{table*}
\centering
\caption{Summary of galaxy properties for the MUSEQuBES and Magellan samples used in this study.}  
\label{tab:galprop}
\begin{tabular}{  p{3cm} p{3cm} p{3cm} p{3cm} p{3cm}  }
 \hline
  & \multicolumn{2}{c}{MUSEQuBES ($N_{\rm gal}=256$)} & \multicolumn{2}{c}{Magellan/IMACS ($N_{\rm gal}=3258$)} \\ 
  \cmidrule(l){2-3} \cmidrule(l){4-5} 
Property & Median & 68\% range & Median & 68\% range \\  
 \hline
$D$ (pkpc) & 140 & 77--196 & 2129 & 922--3551 \\ 
\logm & 8.7 & 7.5--9.8 & 10.3 & 9.5--10.9 \\ 
$D/R_{\rm vir}$ & 1.6 & 0.8--2.7 & 12.6 & 4.4--23.1 \\ 
$z$ & 0.48 & 0.31--0.64 & 0.4 & 0.2--0.6 \\ 
 \hline
\end{tabular}
\end{table*}

In this work, we use a unique, primarily low-mass sample of galaxies from the MUSEQuBES survey at low-$z$ \citep[$z<1$;][]{Dutta1_2024,Dutta2_2025}. \footnote{The MUSEQuBES survey at high-$z$ is focused on the CGM of Ly$\alpha$ emitting galaxies at $z\approx3.3$ \citep[see,][]{Muzahid_2021,Banerjee_2023,Banerjee_2025,Banerjee2_2025}.} 
Exploiting a total of $\approx65$~h of GTO observations, MUSEQuBES conducted a galaxy survey around 16 UV-bright background quasars that have high-quality HST/COS spectra. The deep MUSE observations, with on-source exposure times ranging from 2 to 10 hours, yielded a homogeneous sample primarily composed of low-mass (median $M_{\star} \approx 10^{8.7}~\rm M_{\odot}$), mildly star-forming (median SFR $\approx 0.1$ \Msun\ ${\rm yr}^{-1}$ ), sub-$L_*$ galaxies. 
A significant fraction of these galaxies are located at impact parameters, $D < R_{\rm vir}$ from the background quasars, where $R_{\rm vir}$ is the inferred virial radius{\footnote{Defined as the radius of a spherical region within which the mean mass density in 200 times the critical density of the universe.}}.  Additionally, a MOS-based galaxy survey with the Magellan telescope has been carried out for 6 out of the 16 MUSEQuBES sightlines (Johnson et. al., in prep). These primarily massive galaxies at larger impact parameters are used to investigate environmental dependence on the \HI\ distribution in and around galaxies.

This paper is organized as follows. In Section~\ref{sec:data}, we describe the galaxy sample used in this study. Section~\ref{sec:absorption_analysis} outlines the absorption-line data and the identification of galaxy–absorber pairs. The results are presented in Section~\ref{sec:results}, followed by a discussion in Section~\ref{sec:discussion}. Finally, our key findings are summarized in Section~\ref{sec:summary}. Throughout the paper, we adopt a $\Lambda \rm CDM$ cosmology with $\Omega_{\rm m} = 0.3$, $\Omega_{\Lambda} = 0.7$, and a Hubble constant of $H_0 = 70~ \rm km~ s^{-1}~ Mpc^{-1}$. All distances are in physical units unless specified otherwise.

\begin{figure*}
    \centering
    \includegraphics[width=1\linewidth]{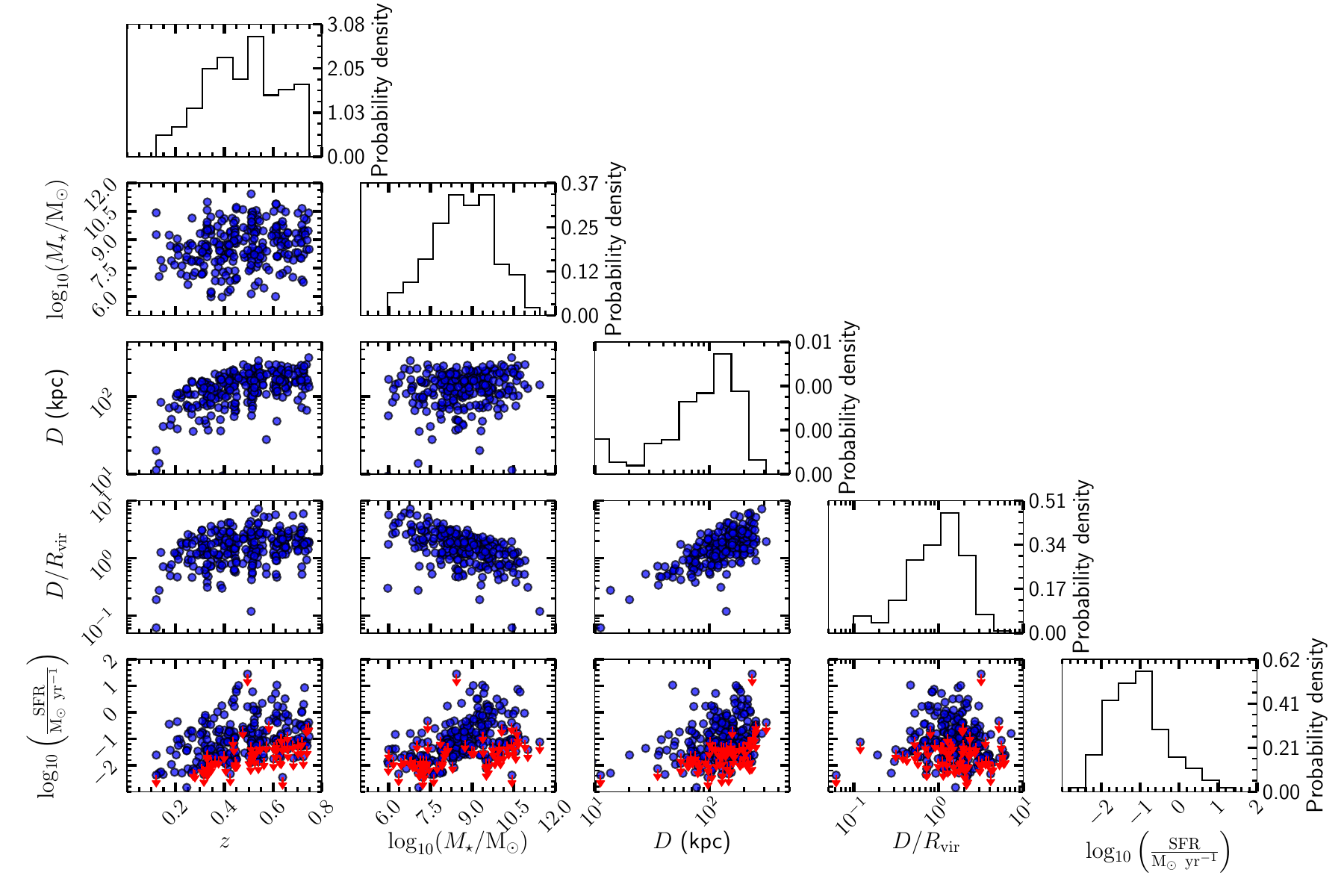}%
    \vskip-0.3cm 
    \caption{The various galaxy properties for the MUSEQuBES sample are plotted against each other with solid blue circles. The red downward arrows in the last row indicate the SFR upper limits. The panels along the diagonal show the probability density of the corresponding galaxy property with black solid lines. The upper limits are considered as measurements for the SFR probability density.}
    \label{fig:gal_prop}
\end{figure*}

\section{Galaxy Data} 
\label{sec:data}

This work is primarily based on the galaxies detected as part of the  low-$z$ MUSEQuBES survey \citep[]{Dutta1_2024,Dutta2_2025}. The survey was conducted around 16 quasar fields with high $S/N$ (15--85 per resolution element) COS spectra available in the HST spectral legacy archive \citep[]{Peeples_2017}. Each MUSE field yields $\approx90,000$ spectra, one from each $0.2'' \times 0.2''$ spaxel, enabling the identification of low-mass galaxies without requiring any photometric pre-selection.

In this work, we select 256 continuum-detected galaxies from the 413 foreground galaxies identified in the MUSEQuBES survey. These galaxies have measured stellar masses ($M_\star$), star formation rates (SFRs), and spectroscopic redshifts, and lie within the redshift range where redshifted \lya\ and/or \lyb\ absorption is covered by the medium-resolution COS spectra. 
We note that pure line emitters—primarily \OII-emitting galaxies at the redshifts relevant to this study—are not considered in our sample, as our selection requires detectable continuum emission to ensure reliable stellar mass estimates. \citet[]{Bouche_2025} reported that up to 15\% of \OII\ emitters may lack detectable continuum counterparts in typical 2-hour deep MUSE data. The implications of such pure line emitters to CGM studies will be investigated elsewhere. 
The galaxy survey is described in Section~2.1 of \citet[]{Dutta2_2025}.
The following selection criteria were applied to identify galaxies from the sample of 413 foreground galaxies in the MUSEQuBES survey: 
\begin{itemize}
    
    \item The redshifted wavelength of \lya\ and/or \lyb\ absorption for the galaxy must fall within the COS spectral coverage. As a result, 141 galaxies with $z \gtrsim 0.75$ are excluded from the analysis.

    \item A lower redshift limit of $z \gtrsim 0.12$ is imposed on the galaxy sample to ensure that at least \lyb\ is covered within the COS spectral range alongside \lya. This criterion enables robust identification of \HI\ absorption features and reliable constraints on their column densities. Two MUSEQuBES fields - TEX0206-048 and Q1354+048 - contain Lyman-limit systems at $z \approx 0.39$ and $z \approx 0.33$, respectively. To avoid spectral regions affected by Lyman continuum breaks in these fields, the redshift cut is adjusted to $z \approx 0.23$ for TEX0206-048 and $z \approx 0.18$ for Q1354+048. A total of 16 galaxies are excluded based on these criteria.    
\end{itemize}

\begin{figure*}
    \centering
    \includegraphics[width=0.5\linewidth]{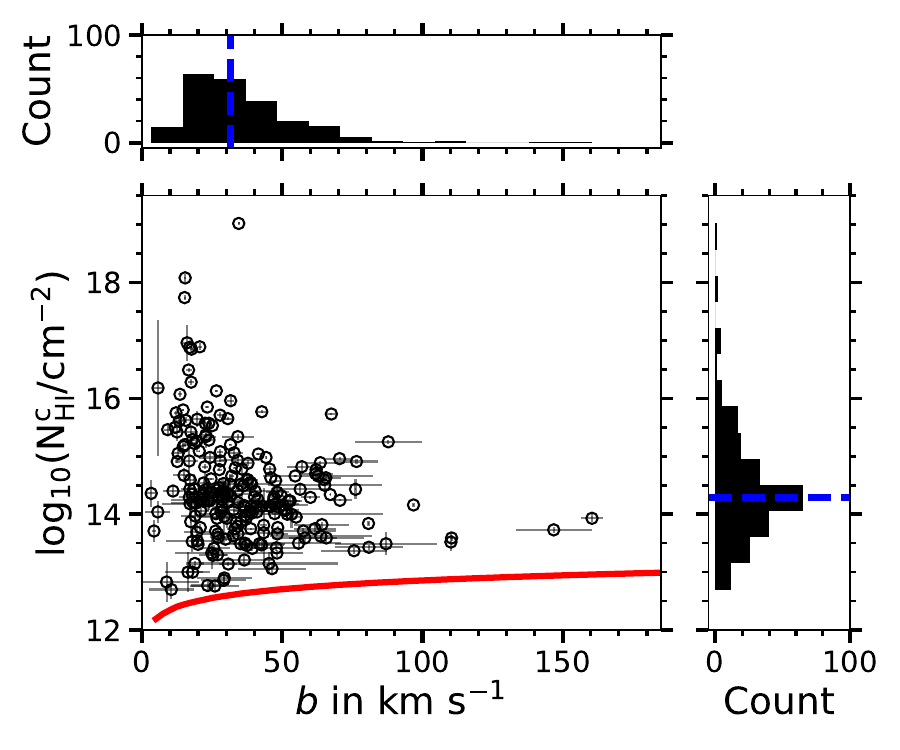}%
    \includegraphics[width=0.5\linewidth]{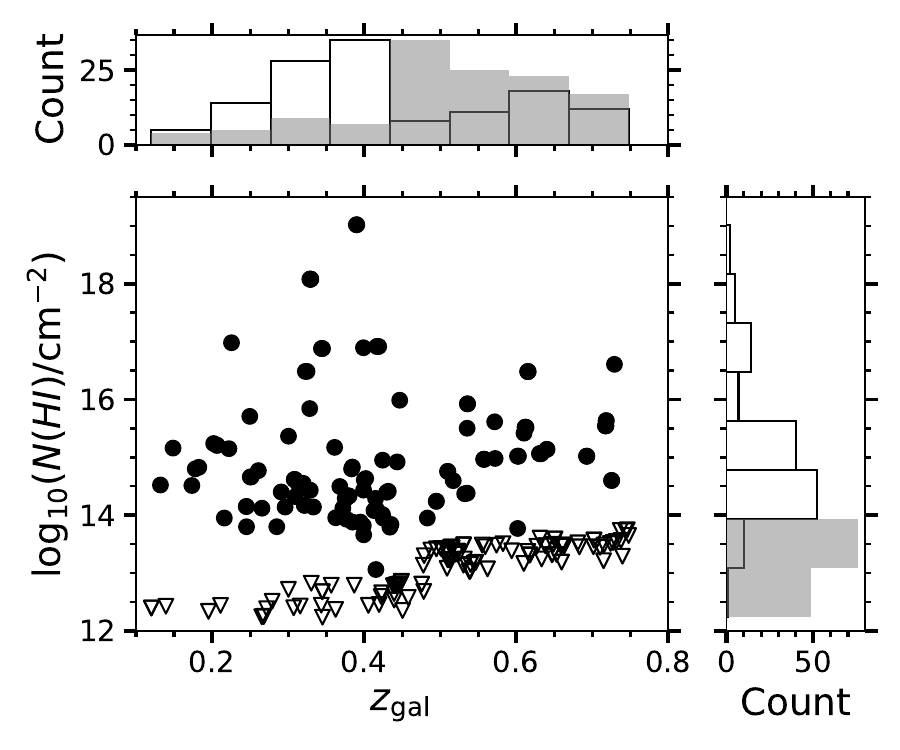}    
    \caption{{\tt Left}: The \HI\ column densities of the 227 components with $Q>1$ used in this work plotted against the $b$-parameters. The red line shows the limiting $N(\HI)$ as a function of $b-$parameter for a fiducial $S/N$ of 10 per pixel. The histograms on the top and right show the distribution of $b$ parameters and component column densities, respectively, with blue dashed lines indicating the corresponding median values. 
    {\tt Right:} The total column densities ($N(\HI)$) of absorption systems, defined by grouping components within $\Delta v = 300$~\kms, are plotted as a function of galaxy redshift using filled circles. The $3\sigma$ upper limits in cases of non-detections are shown by the open triangles. The observed gap between the detections and upper limits arises partly because the upper limits correspond to single-component systems, whereas the detected systems generally consist of multiple \HI\ components. In addition, these detections are located in the vicinity of galaxies and therefore tend to have higher column densities than the general population of \HI\ absorbers.
    The top and side panels show the distributions of galaxy redshift and $N(\HI)$, respectively, with open black histograms representing galaxies with \HI\ detections and filled grey histograms indicating non-detections.}
    \label{fig:abs_prop}
\end{figure*}

The 256 galaxies that meet both selection criteria are classified as star-forming (SF), passive (E), or unclassified (U), based on their position relative to the redshift-dependent star-forming main sequence \citep[SFMS;][]{Boogard_18}, as described in \citet[]{Dutta2_2025}. Briefly, passive galaxies are defined as those lying more than 3$\sigma$ below the SFMS. Among the remaining galaxies, those with measured star formation rates (SFRs) are classified as star-forming, while those with only $3\sigma$ upper limits on the SFR are categorized as unclassified. The median sSFR [68\% range] for star-forming and passive galaxies are $10^{-9.5}~{\rm yr}^{-1}$ [$10^{-10.1} - 10^{-9.0} ~{\rm yr}^{-1}$] and $10^{-11.4}~{\rm yr}^{-1}$ [$10^{-11.7} - 10^{-11.1} ~{\rm yr}^{-1}$], respectively.
In Fig.~\ref{fig:gal_prop}, we show the different galaxy properties of the MUSEQuBES sample plotted against each other with solid blue circles. The SFR upper limits are indicated with red downward arrows in the last row.
The panels along the diagonal show the probability density for the MUSEQuBES sample with black solid lines. The upper limits are considered as measurements for the SFR probability density.

Additionally, a wide-field, shallower galaxy survey has been conducted around 6 of the MUSEQuBES sightlines with the Inamori Magellan Areal Camera and Spectrograph (IMACS) on the Magellan telescope \citep[]{Chen_2009, Johnson_2013, Johnson_15}. Combining the MUSEQuBES galaxies from these 6 sightlines, this sample constitutes a unique dataset spanning a wide dynamic range in both impact parameter and stellar mass. The analysis presented in this paper is primarily based on the MUSEQuBES galaxy sample. We incorporate the Magellan sample solely to investigate the role of galaxy environment in shaping the circumgalactic \HI\ in Section \ref{sec:env}. The overall galaxy properties for the two surveys are tabulated in Table.~\ref{tab:galprop}.

It is evident that MUSEQuBES galaxies have significantly lower stellar masses ($M_{\star}$), smaller impact parameters ($D$), and lower normalized impact parameters ($D/R_{\rm vir}$) compared to the Magellan galaxies, while both samples, by construction, lie at similar redshifts.

\section{Absorption Line Data} 
\label{sec:absorption_analysis}


\subsection{Search for \texorpdfstring{{\mbox{H\,{\sc i}}}}{H I} absorbers around galaxies} 

\label{subsec:abs_cat}

In this study, we investigate the distribution of \HI-absorbing gas surrounding galaxies by utilizing \lya\ and \lyb\ transitions. The \lya\ transition restricts the search to $z\lesssim0.48$ due to the wavelength coverage constraints of the COS G160M grating. Using \lyb\ as the target transition, we extend this search up to $z\approx0.75$. For each of the 256 galaxies, we searched for \HI\ absorbers within a fiducial velocity window of $\pm600$~\kms\ relative to the galaxy redshift, guided by velocity plots of all available Lyman series lines. A velocity window of $\pm600$~\kms\ typically encompasses twice the expected range of circular velocities for the most massive halos in our sample. For the six sightlines with wide-field, Magellan IMACS data, we conducted a blind search for \HI\ absorbers to investigate the dependence of \HI\ absorption on galaxy environment (see Section~\ref{sec:env} for details).

For each identified \HI\ absorber, a quality flag of $Q=3$ is assigned when (1) one or more higher-order transitions are present in addition to \lya\ (for $z\approx 0.12 - 0.48$) or \lyb\ (for $z\approx 0.48-0.75$), and (2) at least one higher-order transition is unblended and unsaturated. A flag of $Q=2$ is assigned when (1) all higher-order transitions are saturated, or (2) no higher-order transitions are detected above the sensitivity limit-- i.e., their expected strength is consistent with the $3\sigma$ column density upper limit set by the $S/N$ of the spectra. The \HI\ absorbers for which {\it all} the available higher-order transitions are blended are assigned a quality flag of $Q=1$. Similarly, potential \lya\ absorbers with no coverage of higher-order transitions in the COS spectra are also assigned $Q=1$. Absorbers with $Q=1$ are {\it not used} in our work.

We used the Voigt profile fitting code {\sc Vpfit} \citep[][]{vpfit} to decompose the \HI\ absorbers with quality flags $Q=2$ and $Q=3$ into individual components. Unblended pixels from all available Lyman series transitions were used to constrain the best-fit redshift ($z_{\rm comp}$), Doppler parameter ($b$), and column density ({$N_{\HI}^{c}$}) of individual Voigt profile components, while accounting for the wavelength- and lifetime position-dependent line spread function of COS.
The Voigt profile fitting exercise yielded a total of 227 \HI\ components. In the left panel of Fig.~\ref{fig:abs_prop}, we show the column density $N_{\HI}^{c}$ of the individual components plotted against the $b$-parameters. The median [68\% range] $N_{\HI}^{c}$ and $b$ of the \HI\ components are $10^{14.3}~{\rm cm}^{-2}$ [$10^{13.6} - 10^{15.2}~{\rm cm}^{-2}$] and 31.4 \kms\ [$17.8-53.2$~\kms ], respectively. The histograms displayed on the top and right panels illustrate the distributions of
${\rm log}_{10}(N_{\HI}^{c})$ and $b$, with blue dashed lines indicating the median values.

\subsection{Constructing galaxy-absorber pairs}
\label{subsec:abs_gal}

For each MUSEQuBES galaxy, we collected all \HI\ components obtained in the previous section within $\pm 600$~\kms, sorted them by velocity, and applied a 1D Friends-of-Friends (FoF) algorithm with a linking velocity of $300$~\kms. Each resulting group of components is treated as a distinct `system'. We obtained a total of 103 unique absorption {\it systems} from the 227 components, with a median of $\approx2$ {\it components} per {\it system}. The system column density, \Nhi, is obtained by summing the $N_{\HI}^{c}$ of individual components contributing to the absorption system.

The galaxy and absorption catalogs are then cross-matched to identify cases where at least one {\it component} lies within $\pm300$~\kms\ of the galaxy redshift. In the case of detection, the absorption {\it system} containing the {\it component} is associated with the galaxy.  
131 out of the 256 galaxies have an associated \HI\ absorption {\it system} within $\pm300$~\kms{\footnote{It is possible for multiple galaxies to be associated with a single \HI\ absorption system, as well as multiple absorption systems to be associated with a single galaxy. However, we found that the latter did not occur.}}. 
For the 125 galaxies with no detectable \HI\ absorption, we estimated   $3\sigma$ upper limits on \Nhi\ using a velocity window of $\pm60$~\kms\ and the standard deviation of the normalized flux. The velocity window used here is twice the median $b$-parameter of the detected components.

The right panel of Fig.~\ref{fig:abs_prop} presents the $N(\HI)$ values as a function of galaxy redshift for the 256 galaxies, along with their corresponding distributions. At redshifts $z \gtrsim 0.48$, measuring \Nhi\ from \lyb\ and higher-order transitions results in reduced sensitivity for a fixed $S/N$,  leading to the apparent increase in the $3\sigma$ upper limits. Note, however, that the COS spectra remain sensitive to column densities of $\gtrsim 10^{14}$~\sqcm\ across the entire redshift range of our study.

We used a 3D FoF algorithm with a linking line-of-sight (LOS) velocity of $\pm500$~\kms\ and a transverse distance of 500~pkpc on the complete MUSEQuBES galaxy sample of 413 galaxies to classify galaxies as either `isolated' or `non-isolated' (`group')\footnote{We note that MUSE FoV is less than 500 pkpc for our redshift range of interest, essentially making this equivalent to 1D FoF with linking velocity of $\pm500$ \kms.}. Of the 256 galaxies used in this work, 125 are classified as isolated, while the remaining 131 have one or more neighboring galaxies and are thus considered non-isolated. These non-isolated galaxies belong to a total of 44 unique groups.

Among the 131 out of 256 galaxies with detected \HI\ absorption, 61 are classified as isolated galaxies and 70 as group galaxies. We note that a single absorption system can be associated with multiple group galaxies.

\section{Results} 
\label{sec:results}

In the first part of this section, we investigate the spatial distribution of \HI\ in the CGM with the aid of \Nhi\ and covering fraction ($\kappa$) profiles. The second part of this section will focus on the connection between galaxy properties and \HI\ kinematics.

\subsection{\texorpdfstring{$N(\HI)$}{N(HI)}-profile for the full sample}
\label{sec:nhi_full}

\begin{figure*}
    \centering
    \includegraphics[width=0.33\linewidth]{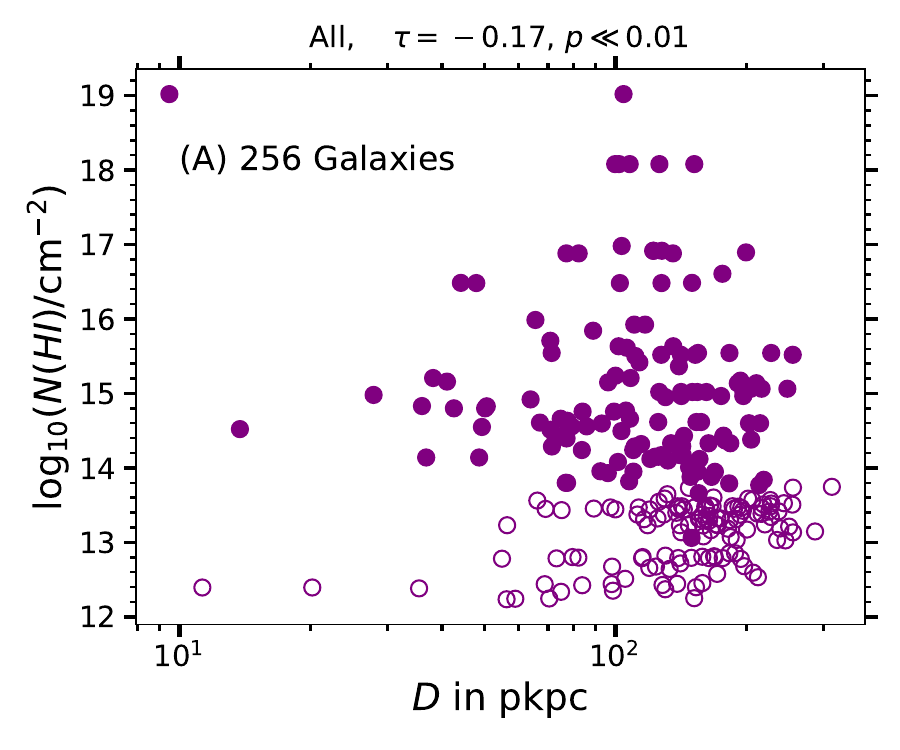}%
    \includegraphics[width=0.33\linewidth]{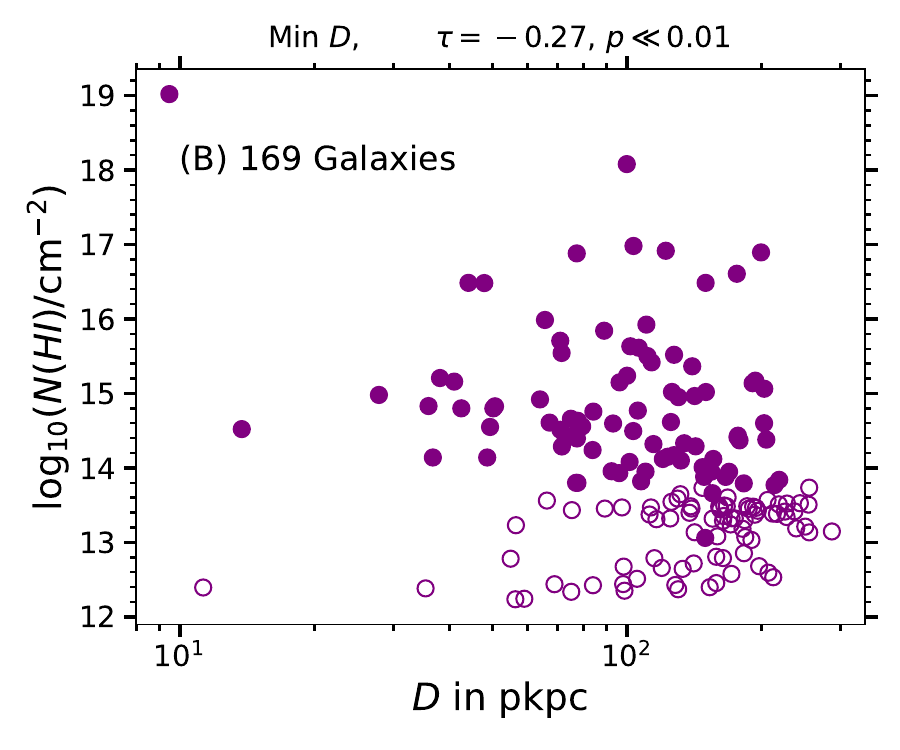}%
    \includegraphics[width=0.33\linewidth]{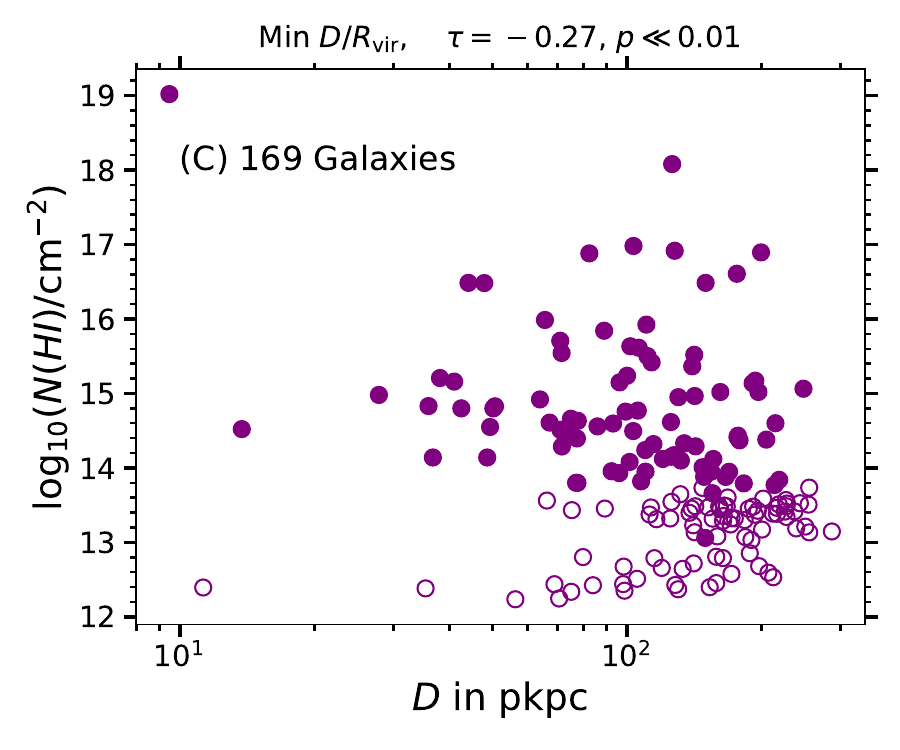}
    \includegraphics[width=0.33\linewidth]{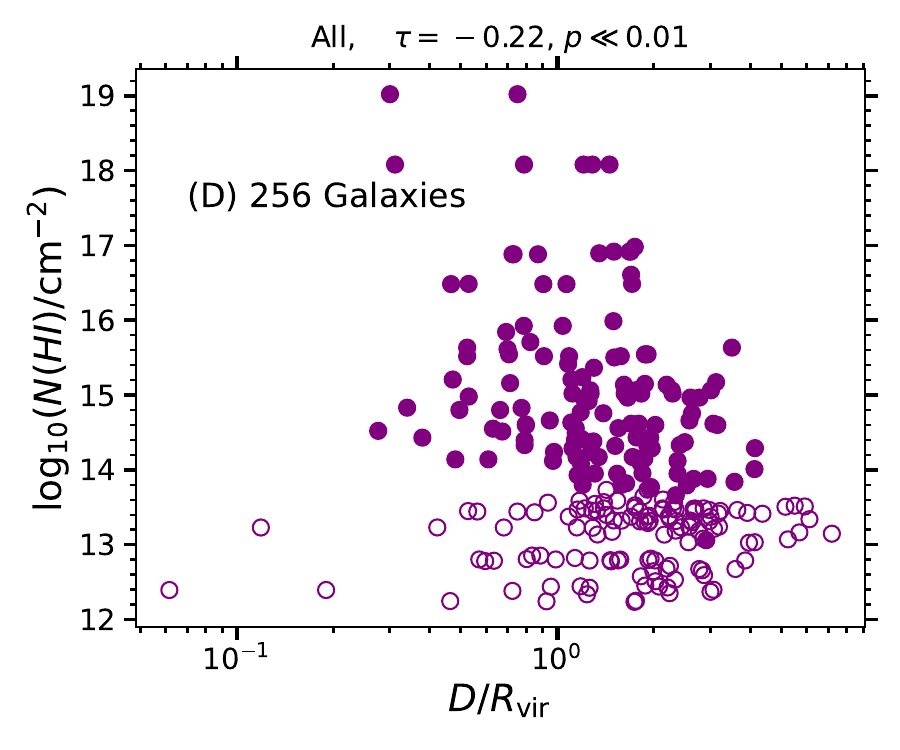}%
    \includegraphics[width=0.33\linewidth]{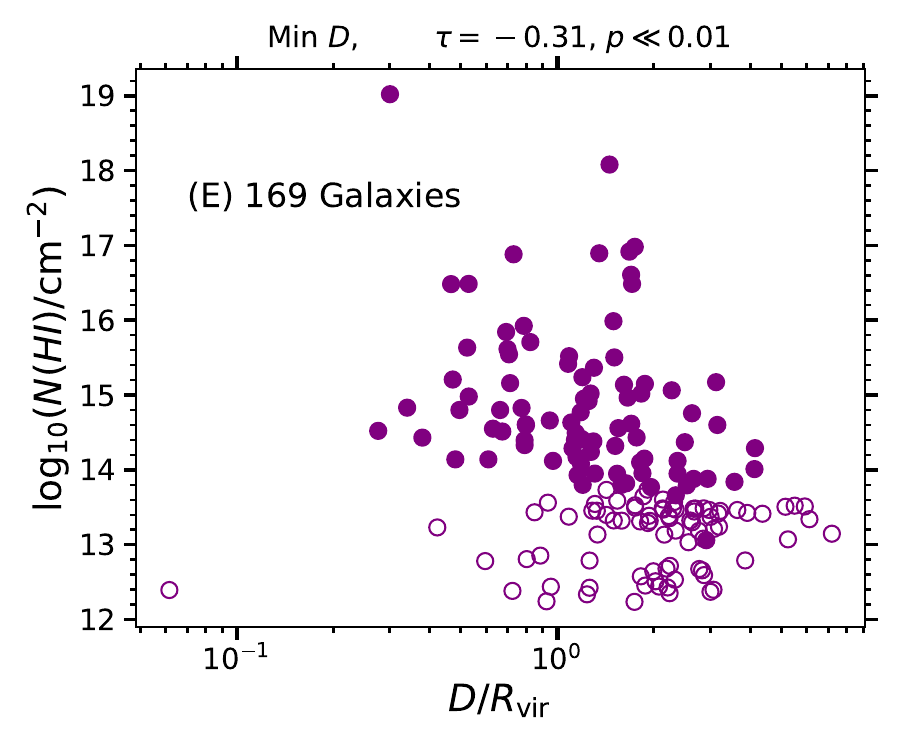}%
    \includegraphics[width=0.33\linewidth]{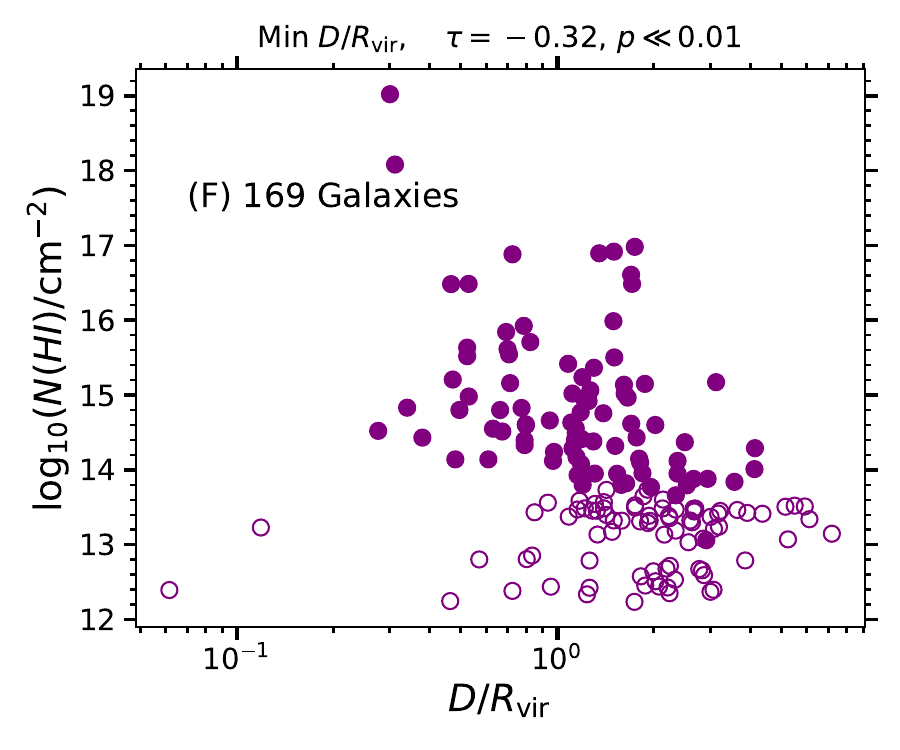}
    
    \caption{ (A):  Total $N(\HI)$ for the 256 MUSEQuBES galaxies plotted against impact parameters $D$. The solid and hollow points represent the detection and 3$\sigma$ upper limits on $N(\HI)$, respectively. (B) Total $N(\HI)$ for the subsample constructed from the isolated and group samples, where the group galaxy with the smallest $D$ is considered as the host.
    (C) Same as panel (B), but considering the galaxy with the smallest $D/R_{\rm vir}$ as the host.
    Panels (D), (E), and (F) are similar to panels (A), (B), and (C), respectively, but plotted against $D/R_{\rm vir}$. 
    The results of generalized Kendall-$\tau$ tests for each case are indicated at the top of the respective panels. The strongest anticorrelation is observed between $N(\HI)$ and $D/R_{\rm vir}$ in panel (F).} 
    \label{fig:Nh_prof}
\end{figure*}

The total circumgalactic $N(\HI)$ for the 256 MUSEQuBES galaxies is plotted against the impact parameter ($D$) and normalized impact parameter ($D/R_{\rm vir}$) in panels A and D of Fig.~\ref{fig:Nh_prof}, respectively. Out of the 256 galaxies, 125 are isolated (no companion galaxy within a LOS velocity of $\pm$500 \kms\ and a projected distance of 500 pkpc). The remaining 131 (non-isolated) galaxies constitute 44 groups. For each group, we consider as the host either the galaxy with (i) the smallest $D$  or (ii) the smallest $D/R_{\rm vir}$. Panels B and E of Fig.~\ref{fig:Nh_prof} present the $N(\HI)$ of 169 (125+44) galaxies, identifying the galaxies with the smallest $D$ as the hosts. Similarly, Panels C and F show the corresponding profiles where the galaxies with the smallest $D/R_{\rm vir}$ are taken to be the host. We note, however, that the term {\it group} in this work does not necessarily imply a virialized structure, but rather refers to a galaxy overdensity with a set of non-isolated galaxies associated with a single absorption measurement. While one could adopt the host galaxy as the most massive member instead of the one with the smallest $D$ or $D/R_{\rm vir}$, we find that our main conclusions are robust to this alternative choice.

A mild anticorrelation between $N(\HI)$ and projected separation ($D$ or $D/R_{\rm vir}$) for the full sample is suggested by the $\tau$ and $p$ values of generalized Kendall-$\tau$ test \citep[]{Isobe_1986}\footnote{Including the upper limits using the {\sc Cenken} package of {\sc R}.}. The $\tau$ and $p$ values are indicated at the top of the respective panels in Fig.~\ref{fig:Nh_prof}. In all cases, a very low $p$-value ($\ll0.01$) indicates a statistically significant anticorrelation between $N(\HI)$ and $D$ or $D/R_{\rm vir}$. 
We find a slightly stronger anticorrelation between $N(\HI)$ and $D/R_{\rm vir}$ ($\tau=-0.22$, panel D) as compared to $N(\HI)$ and $D$ ($\tau$ = $-0.17$, panel A). Considering the smallest $D$ or smallest $D/R_{\rm vir}$ galaxies from the groups enhances the anticorrelation of $N(\HI)$ with both $D$ and $D/R_{\rm vir}$. The strongest anticorrelation is observed between $N(\HI)$ and $D/R_{\rm vir}$ when the smallest $D/R_{\rm vir}$ galaxies from the groups are considered as the hosts ($\tau=-0.32$, panel F). Motivated by the strongest anticorrelation, we use the sample of 169 galaxies-- where galaxies with the smallest $D/R_{\rm vir}$ from the groups are considered as hosts-- in all subsequent analyses, unless stated otherwise.

\subsection{Variation of the \texorpdfstring{$N(\HI)$}{N(HI)}-profile with galaxy properties} 

\begin{figure*}
    \centering
    \includegraphics[width=0.7\linewidth]{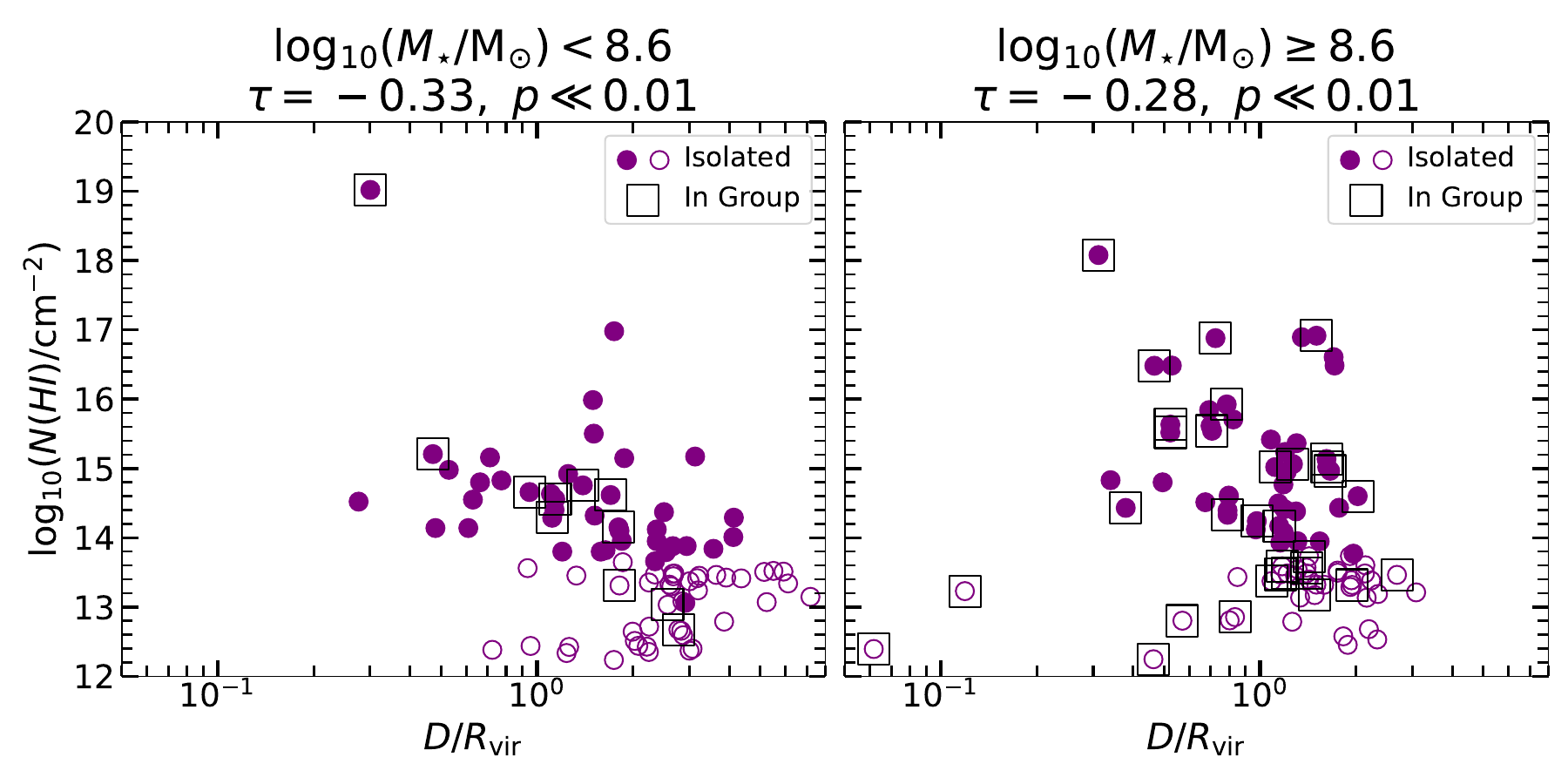}
    \includegraphics[width=0.7\linewidth]{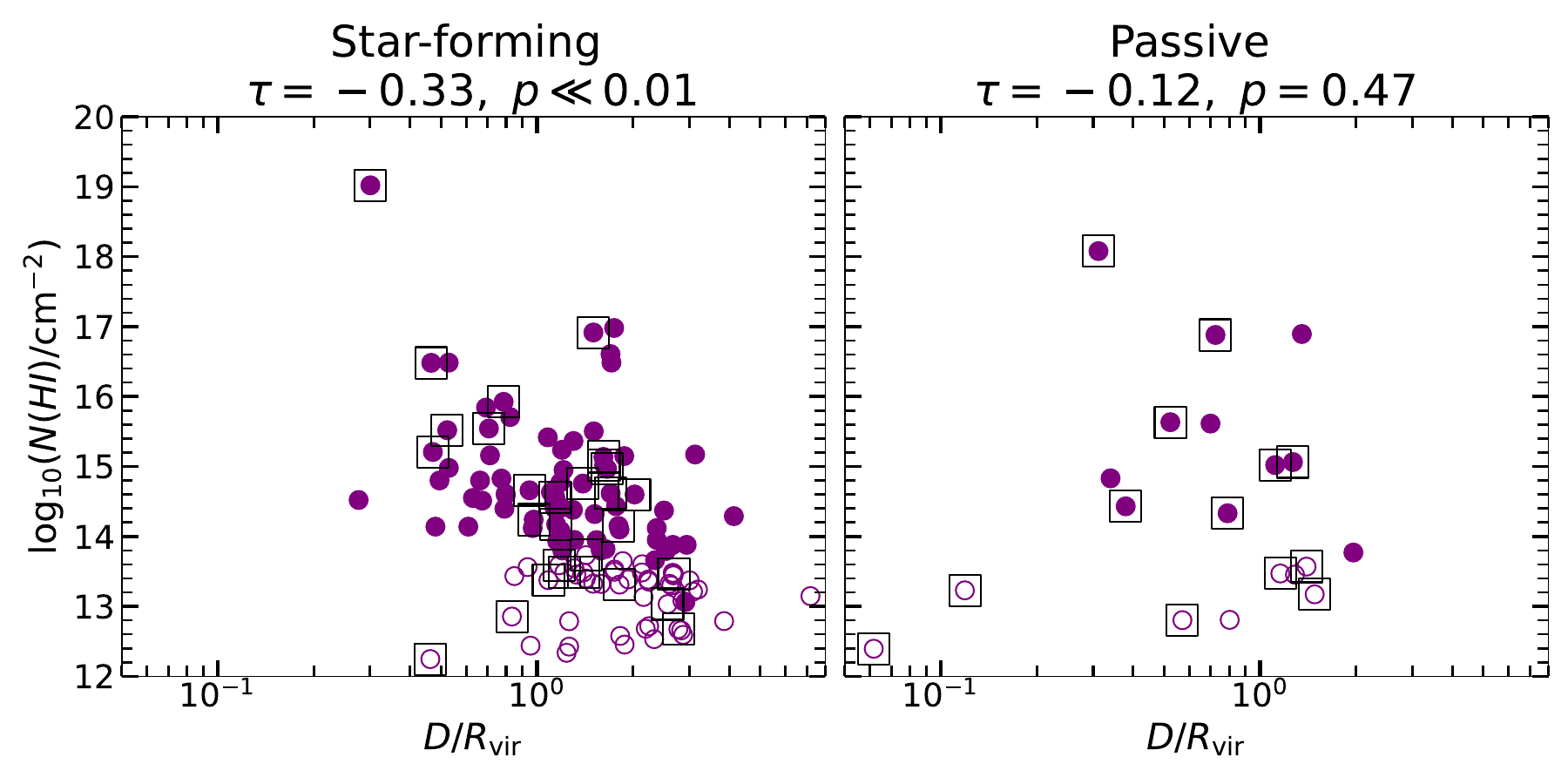}    
     \caption{ {\tt Top}:  Total $N(\HI)$ for low-mass (median \logm$=7.9$; left panel) and high-mass (median \logm$=9.4$; right panel) galaxies plotted against $D/R_{\rm vir}$. The solid and open circles indicate detection and 3$\sigma$ upper limits on $N(\HI)$, respectively. The circles inside black squares indicated the minimum $D/R_{\rm vir}$ galaxies for the groups, and the circles without black squares indicate isolated galaxies. {\tt Bottom}: Similar to the {\tt top} panels, but for star-forming (SF; left) and passive (E; right) galaxies.}  
    \label{fig:Nh_prop_div}
\end{figure*}

The MUSEQuBES galaxies exhibit a large dynamic range in $M_{\star}$ and SFR (see Section~\ref{sec:data}). In order to probe the role of $M_{\star}$ on the $N(\HI)-$profile, we divide our sample into two $M_{\star}$ bins with \logm~$<8.6$ and \logm~$\geq8.6$, where \logm~$=8.6$ is the median of the sample of 169 galaxies described in Section~\ref{sec:nhi_full}. The total $N(\HI)$ is plotted against $D/R_{\rm vir}$ for the two mass bins in the top-left and top-right panels of Fig.~\ref{fig:Nh_prop_div}. The points with the black square envelopes indicate the smallest $D/R_{\rm vir}$ galaxies of groups, and the points without the black squares are isolated galaxies.

The low-mass galaxies exhibit lower $N(\HI)$ (median $10^{14.8}~{\rm cm}^{-2}$) within $R_{\rm vir}$ than the high-mass galaxies (median $10^{15.5}~{\rm cm}^{-2}$). The low-mass galaxy associated with the strongest $N(\HI)$ system ($N(\HI) \geq 10^{19}~{\rm cm}^{-2}$) is part of a group. The high-mass galaxies show a large spread in $N(\HI)$ distribution within $R_{\rm vir}$ as compared to their low-mass counterparts, including non-detections of \HI\ for 7 galaxies. The results of the Kendall-$\tau$ tests for the two subsamples are tabulated in Table \ref{tab:best-fit-post-Novi}. The Kendall-$\tau$ test yields a marginally stronger anticorrelation between $N(\HI)$ and $D/R_{\rm vir}$ for low-mass galaxies ($\tau=-0.33$, $p\ll0.01$) than for the high-mass galaxies ($\tau=-0.28$, $p\ll0.01$).

At this stage, it is worth noting that the number of passive galaxies in our MUSEQuBES sample increases with $M_{\star}$. This trend underscores the importance of investigating the role of SFR in shaping the \HI\ distribution in the CGM. The $N(\HI)-$profiles for star-forming and passive galaxies are shown in the bottom-left and bottom-right panels of Fig.~\ref{fig:Nh_prop_div}. The $N(\HI)$-profile for the SF galaxies largely follows the trend of low-mass galaxies, with a similar median $N(\HI)$ value within $R_{\rm vir}$. The generalized Kendall's-$\tau$ test with $\tau=-0.33$ and $p\ll0.01$ suggests a moderate anticorrelation between $N(\HI)$ and $D/R_{\rm vir}$ for the star-forming galaxies. The trend marginally strengthens ($\tau=-0.34$) when only isolated galaxies are considered. In contrast, passive galaxies exhibit a large scatter in $N(\HI)$. Furthermore, no significant correlation is found between $N(\HI)$ and $D/R_{\rm vir}$, as indicated by a Kendall’s $\tau$ test including censored data points ($\tau = -0.12$, $p \approx 0.47$).

\begin{figure}
\centering
    \includegraphics[width=0.5\textwidth]{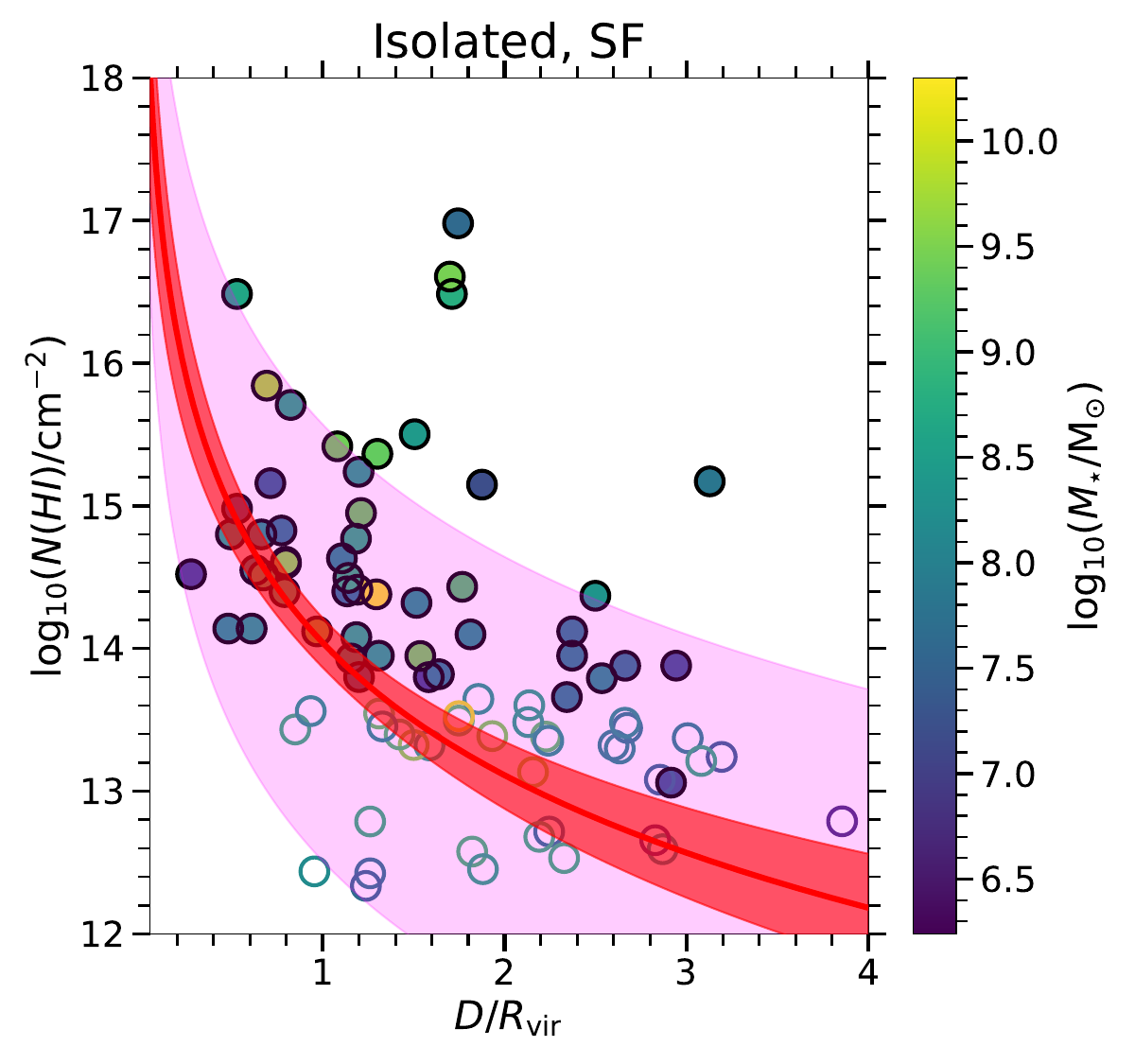}
    \caption{The $N(\HI)-$profile for the isolated, star-forming MUSEQuBES sample. The solid and hollow circles represent the detection and $3\sigma$ upper limits, respectively. The data points are color-coded by the stellar mass of the associated galaxy.     
    The red solid line and shaded region represent the best-fit power law and the corresponding 68\% confidence interval. The shaded magenta region represents the best-fit intrinsic scatter.}  
    \label{fig:nprof_fit_minDn}    
\end{figure}

\begin{table}
\centering
\caption{Results of generalized Kendall-$\tau$ test}   
\label{tab:best-fit-post-Novi}
\begin{tabular*}{\linewidth}{c@{\extracolsep{\fill}}ccccr}  
\hline
Sample  & $\tau$ & $p$ & Slope & Intercept \\ 
\hline
Low-mass & $-0.35$ & $\ll 0.01$ & $-0.8$ & 15.2 \\ 
High-mass & $-0.29$ & $\ll 0.01$ & $-1.7$ & 15.9 \\ 
Star-forming & $-0.33$ & $\ll 0.01$ & $-1.0$ & 15.4 \\ 
Passive & $-0.1$ & 0.57 &  &  \\ 
\\
\hline
\end{tabular*}
\justify
Notes-- $\tau$ and $p$ represent the generalized Kendall-$\tau$ coefficient and $p$ value, including upper limits. The slope and intercept are for the non-parametric Akritas-Theil-Sen (ATS) line \citep[see][]{Akritas} returned by the {\sc Cenken} package. The slope and intercept are quoted only when the anticorrelation is statistically significant. 
\end{table}

\subsection{ \texorpdfstring{$N(\HI)$}{N(HI)}-profile for isolated SF galaxies}

\label{subsec:N_prof_model}

In Fig.~\ref{fig:Nh_prof}, we observe the strongest anticorrelation between \logNhi\ and $D/R_{\rm vir}$ when galaxy–absorber pairs are constructed by selecting the smallest $D/R_{\rm vir}$ galaxy from a group. Furthermore, Fig.~\ref{fig:Nh_prop_div} shows that the strongest \HI\ absorbers are predominantly associated with non-isolated galaxies, contributing to the large scatter in \logNhi\ at a given $D/R_{\rm vir}$. Some non-detections at small $D/R_{\rm vir}$ are also linked to passive galaxies. Indeed, a generalized Kendall's $\tau$ test does not reveal any significant anticorrelation between \logNhi\ and $D/R_{\rm vir}$ for the passive population. Motivated by these findings, we proceed to quantify the spatial distribution of \Nhi\ for the isolated and star-forming galaxies in our sample (see Fig.~\ref{fig:nprof_fit_minDn}).

To model the $N(\HI)$-profile, we use a simple power-law function characterized by a slope $\alpha$ and normalization $N_0$, expressed as:   
\begin{equation}
\label{eq:Ndet_rhoN}
    N(D/R_{\rm vir})=N_0~(D/R_{\rm vir})^{-\alpha}.
\end{equation}
We construct a likelihood function following this joint probability as: 
\begin{equation}
\label{eq:logl}
\begin{split}
    \mathcal{L} \propto \prod_{i=1}^{n}  \frac{1}{\sqrt{2\pi\sigma^2} } {\rm exp} \left(- \frac{ ({\rm log}~N_i-{\rm log}~N)^2}{2\sigma^2} \right)  \times \\
    \prod_{j=1}^{m}   \int_{-\infty}^{{\rm log}~N_{j}^{u}}  \frac{1}{\sqrt{2\pi\sigma_p^2} } {\rm exp} \left(- \frac{ ({\rm log}~\bar{N}-{\rm log}~N)^2}{2\sigma^2} \right) d~{\rm log}~\bar{N}~.
\end{split}    
\end{equation}
Here, $N_i$ stands for the \HI\ column density for the $i^{\rm th}$ detection, $N_{j}^u$ stands for the 3$\sigma$ upper limit on the $j^{\rm th}$ non-detection. 
At a given $D/R_{\rm vir}$, the value of $\log~N(\equiv{\rm log}_{10}(N/{\rm cm}^{-2}))$ is obtained from Eqn.~\ref{eq:Ndet_rhoN}. 
Finally, $\sigma^2=\sigma_i^2+\sigma_p^2$ where $\sigma_p$ represents the intrinsic scatter and $\sigma_i$ is the uncertainty in $\log~N_i$.

Ideally, the likelihood for the non-detections is given by: 
 \begin{equation}
 \begin{split}
  \prod_{j=1}^{m} \frac{1}{\sqrt{2\pi\sigma_p^2} } \Big[  p_s \int_{-\infty}^{{\rm log}~N_{j}^{u}}   {\rm exp} \left(- \frac{ ({\rm log}~\bar{N}-{\rm log}~N)^2}{2\sigma_p^2} \right) d~{\rm log}~\bar{N}      \\
   +(1-p_s)  \int_{{\rm log}~N_{j}^{u}}^{\infty} {\rm exp} \left(- \frac{ ({\rm log}~\bar{N}-{\rm log}~N)^2}{2\sigma_p^2} \right) d~{\rm log}~\bar{N} \Big]~. 
 \end{split}
 \end{equation}
 where $p_s$ corresponds to 0.9973 for $3\sigma$ upper limits. However, this can be approximated as Eqn.~\ref{eq:logl} for all practical purposes.

 We assume uniform priors on the free parameters and construct the posteriors of individual model parameters (i.e., ${\rm log}~N_0$, $\alpha$, and $\sigma_p$) based on Markov chain Monte Carlo (MCMC) samples, which is implemented using the {\sc Emcee} package \citep[]{Foreman-Mackey_2013}. The best-fit parameters are listed in Table~\ref{tab:best-fit-NHIprof}. The best-fit model and its $1\sigma$ uncertainty are shown in Fig.~\ref{fig:nprof_fit_minDn} with the red solid line and shaded region, respectively. The best-fit $N(\HI)$--profile for the isolated, SF galaxies in our sample is characterized by a power-law slope of $\approx-3$ and a normalization of $\log N_0 = 14.0$ with a large intrinsic scatter of $\approx1.5$~dex,  as indicated by the magenta shaded region in Fig.~\ref{fig:nprof_fit_minDn}. Note that these values are consistent, within the allowed range of uncertainties, with those derived for the full sample, including the non-isolated galaxies (Table~\ref{tab:best-fit-NHIprof}).

\begin{table}
\centering
\caption{Best-fit model parameters for the \Nhi-profiles}   
\label{tab:best-fit-NHIprof} 
\begin{tabular*}{\linewidth}{l@{\extracolsep{\fill}}ccr}  
\hline
Sample  & ${\rm log}~(N_0/{\rm cm}^{-2})$ & $\alpha$ & $\sigma_p$  \\
\\  
\hline
Isolated, star-forming & $14.0_{-0.3}^{+0.3}$ & $3.1_{-1.0}^{+1.2}$  & $1.5^{+0.3}_{-0.3}$ \\ \medskip 
All, star-forming & $14.1_{-0.2}^{+0.2}$ & $3.5_{-0.9}^{+1.1}$  & $1.6^{+0.2}_{-0.2}$ \\
\hline
\end{tabular*}
\justify
Notes-- For each parameter, the best-fit estimate represents the median value of the corresponding posterior distributions. The quoted uncertainties are 68\% confidence intervals of the posterior probability distributions.
\end{table}

The best-fit $N(\HI)-$profile, constrained with our \HI\ measurements at $D/R_{\rm vir}\gtrsim0.3$ can be subsequently used to constrain the circumgalactic \HI\ mass, $M(\HI)$. The $M(\HI)$ within $D_{\rm min}$ and $D_{\rm max}$ is given by:
 \begin{equation}
     M(\HI) = m_{\rm H} \times \int_{D_{\rm min}}^{D_{\rm max}} N(\HI)(D) 2\pi DdD~.    
 \end{equation} 
With a variable transfer of $D\rightarrow D/R_{\rm vir}$, this can be written as: 
 \begin{equation}
     M(\HI) = m_{\rm H} \times 2\pi \left<R_{\rm vir}^2\right>\int_{0.3}^{1} N(\HI)(x) xdx~,   
 \end{equation} 
where $x=D/R_{\rm vir}$, $m_{\rm H}$ is mass of a hydrogen atom, and $\left(D/R_{\rm vir}\right)_{\rm min}=0.3$ and $\left(D/R_{\rm vir}\right)_{\rm max}=1$. The lower limit of integration is motivated by the scarcity of MUSEQuBES measurements at $D/R_{\rm vir}\lesssim 0.3$. For the median $R_{\rm vir}$ of $\approx 76$~pkpc, the average ${\rm log}_{10}\left[M(\HI)/{\rm M_{\odot}}\right]$ turns out to be $4.9^{+0.4}_{-0.4}$ for our sample. Extrapolating the best-fit column density profile down to 0.1$R_{\rm vir}$ results in ${\rm log}_{10}\left[M(\HI)/{\rm M_{\odot}}\right] = 5.3^{+0.6}_{-0.6}$ within $D/R_{\rm vir}=0.1-1$.

In Section~\ref{sec:Mhi_musequbes}, we present a detailed discussion on the dependence of $M(\HI)$ with stellar mass, and compared the circumgalactic $M(\HI)$ with the galactic measurements obtained from 21-cm studies.

\begin{figure*}
    \centering
    \includegraphics[width=0.5\linewidth]{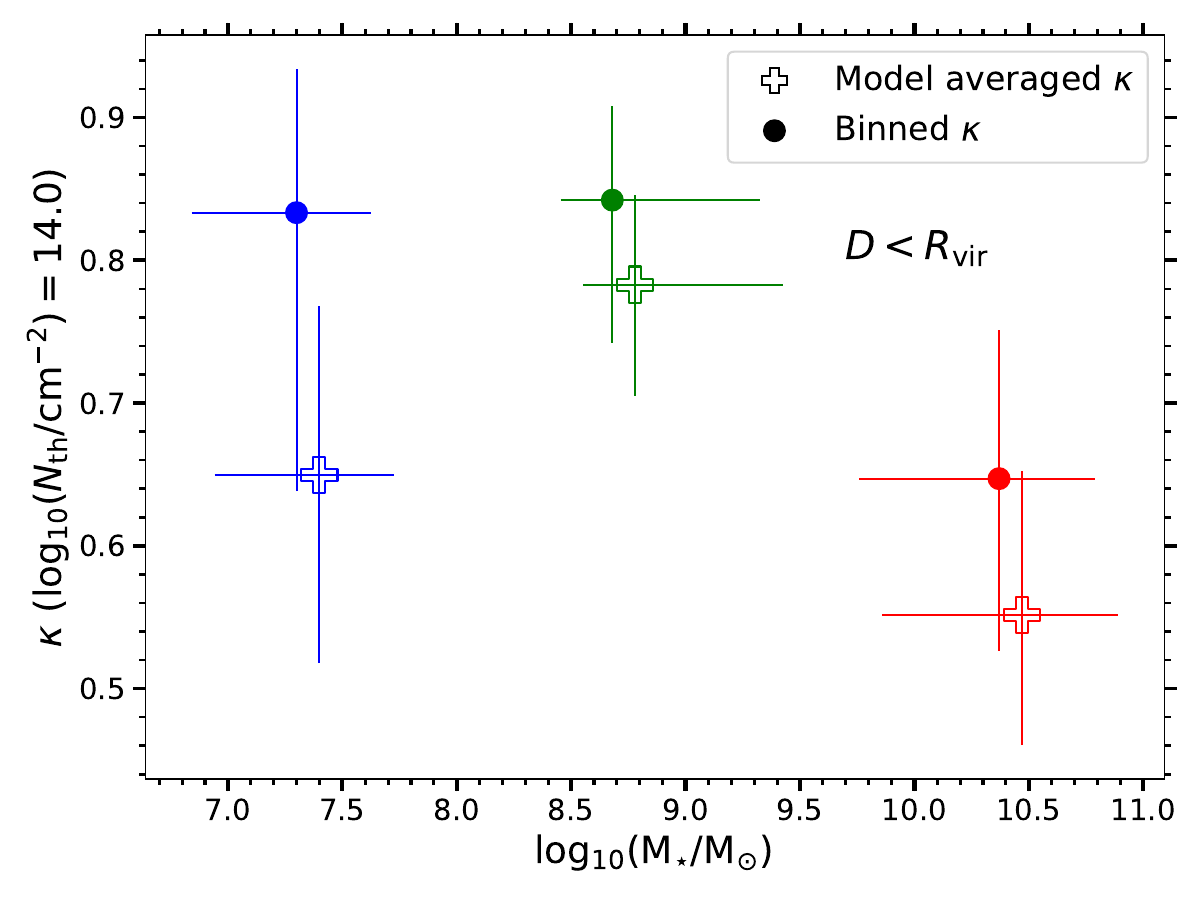}%
    \includegraphics[width=0.5\linewidth]{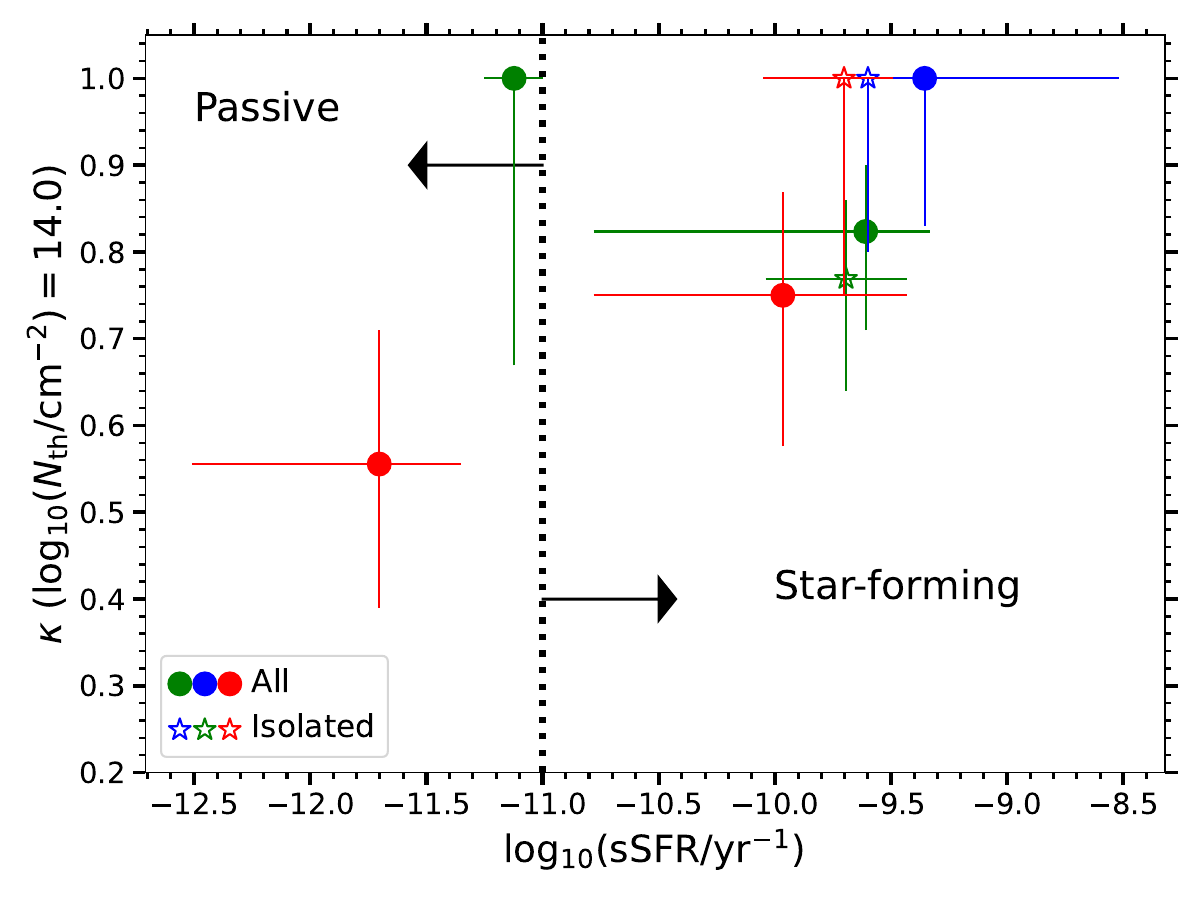}      
    \caption{{\tt Left:} The binned \HI\ covering fraction ($\kappa$) for a  threshold \logNhi~$=14$ plotted against $M_{\star}$ using galaxies with  $D<R_{\rm vir}$. The x and y error bars indicate the 68\% confidence intervals on \logm\ and $\kappa$. The open plus symbols indicate the area-averaged $\kappa$ within $R_{\rm vir}$ obtained from the best-fit $\kappa-$profiles (see text). 
    {\tt Right} $\kappa$ for threshold \logNhi~$=14$ shown for star-forming and passive galaxies with $D<R_{\rm vir}$. The color coding denotes the stellar mass bins shown in the left panel. 
    The solid filled circles and open star symbols indicate $\kappa$ for all galaxies and isolated galaxies, respectively. }
    \label{fig:N_fc_invir}
\end{figure*}


\begin{table}
\centering
\caption{Summary of $\kappa$ measurements within $R_{\rm vir}$ for different subsamples}   
\label{tab:fc_all}
\begin{tabular*}{\linewidth}{l@{\extracolsep{\fill}}ccc}  
\hline
  & Low-mass & Intermediate-mass & High-mass  \\ 
 \logm &  $6-8$ & $8-9.5$ & $9.5-11.4$ \\ \hline \medskip 
All & $0.83_{-0.19}^{+0.10}$ (6) & $0.84_{-0.10}^{+0.07}$ (19) & $0.65_{-0.12}^{+0.10}$ (17) \\ \medskip   
All-SF & $1.00_{-0.20}^{+0.00}$ (5) & $0.82_{-0.11}^{+0.08}$ (17) & $0.75_{-0.17}^{+0.12}$ (8)  \\ \medskip 
All-E & ...(0) & $1.00_{-0.33}^{+0.00}$ (2) & $0.56_{-0.18}^{+0.17}$ (9) \\ \medskip 
Isolated-SF & $1.00_{-0.20}^{+0.00}$ (4) & $0.77_{-0.13}^{+0.09}$ (13) & $1.00_{-0.25}^{+0.00}$ (3)  \\ \medskip 
Isolated-E & ... (0) & ... (1) & ... (2)  \\ 
\hline
\end{tabular*}
\justify
Notes-- The numbers and error bars indicate $\kappa$ for a threshold $N(\HI)$ of $10^{14}~{\rm cm}^{-2}$ and 68\% Wilson-score confidence intervals. The numbers in parentheses indicate the number of contributing galaxies.  
\end{table}

\subsection{ \texorpdfstring{$\HI$}{HI} covering fraction} 

\label{sec:hi_cov}

The covering fraction ($\kappa$), defined as the fraction of sightlines with column density above a given threshold, serves as a quantitative metric for assessing the spatial inhomogeneity of the absorbing medium in the plane perpendicular to the LOS. In the left panel of Fig.~\ref{fig:N_fc_invir}, we show the \HI\ covering fractions for three $M_{\star}$ bins with \logm~$<8$ (median 7.3), $8\leq~$\logm$~<9.5$ (median 8.7) and \logm~$\geq9.5$ (median 10.1) for a threshold\footnote{ We have discarded measurements for which the spectral sensitivity, defined as the $N(\HI)$ upper limit in the line-free region of the quasar spectra, is higher than the threshold.} \logNhi~$=14$ in blue, green, and red colored circles. Only galaxies with $D<R_{\rm vir}$ are considered here. The results are further tabulated in Table \ref{tab:fc_all}.

Instead of the binned $\kappa$ within the $R_{\rm vir}$, the $\kappa-$profile as a function of $D/R_{\rm vir}$ can be modeled based on the unbinned data following the formalism introduced by \citet[][see also \citet{Dutta2_2025}]{Schroetter_2021}. Briefly, the detection probability of an \HI\ absorber above a threshold column density is modeled with a slightly modified logistic function, i.e., 
\begin{equation}
\label{eq:fc_D}
    p[Y=1] = \frac{\kappa_0}{1+e^{t}} + \kappa_1~,  
\end{equation}
in order to produce a smooth transition between $\kappa_1$ and $\kappa_0 + \kappa_1$. Here, $p[Y=1]$ denotes the probability of detection above the threshold (essentially the covering fraction). The parameter $t$ is taken to be a function of the independent variable $D/R_{\rm vir}$. We adopted 
\begin{equation}
\label{eq:fc_tparam}
    t=\alpha({\rm log}_{10}(D/R_{\rm vir})-\beta)~, 
\end{equation}
where $\alpha$ and $\beta$ are two free parameters. Here, $\alpha$ describes the slope of the covering fraction profiles, and $\beta$ represents the distance (i.e., ${\rm log}_{10}(D/R_{\rm vir})$) at which the covering fraction, $p[Y=1] =\kappa_0/2 + \kappa_1 (\equiv \kappa_{50})$. Clearly, $\beta$ is analogous to the zero point at 50 percent covering fraction. At small and large $D/R_{\rm vir}$, the covering fraction ($p[Y=1]$) converges to $\kappa_0+\kappa_1$ and $\kappa_1$, respectively. Both $\kappa_0$ and $\kappa_1$ are treated as free parameters in our model. The parameter space is sampled to generate the dichotomous observable (1 if detected and 0 otherwise) based on the Bernoulli distribution. We use the {\sc pymc3} for the {\sc mcmc} sampling in order to obtain the best-fit parameters. 

We use the \HI\ measurements in the three stellar mass bins of Fig.~\ref{fig:N_fc_invir} to obtain the corresponding best-fit $\kappa-$profiles  (shown in Appendix Fig.~\ref{fig:fc_prof_Dn_model}). The notable lack of any anticorrelation in the highest stellar mass bin is reflective of the lack of $N(\HI)-D/R_{\rm vir}$ anticorrelation reported for massive and passive galaxies in section 4.2. The best-fit $\kappa-$profiles can be used to obtain an area-averaged mean $\kappa$ ($\left<\kappa\right>$)\footnote{Essentially, $\left<\kappa\right>$ is $\int_0^{1} \kappa(x) 2xdx$ where $x=D/R_{\rm vir}$.} within $R_{\rm vir}$. The $\left<\kappa\right>$ for the three stellar mass bins are shown with open plus symbols in the left panel of Fig.~\ref{fig:N_fc_invir}.

The covering fraction shows a tentative increase from the low mass to the intermediate stellar mass bin before undergoing a considerable decline at the highest $M_{\star}$ bin. This trend aligns qualitatively with \citet[]{Dutta1_2024}, who reported a significant decline with mass in stacked \lya\ absorption equivalent width within $\approx R_{\rm vir}$ for galaxies in a similar $M_{\star}$ range. \citet[]{Ramona_2024} also found that more massive galaxies are less likely to harbor large amounts of \HI\ within $\approx120$ pkpc.

\citet[]{Dutta1_2024} further reported a significant suppression in stacked \lya\ absorption equivalent width for passive galaxies compared to their star-forming counterparts within the virial radius. To verify this, we divided the galaxies in each $M_{\star}$ bin into star-forming and passive subsamples. The binned $\kappa$ within $R_{\rm vir}$ for star-forming and passive subsamples for each stellar mass bin is shown against sSFR in the right panel of Fig.~\ref{fig:N_fc_invir}. The color coding is as in the left panel. Additionally, we show $\kappa$ for isolated star-forming galaxies in each bin with open star symbols.  Note that here we have refrained from the logistic function analysis due to the small sample size.

The suppression in $\kappa$ for the high-mass galaxies is seen in both star-forming and passive galaxies. However, the isolated and star-forming galaxies in the highest mass bin exhibit near-unity $\kappa$ $\left(\kappa=1.00^{+0.00}_{-0.25}\right)$. This is consistent with the findings of \citet[]{Tumlinson_2013} for star-forming galaxies in a similar stellar mass range. However, $\kappa$ for star-forming galaxies declines to $0.75^{+0.12}_{-0.17}$ when the isolation condition is not applied. The value of $\kappa$ decreases significantly for the passive galaxies $\left(\kappa=0.56^{+0.17}_{-0.18}\right)$. However, due to the small sample size, the isolation condition could not be imposed on the passive galaxies. The significant suppression in $\kappa$ in the highest \logm\ bin is thus partly contributed to by both the environment and the lack of star-forming activity of the host galaxy. The passive galaxies in the intermediate mass bin show near-unity covering fraction. However, this high covering fraction is uncertain due to the small sample size (only 2 galaxies).

At this stage, we note that a fraction of passive galaxies with detected \HI\ absorption have a star-forming neighbor (at a smaller $D$ but higher $D/R_{\rm vir}$) that could be responsible for the observed \HI\ gas. In particular, two passive galaxies, (one belonging to the intermediate mass bin having \logm=9.49 and the other belonging to the high-mass bin having \logm=11) associated with two of the strongest \HI\ absorption (pLLS with \logNhi$\approx17$ and 18) have a low-mass, star-forming neighbor with a smaller $D$ and smaller LOS velocity separation. Discarding the massive, \logm=11 galaxy from the analysis
 further suppresses $\kappa$ for passive galaxies of the highest mass bin $\left(\kappa=0.50^{+0.16}_{-0.17}\right)$. We note that such strong \HI\ absorption associated with passive galaxies is also reported in \citet[]{Tumlinson_2013}, a fraction of which could be a consequence of missing star-forming dwarf galaxies in their sample.

The presence of cool gas in the halo of passive galaxies has been widely reported in the literature \citep[e.g.,][]{Thom_2012, Chen_2018}. 
\citet[]{Borthakur_2016} argued that these cool clouds near massive red galaxies are transient with a lifetime shorter than a halo crossing time. Using semi-analytical parametric models, \citet[]{Afruni_2019} argued that the cool CGM in these galaxies is the manifestation of cosmological accretion of gas into their dark matter halos, which likely evaporates during their journey due to the interaction with the hot gas. The inability to reach the central regions and feed the galaxy's star formation may explain why these passive objects are no longer forming stars despite the abundance of gas in their CGM.

\begin{figure*}
    \centering
    \includegraphics[width=1\linewidth]{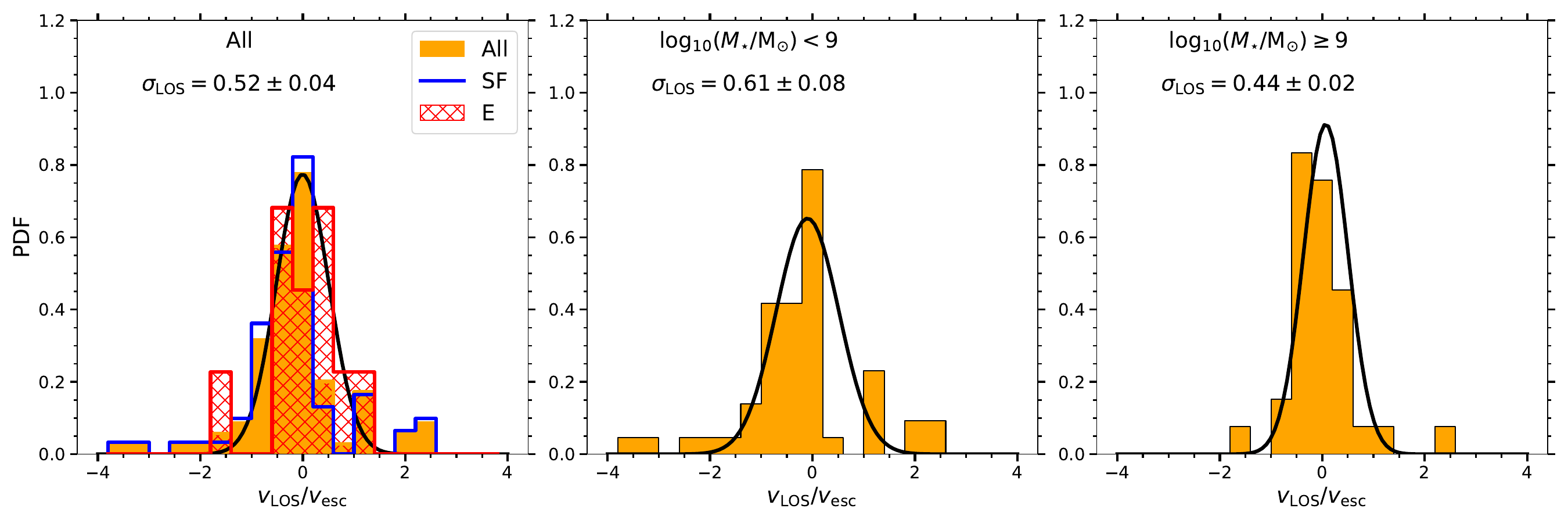}  
    \includegraphics[width=1\linewidth]{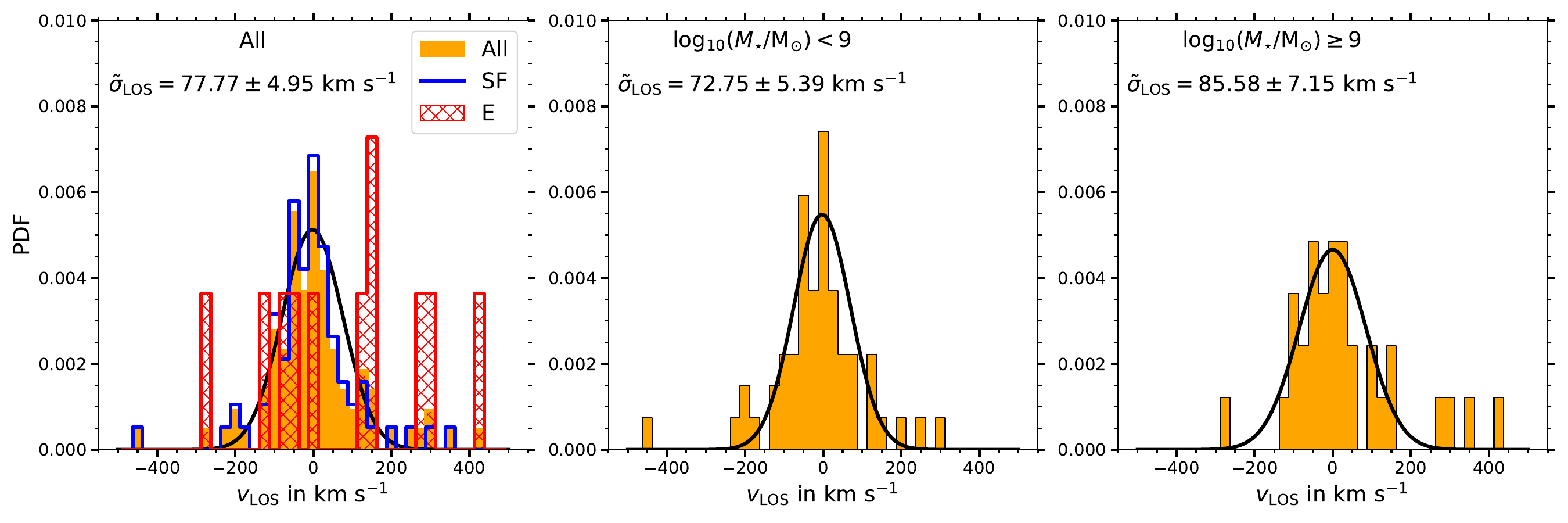}    
    \caption{{\tt Top-Left:} The probability density function (PDF) of $v_{\rm esc}$-normalized LOS velocity separation of \HI\ systems with respect to the host galaxy redshift shown in orange histograms. The black solid line shows the best-fit Gaussian profile to the observed PDF. The PDF for the \HI\ systems associated with SF and E galaxies are indicated by the blue and red histograms, respectively. 
    {\tt Top-Middle:} PDF of $v_{\rm esc}$-normalized LOS velocity separation of \HI\ systems with respect to the host galaxy redshift for galaxies with \logm$\leq9$ shown with orange histograms. The best-fit Gaussian profile is shown with a black solid line. {\tt Top Right:} Same as {\tt Top-Middle} but for galaxies with \logm$>9$. 
    {\tt Bottom} panels are the same as the top panels, but show the PDF for unnormalized LOS velocity separations of \HI\ systems with respect to the host galaxy redshift.}
    \label{fig:dv_dist}
\end{figure*}

\subsection{LOS kinematics of the \texorpdfstring{$\HI$}{HI} absorbers}  

\label{sec:kin_vcent}

In this section, we analyze the LOS kinematics of \HI\ in the CGM of MUSEQuBES galaxies. Consistent with the previous analyses, we use the galaxies with the smallest $D/R_{\rm vir}$ as hosts. 


In the top left panel of Fig.~\ref{fig:dv_dist}, we plot the probability density function (PDF) of LOS velocity offset ($v_{\rm LOS}$) between the \HI\ column-density weighted redshift of the associated \HI\ system and the galaxy redshift normalized by the local escape velocity of the host galaxy ($v_{\rm esc}$) with orange histograms. We assumed a NFW density profile for the host halo mass, with local escape velocity at a given distance $D$ given by $v_{\rm esc}(D)=\sqrt{(2GM_{\rm halo}/D) {\rm ln}[1+c(D/R_{\rm vir})]/[{\rm ln}(1+c)-c/(1+c)]} $. For a given halo-mass, the redshift-dependent concentration parameter $c$ is obtained using the
{\sc Commah} package \citep[]{Correa_2015}.

The PDF can be modeled with a single-component Gaussian as: 
\begin{equation}
     p(x) = \sqrt{1/2\pi\sigma_{\rm LOS}^2}~{\rm exp}\left( -(x-\mu)^2/2\sigma_{{\rm LOS}}^2 \right)~,
\end{equation}
where $x= v_{\rm LOS} / v_{\rm esc}$. $\mu$ and $\sigma_{{\rm LOS}}$ are kept as free parameters. The best-fit Gaussian is shown with a black solid line. The best-fit centroid ($\mu$) is consistent with zero ($0.00\pm0.05$). The best-fit  $\sigma_{{\rm LOS}}\approx 0.52$ indicates that the \HI\ systems are predominantly bound to the host galaxies, as they do not reach the escape velocity. The best-fit $\sigma_{{\rm LOS}}$ reduces to $\approx0.39\pm0.02$ when the analysis is restricted to absorbers at $D/R_{\rm vir}<1$ only. This suggests a higher fraction of absorbers exceeding the escape velocity at larger $D/R_{\rm vir}$. This will be further explored in Sect.~\ref{sec:bound-frac}.

\begin{figure*}
    \centering
    \includegraphics[width=0.5\linewidth]{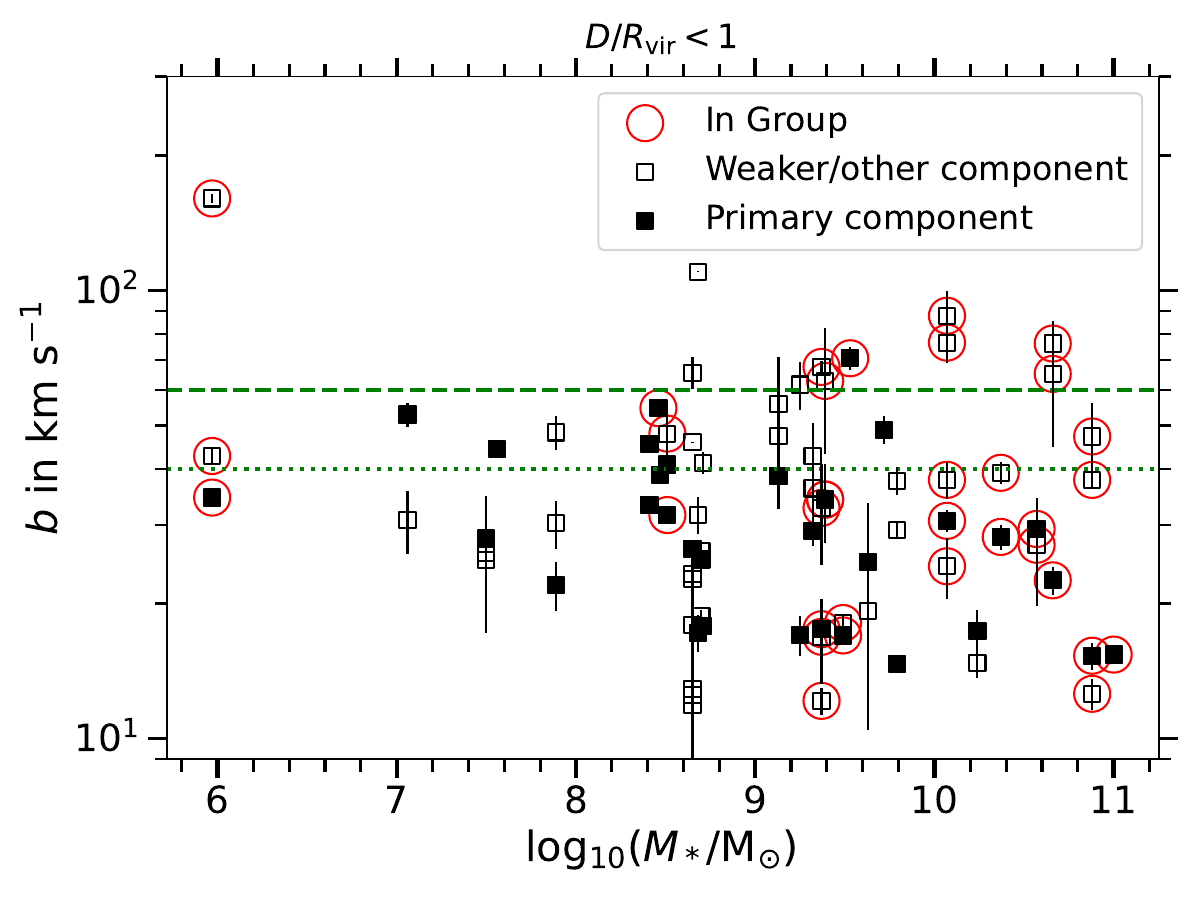}%
    \includegraphics[width=0.5\linewidth]{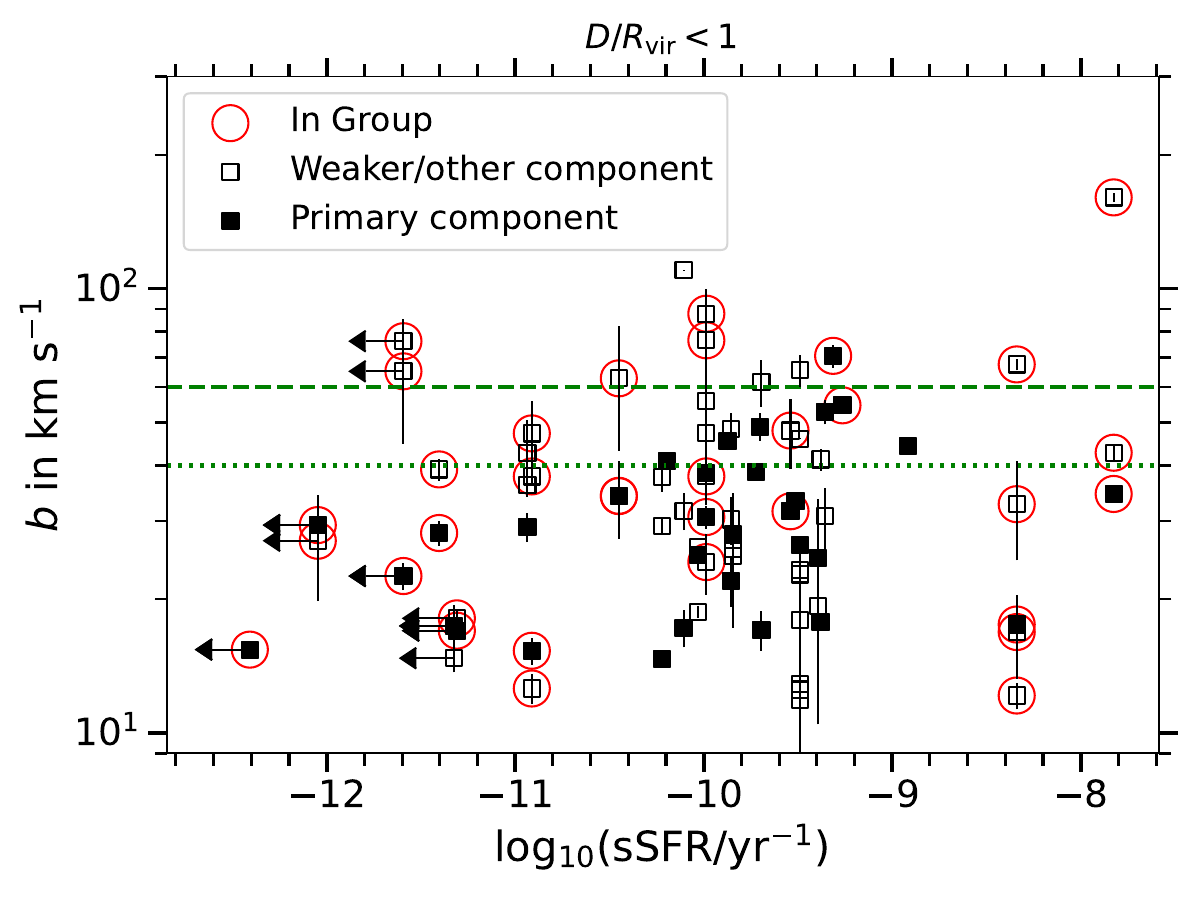}
    \caption{{\tt Left:} Doppler-$b$ parameter plotted against the stellar mass of the host galaxies with $D<R_{\rm vir}$. The solid black and open black squares represent the primary (strongest) and `other' components of a system. A moderate and significant anticorrelation is observed between $b$ and $M_{\star}$ for the primary components. The red circular envelopes indicate components associated with the group galaxies. The green dashed and dotted lines indicate $b=60$ and $40$ \kms, respectively. {\tt Right:} Doppler-$b$ parameter plotted against the sSFR of the host galaxies with $D<R_{\rm vir}$. Other details are similar to those in the left panel. The primary components have a moderate and significant correlation between $b$ and sSFR. }
    \label{fig:b_invir}
\end{figure*}

The majority of \HI\ systems used in this work are hosted by star-forming galaxies. Their PDF is shown by the blue histogram. The handful of \HI\ systems hosted by the passive galaxies (red hatched histograms) show somewhat larger $\mu$ ($0.32\pm0.09$) and $\sigma_{\rm LOS}$ ($0.61\pm0.07$) values. However, both the 2-sample KS test and the Anderson-Darling tests can not conclusively distinguish the normalized velocity distributions of these two populations ($p>0.25$).

The bottom left panel shows the PDF for unnormalized LOS velocity separation. Other details are similar to the top left panel. The velocity dispersion of the best-fit Gaussian to the PDF of the full sample, $\tilde{\sigma}_{\rm LOS}$, is $\approx78$ \kms. Unlike the normalized velocity distributions, both the KS and the Anderson-Darling tests indicate a significant difference between PDFs for the star-forming and passive galaxies ($p\approx0.02$ in both cases).

In the {\tt top-middle} and {\tt top-right} panels of Fig.~\ref{fig:dv_dist}, we show the PDF for two stellar mass bins with \logm~$\leq9$ (median \logm~$=8.2$) and \logm~$>9$ (median \logm~$=9.5$), respectively. The bottom middle and bottom right panels show the PDF for unnormalized LOS velocity separation. The best-fit $\sigma_{\rm LOS}$ of the $v_{\rm esc}-$ normalized PDF of the lower mass subsample is observed to be higher ($0.61\pm0.08$) compared to the higher mass galaxies ($0.44\pm0.02$). However, when restricted to the absorbers at $D/R_{\rm vir}<1$ only, we do not observe any significant difference in the $\sigma_{\rm LOS}$ of the $v_{\rm esc}-$normalized velocity distribution ($\approx0.40\pm0.05$ and $\approx0.39\pm0.03$ for low- and high-mass subsamples, respectively). The $\tilde{\sigma}_{\rm LOS}$ of the unnormalized velocity distribution for the low-mass and high-mass subsamples, $\approx 73\pm5$~\kms\ and $\approx 86\pm7$~\kms,  respectively, are broadly consistent with each other. However, a significant difference ($\approx2.5\sigma$) emerges when the sample is restricted to absorbers with $D<R_{\rm vir}$, yielding $\tilde{\sigma}_{\rm LOS} \approx 53\pm5$~\kms\ and $\approx 85\pm12$~\kms\ for the low- and high-mass subsamples, respectively. The tighter velocity-clustering for low-mass galaxies at $D<R_{\rm vir}$ suggests that a fraction of the absorbers may be unbound at larger impact parameters. We investigate this further in Sect.~\ref{sec:bound-frac}.

Previously, \citet[]{Liang_14} reported the necessity of a 2nd broad Gaussian component to fit the LOS velocity distribution of \lya\ absorbers around galaxies. They argued that the broader component traces the gas in the large-scale structures around the galaxies. \citet[]{Dutta1_2024} showed that such broad wings in the LOS velocity distribution at large impact parameters can be explained by a simple 2-point galaxy-absorber correlation function, tracing the weak \HI\ absorbers in the large-scale structures around the galaxies. We find that the PDF of the unnormalized $v_{\rm LOS}$ is well fit by a two-component Gaussian model, with the inclusion of a broader Gaussian component supported at $4.5\sigma$ significance based on an $F-$test.  In contrast, for the PDF of $v_{\rm esc}-$normalized $v_{\rm LOS}$, we find no statistically significant evidence for a second Gaussian component; the $F-$test yields a significance of $<2\sigma$ in favor of the two-component model.

\subsection{Doppler \texorpdfstring{$b$}{b}-parameters of the \texorpdfstring{$\HI$}{HI} absorbers}

The internal kinematics of the individual \HI\ components can be quantified by the $b$-parameter of individual Voigt components. In the left and right panels of Fig. \ref{fig:b_invir}, we plot the $b$-parameter of individual components against the stellar mass and sSFR of the host galaxies. We restrict to galaxies with $D<R_{\rm vir}$ for this exercise\footnote{We do not observe any correlation between $b-$parameter and $D/R_{\rm vir}$.}. The solid and open black squares represent the primary (strongest) and `other' (weaker) components of a {\it system} associated with the host galaxy, respectively.

We do not see any correlation with $M_{\star}$ or sSFR with $b$-parameter for absorption components detected within $R_{\rm vir}$. However, considering only the strongest component of a given system (solid squares), a weak but significant anticorrelation (correlation) is observed between $b$ and $M_{\star}$ (sSFR) with $\tau=-0.35,~p=0.005$ ($\tau=0.31,~p=0.01$).

The \HI\ $b$-parameter can be decomposed into thermal and non-thermal (i.e., turbulent) components:  
\begin{equation}
\label{eq:b_param}
    b^2 = 2kT/m_{\rm H} + b_{\rm turb}^2~,
\end{equation}
where $b_{\rm turb}$ represents the contribution from non-thermal motions and $k$ and $m_{\rm H}$ are the Boltzmann constant and mass of a hydrogen atom. The correlation between the $b$-parameter and host galaxy sSFR (Fig.~\ref{fig:b_invir}, right) may result from increased $b_{\rm turb}$ in high-sSFR galaxies, likely due to turbulence injected into the CGM by star-formation-driven outflows. In this scenario, a narrow (broad) $b$-parameter may indicate gas tracing a less (more) turbulent medium hosted by low-sSFR (high-sSFR) galaxies. The fact that the most massive galaxies in our sample preferentially exhibit lower sSFR suggests that the observed anticorrelation between $b$ and $M_{\star}$ may be a consequence of an underlying $b-$sSFR correlation.

Ignoring the non-thermal part, a conservative upper limit on the gas temperature can be obtained from Eqn.~\ref{eq:b_param} as:  
\begin{equation}
    T_{\rm max}\approx 10^{5}\left(\frac{b}{40~{\rm km~s^{-1}}}\right)^2 {\rm K}~. 
\end{equation}
The broad \lya\ absorbers (or BLAs), defined as \lya\ absorbers with $b\geq40$~\kms\ \citep[]{Richter_2004, Lehner_2007}, are believed to trace the $\gtrsim10^{5}$~K warm-hot intergalactic medium (WHIM). Among the 26 BLA components in Fig.~\ref{fig:b_invir}, 13 are associated with galaxies having \logm$< 9$, and the remaining 13 with galaxies having \logm$\geq 9$. Of these, 14 BLAs are hosted by isolated galaxies and 12 by group galaxies.

A non-negligible fraction of \HI\ components in this work have $b\gtrsim60$ \kms, inferring a $T_{\rm max}\gtrsim10^{5.4}$~K, which is broadly consistent with the virial temperature of halos hosting galaxies with \logm$\gtrsim9$. Thus, these BLAs may originate from collisionally ionized ambient halo gas in massive galaxies, provided that non-thermal contributions to the $b$-parameter are negligible.

We find that 8 out of the 11 components with $b\gtrsim60$~\kms\ in Fig.~\ref{fig:b_invir} are hosted by galaxies with \logm~$\geq9$. 
Except for one, these BLAs are not the primary components, i.e., each has a higher \Nhi\ companion(s). Hence, the observed anticorrelation between $b$ and $M_{\star}$, based on the primary (i.e., the strongest) components, is not in conflict with this finding. Seven out of these 8 BLA components have \OVI\ coverage within COS spectra.{\footnote{The \OVI\ for the other BLA component falls within the geocoronal \lya\ emission.}} All of these have an associated \OVI\ absorber. Using \HI\ and \OVI\ components aligned in velocity space (within 10~\kms), we constrained the temperature of 4 out of 7 BLAs to be $\approx10^{5.4}$~K, suggesting a collisionally ionized origin for these absorbers. \footnote{However, we note that the robustness of these temperature estimates depends on the assumption that \HI\ and \OVI\ trace the same gas phase.} The median $N(\HI)\approx10^{14.6}~{\rm cm}^{-2}$ of these absorbers, combined with an \HI\ ion fraction of $\sim2\times10^{-6}$ in CIE at $T=10^{5.4}$~K \citep[for a sub-solar metallicity;][]{Oppenheimer_2013} implies a total hydrogen column density of $N(\rm H) \approx10^{20}~{\rm cm}^{-2}$. Assuming that the absorbing column spans at least the size of the halo (i.e., $\approx100$~pkpc), the inferred total hydrogen density is $n_{\rm H} \approx 4.5 \times 10^{-4}~{\rm cm}^{-3}$-- a value typical of virialized CGM gas.

\section{Discussion} 
\label{sec:discussion}

\subsection{Strong \texorpdfstring{$N(\HI)$}{N(HI)} outside \texorpdfstring{$R_{\rm vir}$}{R(vir)} of isolated, SF galaxies}

In Fig.~\ref{fig:nprof_fit_minDn}, we showed the best-fit $N(\HI)-$profile for the isolated and star-forming MUSEQuBES galaxies. Although most of the data points follow the best-fit profile within the intrinsic scatter, there are a few outliers. Particularly, the four partial LLS with $N(\HI)>10^{16}~{\rm cm}^{-2}$. Two out of these four galaxies are separated by 506~\kms\ along the LOS in the same sightline. They are selected as `isolated' galaxies as this just exceeds our adopted linking velocity of $\pm500$~\kms\ for defining `groups'.

Among the other two galaxies, one is at $z\approx 0.226$. Despite having a low-mass (\logm~$=7.6$) and a large impact parameter ($D/R_{\rm vir}>1$), this galaxy exhibits low-, intermediate- and high-ionization metal line absorbers, which is typical of $\approx L_{\star}$ galaxies at $D\lesssim R_{\rm vir}$. We point out here that the MUSE FoV of $1'\times1'$ restricts the maximum impact parameter to $\approx150$ pkpc at this redshift. It is possible for a massive galaxy (\logm$>10$) having $R_{\rm vir}\geq 200$ pkpc to exist just outside the MUSE FoV within $\pm500$ \kms\ of the absorption system.

The expected number of massive galaxies outside the MUSE FoV within $\pm500$~\kms\ of the absorber can be obtained from the stellar mass function and galaxy-absorber clustering. The number of galaxies  within $R_{\rm max}$ to 500~pkpc of the quasar and within $\pm500$~\kms\ of the absorber is given by: 
\begin{equation}
  N_{\rm exp} = \int_{-R_z}^{+R_z}  \int_{R_{\rm max}}^{500~{\rm pkpc}} \bar{n}(M)(1+\xi(r)) 2\pi r_{\perp} dr_{\perp}dr_{z}~,
\end{equation} 
where $R_{z}$ denotes the distance corresponding to the velocity interval of $\pm500$~\kms\ w.r.t absorber redshift, $\xi(r)$ is the galaxy-absorber two-point correlation function, $R_{\rm max}$ is the maximum MUSE FoV at the absorber redshift and $\bar{n}(M)$ is the number density of galaxies with \logm$>M$ obtained from \citet[][]{Tomczak_2014}.

 Using the best-fit galaxy-absorber clustering for star-forming galaxies and \HI\ absorbers (${\rm log}_{10}(N(\HI)/{\rm cm}^{-2})>14$) from \citet[]{Tejos_14}, the number of expected galaxies with \logm~$\geq10$ within 500~pkpc from the quasar (but outside the MUSE FoV at $z\approx0.226$) and within $\pm500$~\kms\ of the absorption system is only $\approx 0.4$. Although the probability is relatively small, this possibility can not be ruled out. The other galaxy with \logm$\approx9.5$ exhibiting pLLS is at a redshift of $0.7$, where this effect is smaller (FoV = 430 pkpc x 430 pkpc at this redshift). The galaxy shows metal absorption for intermediate ions (such as \OIII, \CIII), but unlike the other two cases, this does not show any detectable low-ionization lines, indicative of its origin in a fairly low-density region. \citet[]{Lehner_2019} estimated a metallicity of $[{\rm X/H}]\approx-3$ for this pLLS with a high ionization parameter of $\log~U\approx-0.65$ which is indeed indicative of a fairly low-density gas.

In the CUBS survey, \citet[]{Cooper_2021} found that pLLS can trace complex, multiphase circumgalactic gas that exhibits significant variations in chemical abundances and density on small spatial scales, and arises from a diverse galaxy environments \citep[see also][]{Zahedy_2021}. Of the two pLLS studied in their work, one is associated with a massive, star-forming galaxy at $D\approx55$ pkpc and the other resides in an overdense environment of 11 galaxies including a luminous red galaxy. The MUSEQuBES survey at high $z$ recently reported that two pLLSs with primordial chemical composition arising from cosmic filaments, traced by seven Ly$\alpha$ emitters \citep[][]{Banerjee_2025}.

In summary, the strong \HI\ absorbers observed beyond the virial radius of star-forming and isolated galaxies in our sample may originate from a variety of physical environments, including galaxy overdensities with undetected member galaxies, or low-density, metal-poor gas tracing cosmic filaments. Such diversity in the origin of \HI\ is one of the key contributors to the appreciable scatter in the observed $N(\HI)$-profiles.

\subsection{\texorpdfstring{$M(\HI)$}{M(HI)} measurements for MUSEQuBES galaxies}  
\label{sec:Mhi_musequbes}

\begin{figure}
    \centering
    \includegraphics[width=1\linewidth]{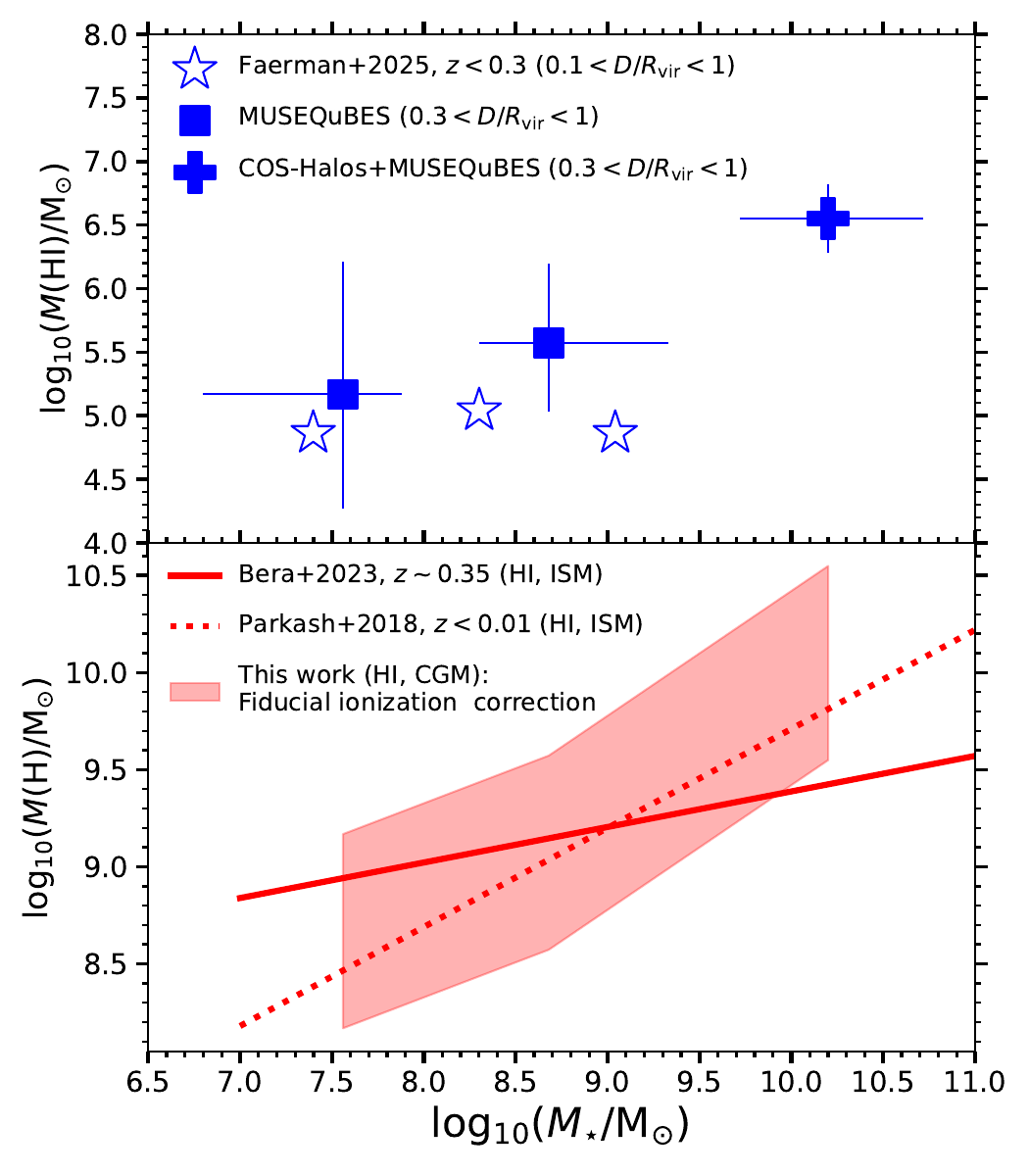}
    \caption{{\tt Top:} The mean mass of cool \HI\ in the outer CGM ($\approx 0.3-1~R_{\rm vir}$) of star-forming galaxies shown as a function of $M_{\star}$ for three stellar mass bins with solid blue squares and plus symbols. The highest mass bin contains galaxies from both MUSEQuBES and COS-Halos. The open star symbols represent $M(\HI)$ estimates of \citet{Faerman_2025}.
    {\tt Bottom:} The red stripe indicates the total hydrogen mass, $M(\mathrm{H})$, obtained by applying a fiducial ionization correction of $10^{-4}-10^{-3}$ to the measured $M(\HI)$ of the {\tt top} panel. 
    The solid and dotted blue lines show the Hydrogen mass estimated from the 21-cm measurements from \citep[][at $z<0.01$]{Parkash_2018} and \citep[][at $z\sim0.35$]{Bera_2023}, respectively.}
    \label{fig:MHI}
\end{figure}

\begin{figure*}
    \centering
    \includegraphics[width=1\linewidth]{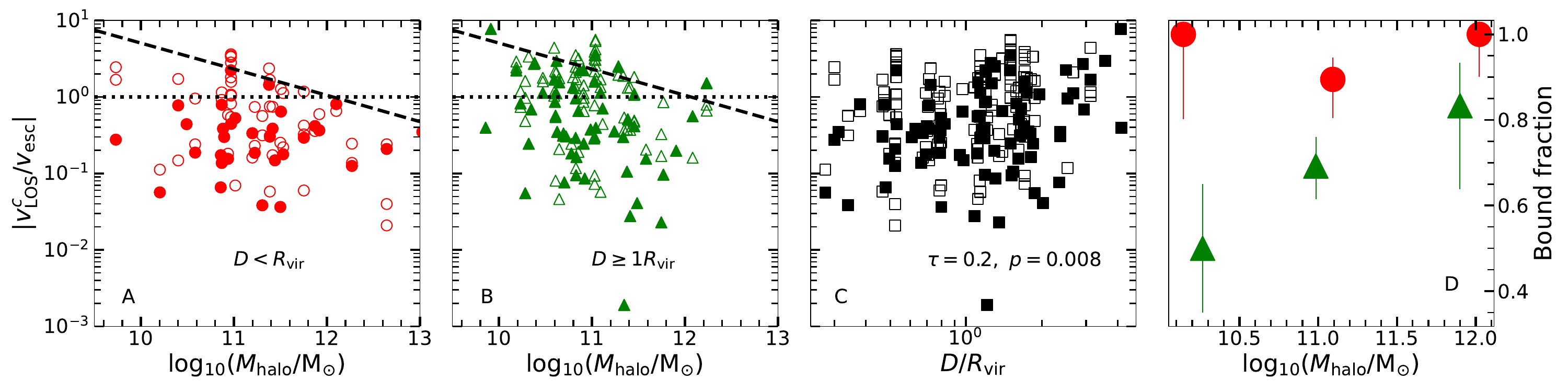}
    \caption{{\tt Panel A:} The velocity of individual \HI\ components ($v_{\rm LOS}^{c}$) scaled by local escape velocity ($v_{\rm esc}$) is plotted against the halo mass of the host galaxy for the MUSEQuBES sample with red circles for galaxies with $D\leq R_{\rm vir}$. {\tt Panel B:} Same as {\tt Panel A} but for galaxies with $D/R_{\rm vir}>1$ shown with green triangles. 
In all panels, the solid and open markers represent the primary (strongest) and other weaker components of an absorption system. The horizontal dotted line indicates $|v_{\rm LOS}^c/v_{\rm esc}|=1$. The dashed line indicates the selection limit, i.e., $|v_{\rm LOS}^{\rm max}/v_{\rm esc}|$, where $v_{\rm LOS}^{\rm max} = 300$ \kms\ is the adopted LOS velocity limit for associating \HI\ absorbers with galaxy. For this, the $v_{\rm esc}$ at a given halo mass is calculated at $D=R_{\rm vir}$ and $z=0.4$. {\tt Panel C:} $v_{\rm LOS}^{c}/v_{\rm esc}$ of individual \HI\ components are plotted against $D/R_{\rm vir}$. A weak and significant correlation is observed between the two quantities for the primary components. {\tt Panel D:} The bound fraction (see text) plotted against the halo mass in three halo
mass bins for galaxies with $D\leq R_{\rm vir}$ and $D>R_{\rm vir}$. The color and markers are the same as in the {\tt Panels A} and {\tt B}.}
    \label{fig:v_esc}
\end{figure*}

In Section~\ref{subsec:N_prof_model}, we estimated $M(\HI)\approx 10^{5}~\rm M_{\odot}$ in the CGM of isolated, star-forming MUSEQuBES galaxies with median \logm~$\approx8.5$.
In order to probe the effects of the stellar mass on the mass of the circumgalactic \HI-bearing gas, we combine the star-forming COS-Halos galaxies with our sample of 118 SF galaxies to increase the dynamic range in $M_{\star}$. We divide the combined sample of star-forming galaxies into three stellar mass bins with \logm = $6-8$ (median 7.6), $8-9.5$ (median 8.7), and $9.5-11.1$ (median 10.2). The average $M(\HI)$ within $D/R_{\rm vir}=0.3-1$ for the three bins of galaxies is obtained following the same exercise described in Section~\ref{subsec:N_prof_model}, and is shown with solid blue squares and plus symbols in the top panel of Fig.~\ref{fig:MHI}. Note that the COS-Halos galaxies contribute to the highest stellar mass bin only. The x error bars represent the \logm\ ranges that encompass 68\% of the galaxies. The y error bars represent 68\% confidence interval on the $M(\HI)$,  calculated by propagating the error in the $N(\HI)$-profile. The $N(\HI)-$profiles of the three mass bins are shown in Appendix Fig.~\ref{fig:App1}.

Fig.~\ref{fig:MHI} shows a positive trend between $M_{\star}$ and $M(\HI)$ in the outer CGM ($0.3-1~R_{\rm vir}$) with  $M(\HI)$ increasing from $\approx10^{5.0}$~\Msun\ for the lowest mass bin to $\approx10^{6.5}~\rm M_{\odot}$ for the highest mass bin. 
Based on a study of strong \MgII\ ($W_{2796}>1$\AA) absorbers, \citet[]{Lan_2020} reported $M(\HI)\sim3\times10^{8}~$\Msun\ at $10~{\rm pkpc}\leq D < R_{\rm vir}$ for galaxies with \logm$\approx 9-10$ at $z\sim0.5$. This is at least $\approx2$ orders of magnitude higher compared to our $M(\HI)$ estimates. For a proper comparison, we calculated $M(\HI)$ for \citet[]{Lan_2020} using their equation 11 within $0.3R_{\rm vir}$ to $R_{\rm vir}$ as the integration limits. However, the $M(\HI)$ did not change appreciably. We emphasize that \citet{Lan_2020} did not directly measure the \HI\ column densities around galaxies, but used empirical scaling relations between \MgII\ equivalent width and $N(\HI)$ from \citet{Lan_2017} which has a significant scatter (several orders of magnitude in $N(\HI)$ at a given $W_{2796}$). They multiplied the \MgII\ covering fraction (for $W_{2796}>1$~\AA) and the average $N(\HI)$ expected from this relation to obtain a proxy for the $N(\HI)$-profile. In Appendix Fig.~\ref{fig:Nhi_LAN_comp} we show that this is not a good proxy of the $N(\HI)$-profile, but rather a systematically biased estimate of it.

Recently, \citet[]{Faerman_2025} has estimated the $M(\HI)$ using $N(\HI)$ measurements around $M_{\star} \sim 10^{6.5}-10^{9.5}~\rm M_{\odot}$ galaxies at $z<0.3$. In the top panel of Fig.~\ref{fig:MHI}, the open blue star symbols show their estimated $M(\HI)$ within $0.1-1~R_{\rm 200m}$ in three stellar mass bins.
Their inferred $M(\HI)$ is somewhat lower as compared to our findings. Since a significant fraction of the galaxies in their sample (i.e., 20 out of 40) have $z<0.1$, the $N(\HI)$ values, determined solely from the \lya\ line, are likely underestimated due to the absence of higher-order Lyman series lines. This could explain the difference in the $M(\HI)$. In contrast, the relatively higher redshift of our sample ensures the coverage of higher-order transitions, enabling robust measurement of $N(\HI)$, and subsequently the $M(\HI)$.

The inferred $M(\HI)$ at $D/R_{\rm vir} = 0.3 - 1$ in this work is $\sim 10^5 - 10^{6.5}$~\Msun. The $M_{\star} - M(\HI)$ relations in the local universe, using 21-cm measurements \citep[see][]{Parkash_2018}, show that $M(\HI)$ within galaxies spans $\sim 10^8 - 10^{10}$~\Msun\  for the stellar mass range considered in this work. A negligible ionization correction in the ISM renders $M(\HI)\approx M({\rm H})$ for the 21cm measurements. In the bottom panel of Fig.~\ref{fig:MHI}, the blue dotted line shows the $M({\rm H})$ against $M_{\star}$ from \citet[]{Parkash_2018}.
Additionally, at a higher redshift of $z \approx 0.35$, stacked 21-cm measurements revealed a significant cool neutral gas mass of $M(\HI)\approx M({\rm H})$~$\sim 10^9 - 10^{9.5}~\rm M_{\odot}$ within galaxies of the same stellar mass range \citep[][blue solid line in the bottom panel of Fig.~\ref{fig:MHI}]{Bera_2023}. Clearly, the \HI\ mass in the outer CGM ($>0.3R_{\rm vir}$) is several orders of magnitude lower than that inside galaxies. Note, however, that the bulk of the baryons (hydrogen) in the CGM is highly ionized, and cannot be probed via Lyman series lines.

The scaling relation between the $\log U$ and $D/R_{\rm vir}$ from \citet[]{Werk_2014} suggests a mean $U\approx10^{-2.5}$ at $D/R_{\rm vir}=0.3-1$, implying an \HI/H fraction of $\sim2-3\times10^{-4}$ for the HM01 UV background \citep[]{Haardt_2001}. This \HI\ fraction implies that the total gas mass (\HI\ + \HII) in the outer CGM (i.e., $D/R_{\rm vir}=0.3-1$) of low-mass galaxies can be as large as $\approx10^{5.5}/(2\times10^{-4})\approx 10^{9.2}~\rm M_{\odot}$ (ignoring the contribution of He). In the bottom panel of Fig.~\ref{fig:MHI}, we show the range of inferred $M(\rm H)$ from the measured $M(\HI)$ using \HI/H fraction of $10^{-4} - 10^{-3}$ (corresponding to $U\approx10^{-2}-10^{-3}$) with the red stripe.
The similarity between these measurements and those from 21-cm observations indicates that the total hydrogen mass of the outer CGM is comparable to that within the galaxies themselves, \footnote{The \HI\ within galaxies, traced by the 21-cm line, is predominantly neutral and generally does not require an appreciable ionization correction.} across the stellar mass range probed in this study. However, we note that the scaling relation of $\log U-D/R_{\rm vir}$ from \citet[]{Werk_2014} is valid for $\approx L_{\star}$ galaxies, and may not hold true for galaxies with \logm$\lesssim9.5$. A detailed photoionization modeling will be presented in a future paper for the MUSEQuBE sample to obtain a robust estimate of $\log~U$ and its dependence on $D/R_{\rm vir}$ and calculate the CGM baryon budget with appropriate ionization corrections.


\subsection{Bound fraction of \texorpdfstring{$\HI$}{HI} absorbers}

\label{sec:bound-frac}

In Fig.~\ref{fig:dv_dist}, a significant clustering of \HI\ absorbers around the galaxy redshifts is observed. Although the dispersion of the distribution ($\sigma_{\rm LOS}\approx0.44$) suggests that \HI\ absorbers are primarily `bound' to the associated galaxy, in this section we investigate how the line of sight velocities of individual \HI\ components compare with respect to the local escape velocities of the halos when galaxies of different stellar mass at different $D/R_{\rm vir}$ are considered.

In Fig. \ref{fig:v_esc}, we show the absolute LOS velocity of individual \HI\ components w.r.t. the host galaxy, normalized by the local escape velocity ($|v_{\rm LOS}^{c}/v_{\rm esc}|$), plotted against the halo mass with red circles (for $D\leq R_{\rm vir}$), and green triangles (for $D/R_{\rm vir}>1$),
in {\tt Panels A} and {\tt B}, respectively, for the galaxies in our sample. The solid and open symbols represent the primary (i.e., strongest) and other weaker components in an absorption system. A fraction of the \HI\ components clearly lie above the $|v_{\rm LOS}^{c}/v_{\rm esc}| = 1$ threshold, indicating that they are unbound from their associated galaxies. However, this effect is less pronounced for absorbers detected within the virial radius, suggesting that most of the inner CGM gas remains gravitationally bound. A few \HI\ components appear to lie beyond the nominal selection limit (dashed line). This is because the limit is computed assuming a fixed impact parameter of $D = R_{\rm vir}$ and a redshift of $z = 0.4$. 
 In {\tt Panel C}, we show the $|v_{\rm LOS}^{c}/v_{\rm esc}|$ of the \HI\ components plotted against the $D/R_{\rm vir}$. The two quantities exhibit a weak but significant correlation for the primary components ($\tau=0.2,~p=0.008$).

 In {\tt Panel D}, we show the `bound fraction' ($f_{\rm bound}$) of \HI\ components as a function of halo mass for the two $D/R_{\rm vir}$ bins. 
In a given halo mass bin, the bound fraction is defined as $f_{\rm bound} = n_1/(n_1+n_2)$, where $n_1$ and $n_2$ are the number of components with
$|v_{\rm LOS}^{c}/v_{\rm esc}|\leq1$ and $>1$, respectively.
The colors and markers are similar to the {\tt Panels A} and {\tt B}. 
At all halo (stellar) masses, the majority of the absorbers are consistent with being `bound' when $D\leq R_{\rm vir}$ (i.e, $f_{\rm bound} \approx 1$). However, at larger $D/R_{\rm vir}$, the bound fraction decreases significantly, particularly at the low mass end ($f_{\rm bound} \lesssim 0.5$). Here we note that $f_{\rm bound}$ relies on quantities with large uncertainties (such as $M_{\rm halo}$). Additionally, the impact parameter represents the minimum possible three-dimensional (3D) distance of the component, implying that the corresponding $v_{\rm esc}(D)$ serves as an upper limit. Conversely, $v_{\rm LOS}^{c}$ provides only a lower limit on the true 3D velocity. Therefore, the derived bound fractions should be interpreted as strict upper limits. A larger fraction of `unbound' absorbers at larger $D/R_{\rm vir}$, primarily traced by low-mass galaxies, likely stems from galaxy-absorber clustering and/or the two-halo contributions, as the projection effects become increasingly important at larger $D/R_{\rm vir}$ in the galaxy-absorber association \cite[]{Ho_2021}.

\begin{figure*}
    \centering
    \includegraphics[width=0.5\linewidth]{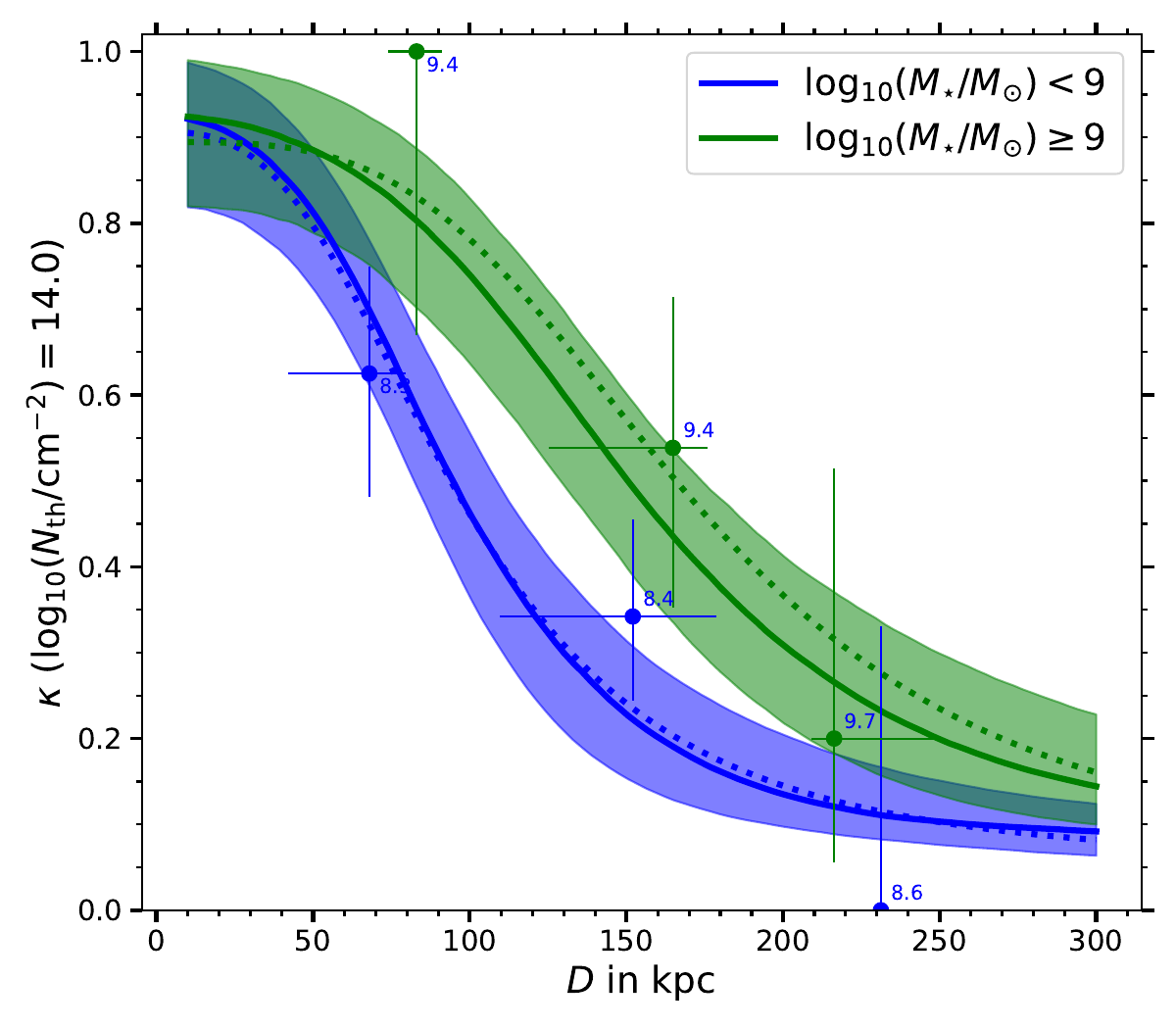}%
    \includegraphics[width=0.5\linewidth]{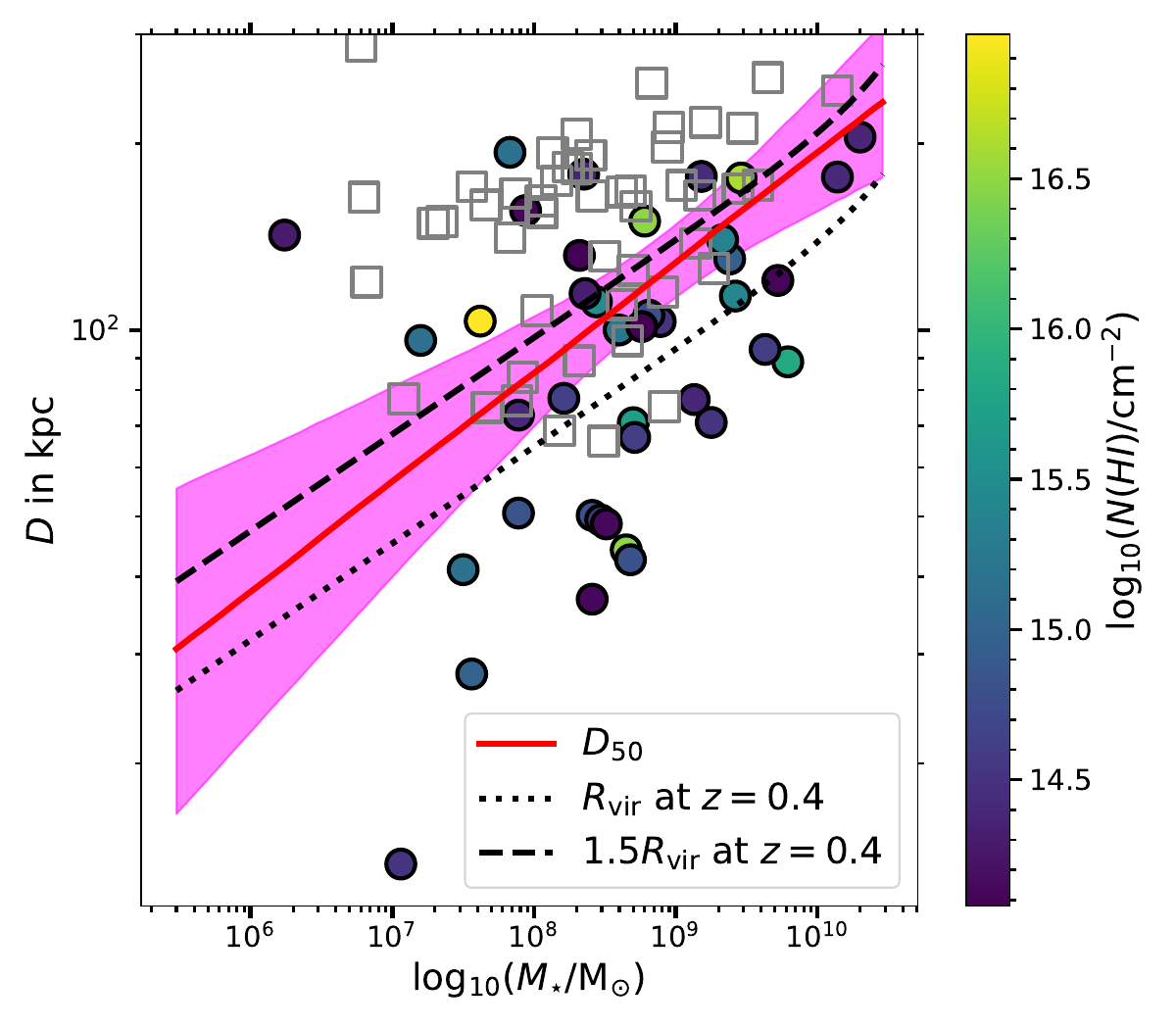}    
    \caption{{\tt Left:} The blue and green solid circles represent the directly measured, binned $\kappa$ values for a threshold of \logNhi~$=14$, plotted as a function of $D$ for the two stellar mass bins (\logm~$=6.3-9$ and \logm~$=9-10.3$). The error bars represent 68\% confidence intervals.     
    The corresponding best-fit $\kappa$-profiles, derived using Eqs.~\ref{eq:fc_D} and \ref{eq:fc_tparam}, are shown as blue and green dotted lines.
    The blue and green solid lines represent the same but derived using Eqns.~\ref{eq:fc_D} and \ref{eq:k-mass} (see text).   
    The shaded regions represent the 68\% confidence intervals in the latter case. The best-fit models are in good agreement with the binned measurements.  
    {\tt Right}: The $N(\HI)$ measurements for the isolated, star-forming MUSEQuBES galaxies above the threshold \logNhi=14 shown with filled colored circles as a function of \logm\ and $D$. The points are color-coded by their \logNhi. The unfilled grey squares represent the measurements below the threshold. The red solid line represents the best fit $D_{50}$ (see Eqn. \ref{eq:D50}) as a function of \logm. The black dotted and dashed line represents the virial radius and $1.5$ times the virial radius as a function of stellar mass at $z=0.4$.}
    \label{fig:fc_m_dep}
\end{figure*}

\subsection{The extent of the \texorpdfstring{$\HI$}{HI}-rich CGM}
\label{sec:extent}

Despite multiple observational efforts, the spatial extent of the CGM has remained poorly constrained. \citet[]{Wilde_2021} suggested a working definition for the \HI-rich CGM as the region within which the \HI\ covering fraction, for a threshold column density of $10^{14}$~\sqcm, drops to 50\%. They found that the extent of \HI-rich CGM varies with the stellar mass; ranging from $1.3-2.4~R_{\rm vir}$ for $M_{\star}$ varying from $10^7 - 10^9$~\Msun\ to  $10^{9.9}-10^{11}$~\Msun. In our earlier work, we analyzed stacked \lya\ absorption and identified a clear transition in the equivalent width profile from a log-linear form within the inner CGM to a power-law behavior beyond $\approx R_{\rm vir}$ \citep[see][]{Dutta1_2024}. We interpreted this change as marking the transition from a regime dominated by circumgalactic gas to one governed by galaxy-absorber clustering, implying that the \HI-rich CGM is largely confined within the virial radius. \citet[]{Wilde_2023} reported a similar transition in the \HI\ covering fraction profile but at a larger distance of $\approx 2R_{\rm vir}$.

In this section, we revisit the extent of \HI-rich CGM and its dependence on the stellar mass for the isolated, star-forming galaxies in our MUSEQuBES sample using covering fraction profiles derived through the formalism introduced in Section~\ref{sec:hi_cov}. Instead of the normalized impact parameter, we have modified Eqn.~\ref{eq:fc_tparam} as 
\begin{equation}
    \label{eq:fc_tparam_D}
    t=\alpha({\rm log}_{10}(D/{\rm pkpc})-\beta)~
\end{equation}


 In the left panel of Fig.~\ref{fig:fc_m_dep}, the directly measured $\kappa$ values for $\log N_{\rm th}/\rm cm^{-2}>14$ are plotted as a function of $D$ for two stellar mass bins -- \logm~$=6.3-9$ (median 8.3) and \logm~$=9-10.3$ (median 9.4) -- shown with blue and green circles, respectively. The best-fit $\kappa-$profiles using Eqn.~\ref{eq:fc_tparam_D} for the two mass bins are shown by the blue and green dotted lines, which are consistent with the binned measurements within the $1\sigma$ uncertainties. Furthermore, the best-fit $\kappa-$profiles indicate that although the covering fraction decreases monotonically with distance for both mass bins, it remains significantly elevated at $>100$~pkpc for the high mass galaxies as compared to their low-mass counterparts.

Motivated by the observed behavior above, we incorporated an explicit mass dependence in the parametrization of $\beta$ as follows: 
\begin{equation}
\label{eq:k-mass}
  t=\alpha\left[{\rm log}_{10}(D/{\rm pkpc})-\beta_M{\rm log}_{10}(M_{\star}/{\rm M_{\odot}}) - \gamma\right]~,   
\end{equation}
where $\beta_M$ is an additional free parameter, representing the stellar mass dependence of the $\kappa-$profile. Here, we define the extent of the CGM, $D_{50}$ which corresponds to $t=0$ and hence $\kappa=\kappa_0/2 + \kappa_1$, as 
\begin{equation}
\label{eq:D50}
  {\rm log}_{10}~D_{50} = \beta_M{\rm log}_{10}(M_{\star}/{\rm M_{\odot}}) + \gamma~.    
\end{equation}
In order to validate our model, in the left panel of Fig.~\ref{fig:fc_m_dep}, we show the best-fit $\kappa-$profile for \logm$<9$ and \logm$\geq9$ with blue and green solid lines by using the median \logm\ of the two bins in Eqn.~\ref{eq:k-mass}. The shaded regions represent the 68\% confidence intervals. Our best-fit models are in good agreement with the binned measurements, as well as the best-fit models without introducing the mass dependence (dotted lines).

In the right panel of Fig.~\ref{fig:fc_m_dep}, the best-fit $D_{50}$ is shown against \logm\ by the solid red line. The shaded magenta region shows the corresponding 68 percentile range. The filled circles represent the galaxies with detected \HI, color-coded by their $N(\HI)$. The open grey squares represent galaxies without detected \HI\ absorption. The dotted line represents the $R_{\rm vir}$ as a function of \logm\ at the median $z=0.4$. It is evident that the best-fit $D_{50}$ is $\approx 1.5R_{\rm vir}$ (dashed line) for stellar masses in the range \logm~$\approx6-10$. The observed trend between $D_{50}$ and stellar mass suggests that more massive galaxies exhibit more (physically) extended $\kappa$-profiles compared to their low-mass counterparts, as seen in the left panel of Fig.~\ref{fig:fc_m_dep}.

The best-fit model parameters are tabulated in Table~\ref{tab:best-fit-fc_M}. The best-fit $\beta_M\approx0.18$ is in good agreement with \citet[]{Wilde_2023}, who reported that the stellar-mass dependence of the 1-halo extent for galaxies traced by \HI\ absorbers with $N(\HI)>10^{14}~{\rm cm}^{-2}$ can be represented with a power-law index of $\approx0.14$. \citet[]{Wilde_2021} reported the extent of the CGM ($R_{\rm CGM}$) traced by \HI\ absorbers with ${\rm log}_{10}(N(\HI)/{\rm cm}^{-2})>14$ to be $\approx1.5R_{\rm 200m}$. However, carrying out the analysis in three stellar mass bins, they found the extent of CGM to vary from $\approx1.2R_{\rm 200m}$ (\logm$>9.9$) to $\approx2.4R_{\rm 200m}$ ($9.2<$\logm$<9.9$), with low-mass galaxies (\logm$ < 9.2$) exhibiting an intermediate extent of $\approx 1.6R_{\rm 200m}$. In contrast, leveraging the combined CGM$^2$+CASBaH galaxy sample, \citet[]{Wilde_2023} reported the CGM extent (quantified by $R_{\rm cross}$ beyond which the 2-halo contribution dominates over the 1-halo contribution) to be $\approx2R_{200m}$ across the stellar mass range of \logm~$=8-10.5$. A sharp decline of $\kappa\approx90\%$ within $D\leq R_{\rm vir}$ to $\kappa\approx16\%$ at $1<D/R_{\rm vir}\leq 3$ is reported in \citet[]{Johnson_15}. These observations are broadly consistent with our findings. Next, in order to mitigate any possible effects due to sample incompleteness, we selected a ``volume-limited'' sample (see Appendix Fig.~\ref{fig:vol_lim_sample}) of isolated and star-forming MUSEQuBES galaxies. The best-fit parameters for this volume-limited sample remain consistent with the full sample of isolated and SF galaxies within the $1\sigma$ uncertainties (see Table \ref{tab:best-fit-fc_M}).

While we do not invoke redshift dependence in $\kappa$-profile modeling, we have verified that an additional redshift dependence of the form: 
\begin{equation}
\label{eq:k-mass-z}
  t=\alpha\left[{\rm log}_{10}(D)-\beta_M{\rm log}_{10}(M_{\star}/M_{\odot}) - \beta_z(1+z) - \gamma\right]~,   
\end{equation}
 reveal marginal redshift-evolution ($\beta_z = -1.7^{+0.8}_{-1.5}$).
 We note that the limited FoV of MUSE restricts the low-$z$ galaxies to lower impact parameters (see Fig. 1). This systematic is especially crucial at $z\lesssim0.4$. The scarcity of large impact parameter galaxies at low-$z$ from MUSEQuBES alone may result in this marginal redshift evolution. Indeed, carrying out this analysis incorporating isolated galaxies from the Magellan/IMACS survey beyond the MUSE FoV and within 500 pkpc results in a weaker redshift-dependence with $\beta_z=-0.7^{+0.4}_{-0.4}$.

\begin{table}
\centering
\caption{Best-fit parameters for eqn. \ref{eq:fc_D} and \ref{eq:k-mass}}   
\label{tab:best-fit-fc_M}
\begin{tabular*}{\linewidth}{c@{\extracolsep{\fill}}ccccr}  
\hline
Sample & $\alpha$ & $\beta_M$ & $\gamma$ & $\kappa_0$ & $\kappa_1$ \\ 
\hline
All & $7.7^{+2.5}_{-3.2}$ & $0.18^{+0.07}_{-0.07}$ & $0.5^{+0.7}_{-0.7}$ & $0.9^{+0.1}_{-0.1}$ & 0.07[0-0.1]$^a$ \\  

volume- & & & & & \\
limited & $8.3^{+2.4}_{-3.4}$ & $0.24^{+0.12}_{-0.11}$ & $0.0^{+1.0}_{-1.0}$ & $0.8^{+0.1}_{-0.1}$ & 0.05[0-0.1]$^a$ \\ 
\\
\hline
\end{tabular*}
\justify
Notes-- $^a$ The $\kappa_1$ is unconstrained within the flat prior [0,0.1] used in this work. 
\end{table}

\begin{figure*}
\centering   
     \includegraphics[width=0.5\linewidth]{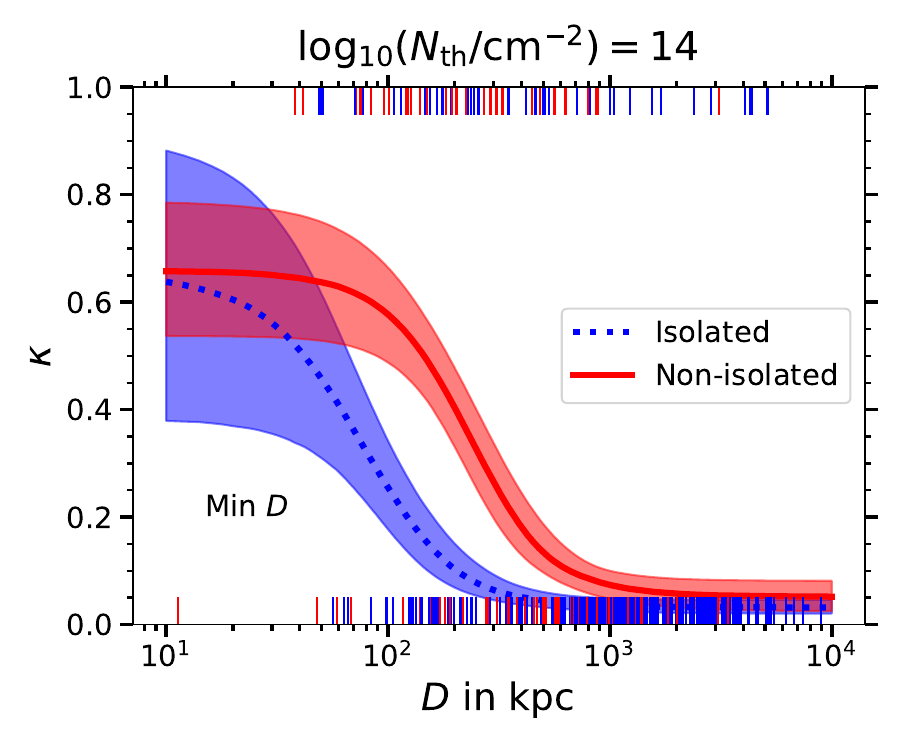}%
     \includegraphics[width=0.5\linewidth]{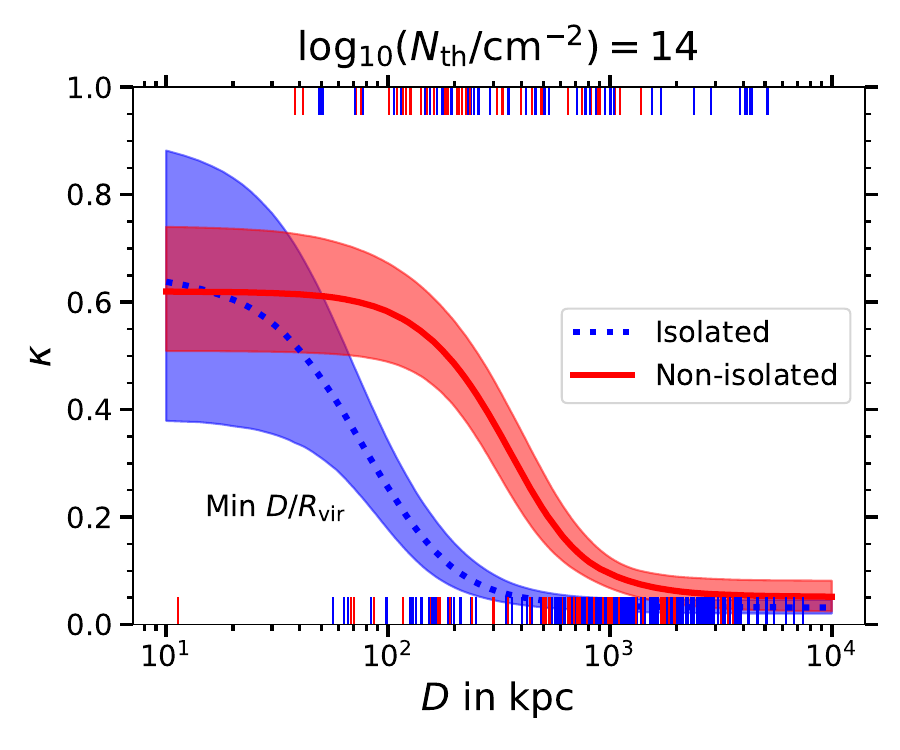}      
      \caption{{\tt Left:} The unbinned \HI\ covering fraction profiles (following Eqns. \ref{eq:fc_D} and \ref{eq:fc_tparam}) as a function of impact parameter for isolated and group MUSEQuBES + Magellan MOS galaxies shown with dotted blue and solid red lines for a threshold \logNhi~$=14$.  The shaded region represents the 68\% confidence interval on the best-fit covering fraction profile. Only the minimum impact parameter galaxy is considered for the non-isolated cases. The vertical lines indicate individual cases of detection (top) and non-detection (bottom). {\tt Right:} As the left panel but the closest $D/R_{\rm vir}$ galaxy is used for the analysis. In both panels, the cool, neutral gas is found to be significantly more extended around group galaxies compared to their isolated counterparts.}          
      \label{fig:env_muse&ext}
\end{figure*}

\subsection{Environmental dependence of \texorpdfstring{$\HI$}{HI} absorption}
\label{sec:env}

The MUSE FoV limits the maximum impact parameter of MUSEQuBES galaxies from the quasar sightline to $\approx 300$~pkpc. It is, therefore, not well-suited for conducting a comprehensive environmental study. As mentioned in Section~\ref{sec:data}, we use galaxies obtained from the Magellan/IMACS survey around six of the MUSEQuBES sightlines, which provide a complementary dataset extending out to $\approx20$ arcmin. Note that, we only included Magellan galaxies located beyond the MUSE FoV in projected physical distance (kpc). This strategy ensures that the primarily low-mass galaxies observed with MUSE contribute at smaller impact parameters, while the more massive galaxies, located at larger projected distances from the sightline, are captured by the Magellan survey (see Table~\ref{tab:galprop} \& Fig.~\ref{fig:gal_prop}). The combination of the two surveys enables us to characterize the large-scale environment surrounding the low-mass galaxies closer to the quasar sightlines. A simple 3D FoF algorithm with a linking velocity of $\pm500$~\kms\ and a projected linking distance of $500$~pkpc is applied to this combined galaxy dataset to identify the isolated and non-isolated (``group'') galaxies.

For the six quasar fields surveyed with Magellan, we conducted a blind search for \lya\ and \lyb\ lines in the HST/COS spectra to compile a comprehensive catalog of \HI\ absorbers. First, we masked the Galactic absorption features and previously identified \OVI\  absorbers \citep[i.e.,][]{Dutta2_2025}. We then accounted for the \HI\ lines already identified around MUSEQuBES galaxies (see Section \ref{sec:absorption_analysis}). Next, we masked the prominent metal lines arising from the strong \HI\ absorbers.
The remaining unidentified lines could be (i) \lya\ with \lyb\ too weak to be detected (i.e., $N(\HI)\lesssim 10^{13.5}~{\rm cm}^{-2}$ for the typical $S/N$ of the quasar spectra in this work), (ii) \lyb\ with \lya\ outside the FUV coverage of COS, and \lyg\ too weak to be detected (i.e., $N(\HI)\lesssim 10^{14}~{\rm cm}^{-2}$ for the typical $S/N$ of the quasar spectra used in this work), which is possible for $z\approx 0.47-0.75$, (iii) Higher order Lyman series lines for strong \HI\ absorbers at $z\geq0.75$, or (iv) extreme-UV (EUV) metal lines (e.g., \OIII~$\lambda\lambda702,832$, \OIV~$\lambda787$) for which the Lyman series lines fall outside the COS FUV coverage. These last two scenarios are applicable to four of the six sightlines with background quasars at redshifts $z_{\rm qso} > 1$. For these sightlines, we performed a systematic search for EUV metal lines using the doublet-matching technique. After eliminating the EUV metal lines, we classified the remaining unassociated absorption features as \lya. The column densities of all identified \HI\ absorbers were constrained using {\sc vpfit}, utilizing higher-order Lyman series transitions when available. We emphasize, however, that eliminating interlopers arising from scenario (ii) above remains challenging without the NUV coverage, particularly for low column density absorbers with $N(\HI)<10^{14}~{\rm cm}^{-2}$. In order to minimize its effects, we only consider covering fractions for a threshold $N(\HI)$ of $10^{14}~{\rm cm}^{-2}$ or above. 
 The column density distribution function (CDDF) of our galaxy-blind \HI\ components detected towards the 6 sightlines is shown in Fig.~\ref{fig:cddf}. Our measurements at $z<0.47$ are broadly consistent with \citet[]{Danforth_2016}.


The blind \HI\ absorber catalog, complete down to $\approx10^{14}~{\rm cm}^{-2}$, is cross-matched with the combined galaxy sample with a velocity window of $\pm300$~\kms\ to construct galaxy-absorption pairs.  Owing to the larger FOV of the Magellan survey, it is possible to detect one or more isolated and group galaxies at the same redshift. In such cases, we select only the smallest $D$ (or $D/R_{\rm vir}$) galaxy, discarding all other galaxies regardless of their environments (isolated or group). This ensures a unique galaxy-absorption association, i.e., one galaxy is associated with one and only one absorption measurement. This resulted in 492 unique quasar-galaxy pairs, with 191 and 301 pairs associated with isolated and non-isolated galaxies, respectively.

Fig.\ref{fig:env_muse&ext} shows the detections and non-detections of \HI\ absorption systems above the threshold \logNhi~$=14$, indicated by the top and bottom vertical ticks, respectively, as a function of $D$. Isolated and non-isolated galaxies are represented in blue and red, respectively. The left and right panels present the cases where host galaxies are selected based on minimum projected distance ($D$) and minimum scaled distance ($D/R_{\rm vir}$),  respectively. In order to obtain the covering fraction profiles without binning the data, we used the logistic regression approach introduced in Section~\ref{sec:extent}. The best-fit modeled $\kappa$ profiles, determined using Eqns.~\ref{eq:fc_D} \& \ref{eq:fc_tparam} for the isolated and non-isolated samples, are shown with the dotted blue and solid red lines, respectively. A two-sided KS-test confirms that the $M_{\star}$ distributions of the two samples are not significantly different. Therefore, we do not incorporate stellar mass dependence in modeling the $\kappa$-profiles.

From Fig.~\ref{fig:env_muse&ext} it is evident that the covering fraction profile of group galaxies is significantly more extended than that of the isolated galaxies. The $D_{50}$ values measured for isolated and non-isolated galaxies are $\approx80$ and $\approx230$ pkpc, respectively, indicating the presence of cool, \HI-rich gas over much larger regions around group galaxies. Given that the stellar mass distributions of the two samples are statistically indistinguishable, the observed differences in the $\kappa$-profiles can be robustly attributed to environmental effects alone.

The mean \HI\ column density of the detected absorbers associated with group galaxies within the virial radius ($10^{15.5\pm0.8}$~\sqcm) is somewhat higher than that of isolated galaxies ($10^{14.7\pm0.5}$~\sqcm). However, a censored {\tt log-rank} test is unable to distinguish between the $N(\HI)$ distributions when the upper limits are taken into account (see the left panel of Fig.~\ref{fig:env_Ndist} in the Appendix). However, for galaxies outside the virial radius (i.e., $1<D/R_{\rm vir}<3$), the censored {\tt log-rank} test indicates a statistically significant difference between the $N(\HI)$ distributions of isolated and non-isolated galaxies ($p\approx0.01$; see the right panel of Fig.~\ref{fig:env_Ndist}), although the mean \logNhi\ of the {\it detected} absorbers are consistent with each other ($14.3\pm0.8$ and $14.7\pm1.1$ for isolated and non-isolated galaxies, respectively).

Multiple high-redshift ($z\gtrsim3$) studies have reported that non-isolated (group) galaxies tend to show stronger absorption, larger spatial extent, and/or flatter radial profiles in equivalent width, column density, and covering fraction for \HI\ \citep[e.g.,][] {Muzahid_2021,Lofthouse_2023,Banerjee_2025} and metal ions— such as \MgII\ \citep[e.g.,][]{Galbiati_2024} and \CIV\ \citep[e.g.,][]{Muzahid_2021,Banerjee_2023,Galbiati_2024}.

At low redshift, \citet[]{Chen_2010} reported a lack of anticorrelation between \MgII\ REW and $D$ for group galaxies, in contrast to isolated systems, indicating a flatter REW profile in group environments \citep[see also][]{Bordoloi_2011,Fossati_2019,Huang_2021,Dutta_2021,Cherrey_2025}. A more extended $\kappa$-profile for group galaxies has also been observed for the highly ionized \OVI\ ion \citep[see][]{Johnson_15, Tchernyshyov_2022}, and low-ionized \MgII\ ion \citep[see][]{Cherrey_2024} suggesting that environmental effects similarly influence the distribution of both cool and warm-hot circumgalactic gas.

Several physical mechanisms have been proposed to explain these observations. For example, \citet{Bordoloi_2011} suggested that the excess \MgII\ absorption observed in group environments arises from the superposition of multiple halos along the line of sight. While this simple model adequately reproduces the \MgII\ REW profile, it fails to account for the observed kinematic complexity of the \MgII\ absorbers \citep[]{Nielsen_18}. \citet{Fossati_2019} and \citet[]{Dutta_2020} argued that the enhanced \MgII\ absorption in group environments arises primarily from gravitational and hydrodynamic interactions among member galaxies, which increase the effective absorption cross-section, rather than from a widespread, diffuse intragroup medium.

To our knowledge, this is the first clear evidence of significantly extended \HI\ covering fraction profiles for non-isolated galaxies at low redshift. However, within $\approx100$~pkpc ($\approx R_{\rm vir}$), we do not observe a significant difference in the \HI\ covering fraction or the mean $N(\HI)$ between isolated and non-isolated galaxies. The difference becomes apparent only beyond the virial radius, where non-isolated galaxies exhibit a systematically higher \HI\ covering fraction. This suggests that enhanced galaxy–absorber clustering—likely driven by the large-scale structures in which these galaxies are embedded—is the dominant factor shaping the observed trend \citep[see also,][]{Muzahid_2021, Lofthouse_2023}.

 \citet[]{Cunnama_2014} predicted higher \HI\ column densities at larger impact parameters for group galaxies compared to isolated galaxies in the  Galaxies-Intergalactic Medium Interaction Calculation ({\tt GIMIC}) cosmological hydrodynamic simulation at $z=0$. They interpreted this enhancement as a consequence of environmental processes such as ram-pressure stripping and tidal interactions within group environments. However, dynamical interactions between group members are a less likely explanation, as such processes are expected to enhance the cross-section of gas and metals in the inner regions (i.e., within the virial radius)-- a trend not supported by our observations.

\section{Summary} 
\label{sec:summary}

In this work, we presented measurements of the column density, covering fraction, and mass of cool gas traced by Lyman series absorption around 256 low-redshift galaxies ($z\lesssim0.75$) from the MUSEQuBES survey. This sample has a median $z$ [68\% range]~$ = 0.48~[0.3-0.6]$, median \logm\ [68\% range]~$ = 8.7~[7.5-9.8]$, and median $D$ [68\% range] = 140~pkpc $[77-196]$~pkpc. Additionally, we incorporated galaxy data from the Magellan/IMACS survey for 6 of the 16 MUSEQuBES sightlines to characterize the impact of environment and large-scale structure on the distribution of cool \HI\ gas. The main findings of this study are as follows: 
\newline

1. The total \HI\ column density, $N(\HI)$, exhibits a significant anticorrelation with projected distance from the galaxies. The anticorrelation is strongest ($\tau=-0.32,~p\ll0.01$) when plotted against the normalized impact parameter $D/R_{\rm vir}$, considering the smallest $D/R_{\rm vir}$ galaxies as hosts (Fig.~\ref{fig:Nh_prof}). 
Upon dividing the sample by stellar mass and star-formation rate, we find that this trend is primarily driven by star-forming galaxies (Fig.~\ref{fig:Nh_prop_div}). In contrast, passive galaxies in our sample do not exhibit any statistically significant anticorrelation between $N(\HI)$ and $D/R_{\rm vir}$.
\newline


2. The non-isolated/group galaxies having neighbours within $\pm500$~\kms\ along the line-of-sight (LOS) and $500$~pkpc projected separation primarily exhibit the strongest $N(\HI)$ absorbers within $R_{\rm vir}$, contributing to a larger scatter in the column density profile. The $N(\HI)$-profile for the isolated, star-forming galaxies can be described with a power-law of slope $\approx-3$ when plotted against $D/R_{\rm vir}$, although a significant fraction of galaxies exhibit much larger \HI\ column densities (Fig.~\ref{fig:nprof_fit_minDn}). Adopting the model profile, we found that the mean \HI\ mass in the outer CGM (0.3-1$~D/R_{\rm vir}$) is $\approx10^{5}~$\Msun\ for the isolated, star-forming MUSEQuBES galaxies.  
\newline

3. Relaxing the isolation condition, and including more massive star-forming galaxies from the COS-Halos survey, we show that the mean \HI\ mass in the outer CGM (0.3-1 $D/R_{\rm vir}$) increases with stellar mass, ranging from $\approx10^5~$\Msun\ to $10^{6.6}~$\Msun\ for the stellar mass range of $\approx10^7~$\Msun\ to $\approx10^{11}~$\Msun\ (Fig.~\ref{fig:MHI}). A fiducial ionization correction indicates that the baryon budget in this highly ionized gas in the outer CGM is comparable to that within the galaxies themselves, across the stellar mass range probed in this study. 
\newline  
 
4. We observe a smaller \HI\ covering fraction (with a threshold \Nhi~$=10^{14}~{\rm cm}^{-2}$) inside the virial radius of high-mass, \logm$>9.5$ galaxies. Dividing the sample based on their star-forming activity and environment, we find that this suppression of the covering fraction for the high mass galaxies can be attributed to the larger fraction of passive and group galaxies in this bin. On the contrary, the isolated and star-forming massive galaxies exhibit a near-unity covering fraction (Fig.~\ref{fig:N_fc_invir}).
\newline

5. The best-fit standard deviation of the line-of-sight velocity distribution, normalized by the escape velocity, suggests that the majority of absorbers are gravitationally bound to their associated galaxies ($\sigma_{\rm LOS} \approx 0.5$). Dividing our sample into two stellar mass bins, we observe a higher $\sigma_{\rm LOS}$ for low-mass galaxies. The difference largely disappears when restricting our analyses to galaxies with $D<R_{\rm vir}$, indicating that a larger fraction of unbound absorbers contribute at large $D/R_{\rm vir}$ (Fig.~\ref{fig:dv_dist}). By plotting the local escape-velocity-normalized velocity centroids of the \HI\ components against the stellar mass of galaxies in three $D/R_{\rm vir}$ bins, we find that the \HI\ components remain primarily bound to their associated galaxies—regardless of stellar mass— when located within $D < R_{\rm vir}$ (Fig.~\ref{fig:v_esc}).
\newline

6. We observe a moderate but significant correlation (anticorrelation) between the \HI\ $b$-parameters of the strongest \HI-components and the sSFR (\logm) of the host galaxy when confined within $D<R_{\rm vir}$ (Fig.~\ref{fig:b_invir}). We speculated that excess turbulence driven by star formation–induced outflows is responsible for the observed trend. However, the weaker (non-primary) and broader ($b\geq60$ \kms) \HI\ components are predominantly hosted by massive (\logm$\geq9$) galaxies. Together with the line-of-sight velocity-aligned \OVI\ components, these broad \lya\ absorbers suggest a collisionally ionized origin with $T\approx10^{5.4}$~K, likely tracing virialized halo gas. 
\newline

7. Using a modified logistic function, we quantified the \HI\ covering fraction profile for isolated, star-forming MUSEQuBES galaxies. We found that the extent of the \HI-rich CGM of galaxies in our sample, quantified by the $D_{50}$ parameter, indicating the projected distance at which $\kappa$ falls to $\approx50\%$ of its peak, increases with stellar mass, with a typical scaling of $D_{50}/R_{\rm vir}\approx1.5$ (Fig.~\ref{fig:fc_m_dep}). 
We did not find any significant redshift evolution of the CGM extent within the redshift range probed in this work.   
\newline

8. Leveraging a sample of over $3000$ galaxies with spectroscopic redshifts from the Magellan/IMACS survey across six MUSEQuBES fields, we found that non-isolated galaxies exhibit a significantly more extended \HI\ covering fraction profile compared their isolated counterparts with comparable stellar masses (Fig.~\ref{fig:env_muse&ext}). The best-fit $D_{50}$ for non-isolated galaxies ($\approx230$~pkpc) is nearly three times larger than that of isolated galaxies ($\approx80$~pkpc).

\section*{Acknowledgments}
SD acknowledges Prof. R. Srianand and Prof. Aseem Paranjape for insightful discussions.
SD, SC, and SM thank DST for Indo-Italy travel grant under INT/Italy/P-35/2022(ER) program.  
SC gratefully acknowledges support from the European Research Council (ERC) under the European Union's Horizon 2020 research and innovation program grant agreement No 864361. RA acknowledges funding from the European Research Council (ERC) under the European Union's Horizon 2020 research and innovation program (grant agreement 101020943, SPECMAP-CGM).

\section*{Data Availability} 
The data used in this article are available in the ESO (https://archive.eso.org/) and $HST$ (https://hla.stsci.edu/) public archives.

\bibliography{all_ref_hi_apj}
\bibliographystyle{mnras}

\appendix

\section{\texorpdfstring{$\kappa$}{kappa}-profiles for three \texorpdfstring{$M_{\star}$}{Mstar} bins}

\begin{figure}
    \centering
    \includegraphics[width=1\linewidth]{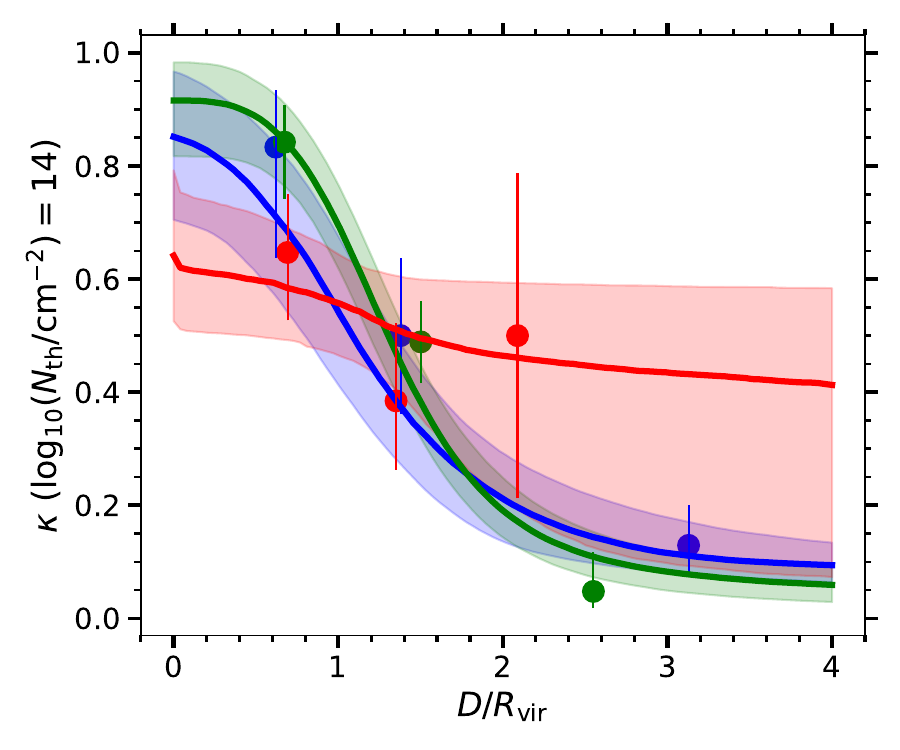}    
    \caption{The best-fit $\kappa$ profiles and the 68\% confidence intervals with threshold $N(\HI)=10^{14}~{\rm cm}^{-2}$ for the three stellar mass bins (shown in Fig.~\ref{fig:N_fc_invir}) are shown with the solid lines and shaded region with the same color-coding. The solid circles and error bars indicate the binned $\kappa$ and 68\% Wilson-score confidence intervals in three $D/R_{\rm vir}$ bins for the same threshold. } 
    \label{fig:fc_prof_Dn_model}
\end{figure}

The best-fit covering fraction ($\kappa$) profiles for three stellar mass bins generated in Sect.~\ref{sec:hi_cov} is shown in Fig.~\ref{fig:fc_prof_Dn_model}. 

\section{\texorpdfstring{$N(\HI)$}{N(HI)}-profiles for three \texorpdfstring{$M_{\star}$}{Mstar} bins} 

\begin{figure*}
    \centering
    \includegraphics[width=0.333\linewidth]{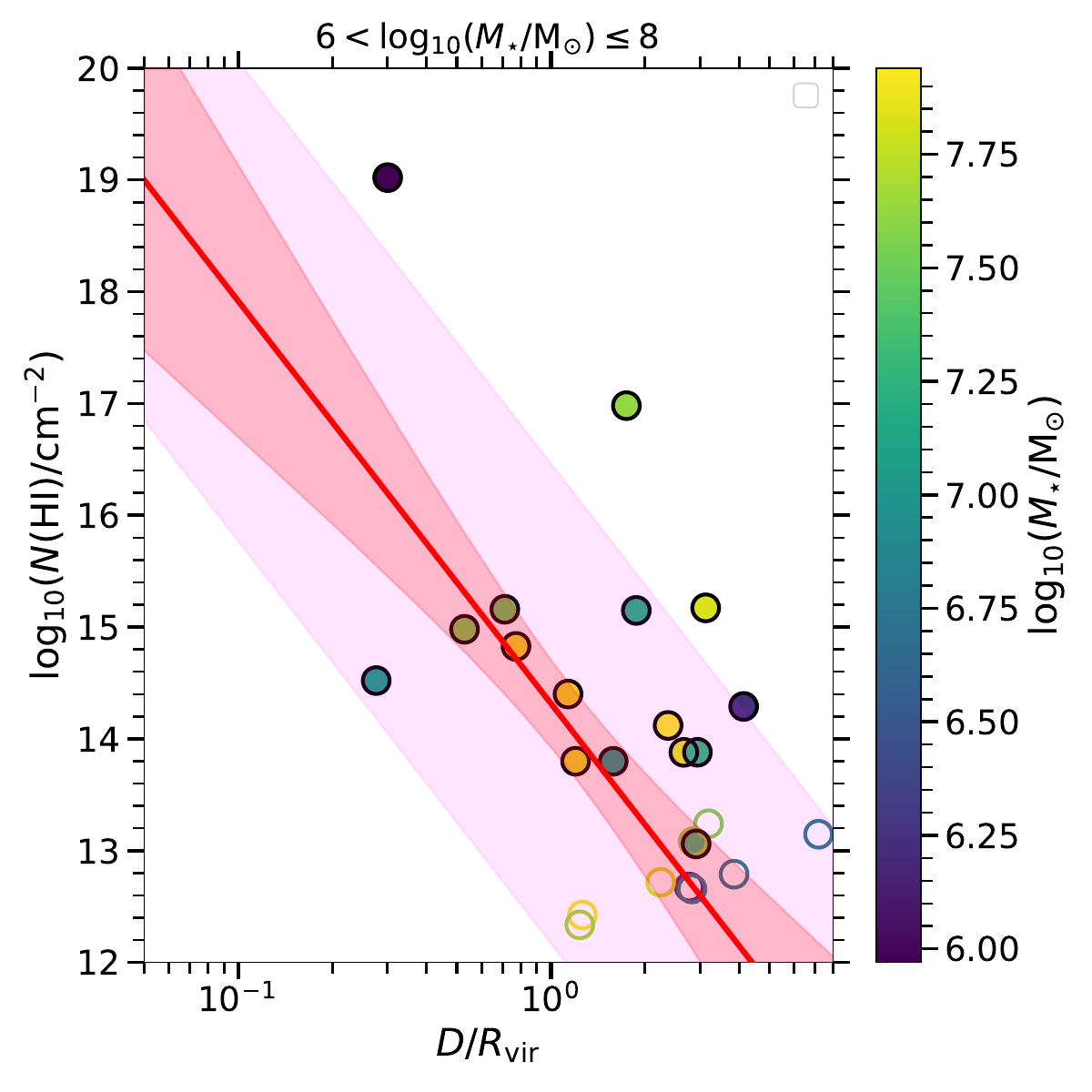}%
    \includegraphics[width=0.333\linewidth]{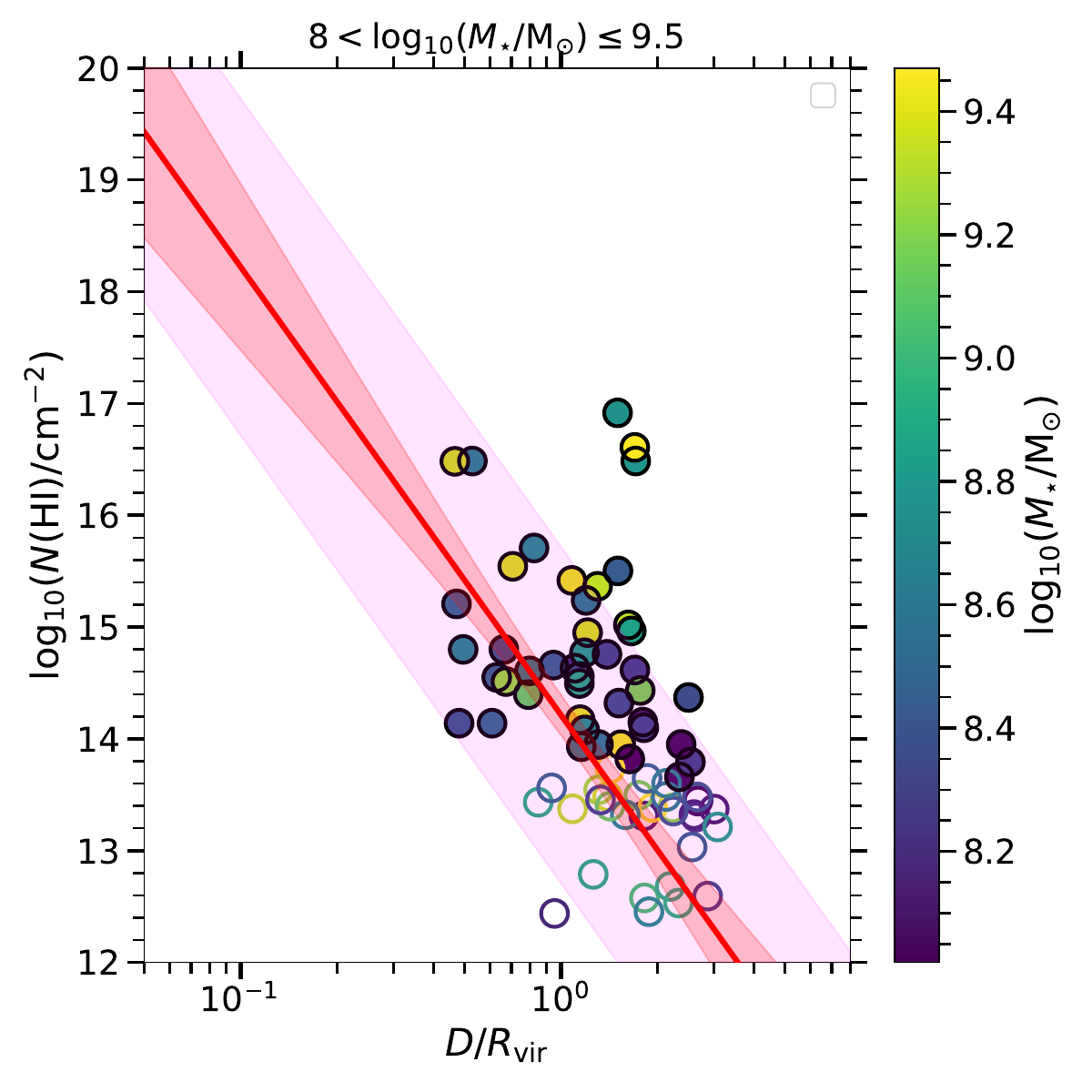}%
    \includegraphics[width=0.333\linewidth]{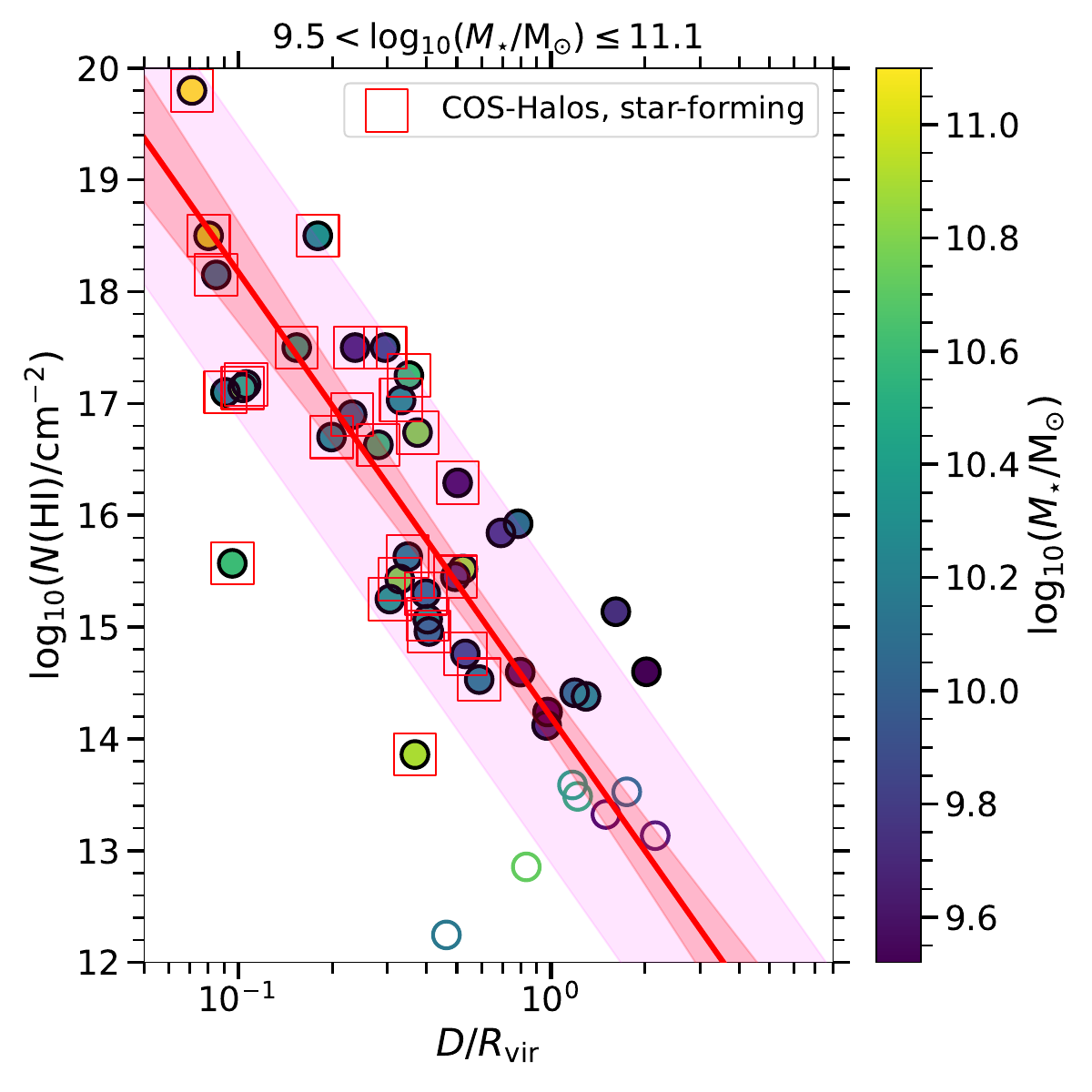}  
    \caption{The $N(\HI)-$profile for star-forming galaxies from MUSEQuBES and COS-Halos survey shown for three stellar mass bins (with \HI\ measurements from COS-Halos survey contributing in the most massive bin only).
    We show star-forming COS-Halos galaxies (circles with red square envelopes) alongside our MUSEQuBE sample of star-forming galaxies. Other details are same as Fig.~\ref{fig:nprof_fit_minDn}.} 
    \label{fig:App1}
\end{figure*}

In Fig.~\ref{fig:MHI}, we have shown the $M(\HI)$ in the outer CGM ($D/R_{\rm vir}\approx 0.3-1$) of star-forming galaxies in three stellar mass bins with \logm~$=6-8$, \logm~$=8-9.5$, and \logm~$=9.5-11.1$. Fig.~\ref{fig:App1} shows the $N(\HI)$ profiles for star-forming galaxies in these stellar mass bins. \HI\ measurements for the COS-Halos galaxies are included in the highest stellar mass bin, and are indicated by the red square envelopes. The solid and hollow points indicate the detection and 3$\sigma$ upper limits in cases of non-detections, respectively. The points are color-coded by the stellar mass of the associated galaxies. The best-fit power-law and the 68\% confidence interval in each stellar mass bin are shown with the red solid line and shaded region. The magenta shaded region indicates the best-fit intrinsic scatter.

\section{Comparison of \texorpdfstring{$N(\HI)$}{N(HI)}-profile with Lan et al.}  

\begin{figure}
    \centering
    \includegraphics[width=1\linewidth]{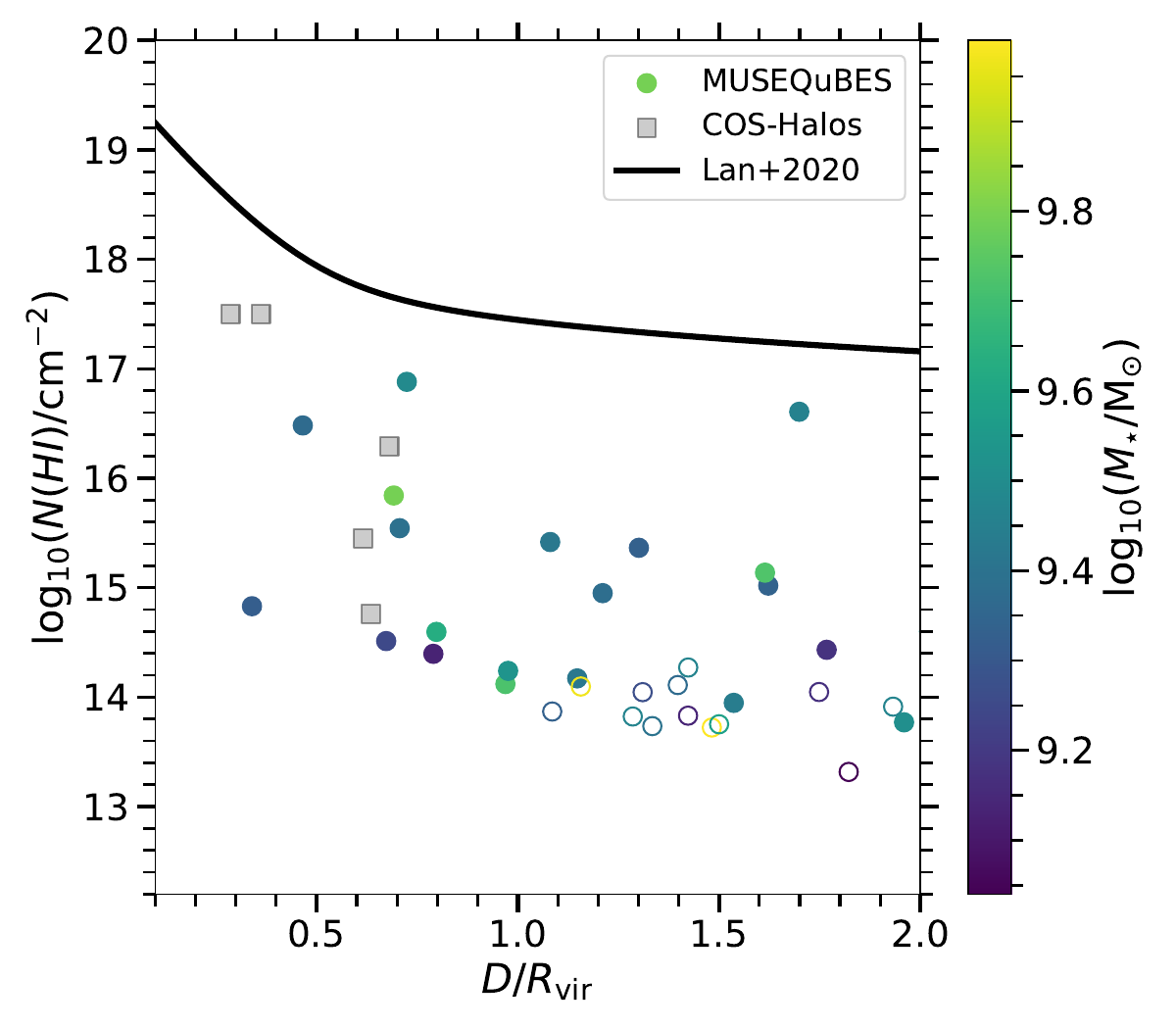}
    \caption{ The $N(\HI)$ measured around MUSEQuBES galaxies with \logm~$=9-10$ are plotted against $D/R_{\rm vir}$, with filled and open circles indicating detections and $3\sigma$ upper limits, respectively. The black squares indicate $N(\HI)$ measurements from the COS-Halos survey within the same stellar mass range. The black solid line represents the best-fit $N(\HI)$-profile of \citet[]{Lan_2020} for star-forming galaxies (adopting $z=0.5$, \logm~$=9.5$, and the parameters for star-forming galaxies in their Table~3). 
    Clearly, their inferred $N(\HI)$-profile is biased high relative to the directly measured values.} 
    \label{fig:Nhi_LAN_comp}
\end{figure}

Fig.~\ref{fig:Nhi_LAN_comp} shows the comparison between directly measured $N(\HI)$ from MUSEQuBES and COS-Halos surveys with the  $N(\HI)$ inferred for strong \MgII\ absorbers by \citet[]{Lan_2020}.

\section{Extent of the CGM for a volume-limited sample}

\begin{figure}
    \centering
    \includegraphics[width=1.0\linewidth]{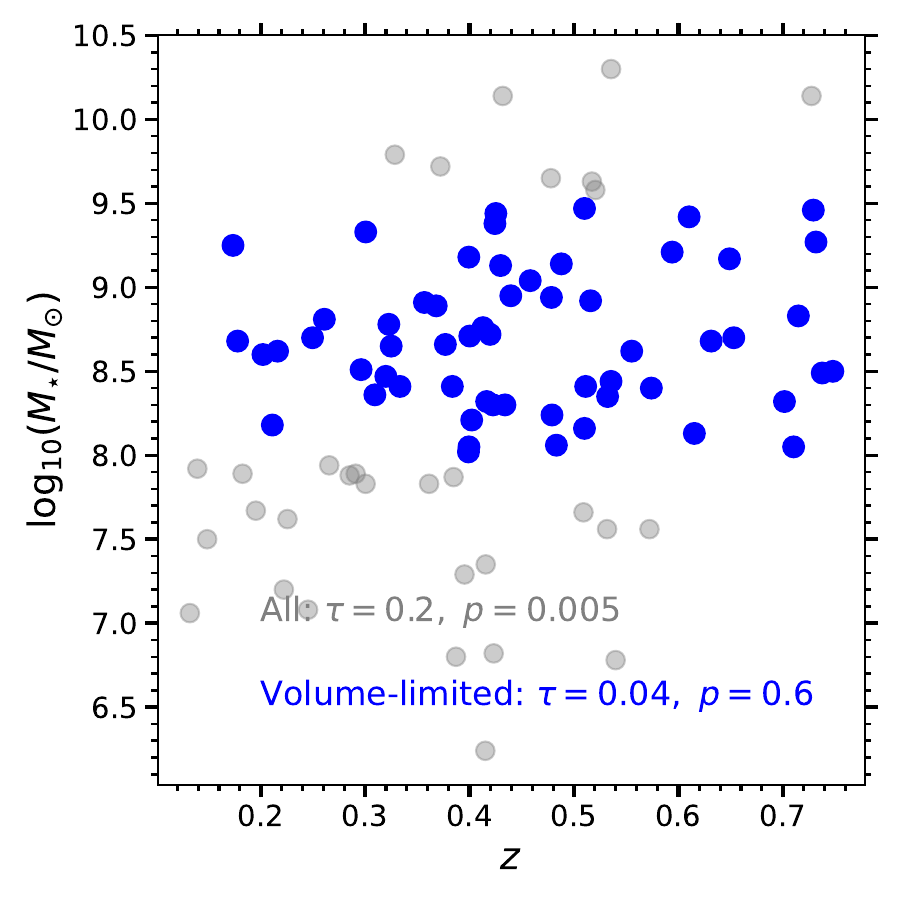}
    \caption{The stellar mass of isolated, star-forming MUSEQuBES galaxies plotted against $z$ with filled circles. A volume-limited subsample selected from this parent sample is shown with blue filled points. The sample is selected to obtain the maximum number of galaxies while ensuring a lack of correlation between $z$ and \logm.} 
    \label{fig:vol_lim_sample}
\end{figure}

The analyses of section \ref{sec:extent} have been repeated with a volume-limited sample of isolated and star-forming MUSEQuBES galaxies. The sample is shown in Fig.~\ref{fig:vol_lim_sample}. The best-fit parameters $\alpha$, $\beta_M$, $\gamma$, $\kappa_0$ remains consistent within $1\sigma$ of the reported values in section \ref{sec:extent}.  

\section{Comparison of our blind \texorpdfstring{$\HI$}{HI} catalog with literature}   

\begin{figure}
    \centering
    \includegraphics[width=1.0\linewidth]{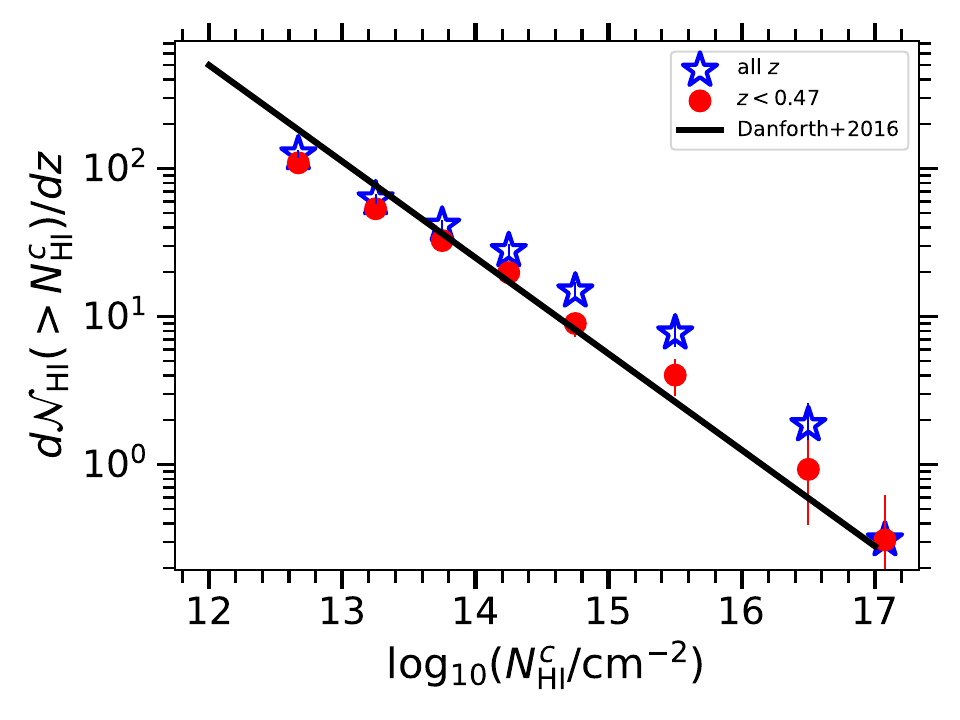}    
    \caption{The \HI\ CDDF for the galaxy-blind \HI\ components detected towards the six sightlines with Magellan/IMACS galaxy data is shown by the open blue star symbols. The solid red circles represent the same, but for components at $z<0.47$. For comparison, the CDDF from \citet{Danforth_2016} at $z < 0.47$ is shown as the solid black line.}   
    \label{fig:cddf}
\end{figure}

We show the \HI\ CDDF for the galaxy-blind \HI\ components detected towards the 6 sightlines with Magellan/IMACS coverage in Fig.~\ref{fig:cddf} with open blue star symbols. The solid red circles show the same, but for components with $z<0.47$. Our measurements are largely consistent with the \HI\ CDDF from \citet[]{Danforth_2016} for $z<0.47$ (solid black line).

\section{Environmental dependence of \texorpdfstring{$N(\HI)$}{N(HI)} distribution} 

\begin{figure*}
    \centering
    \includegraphics[width=0.5\linewidth]{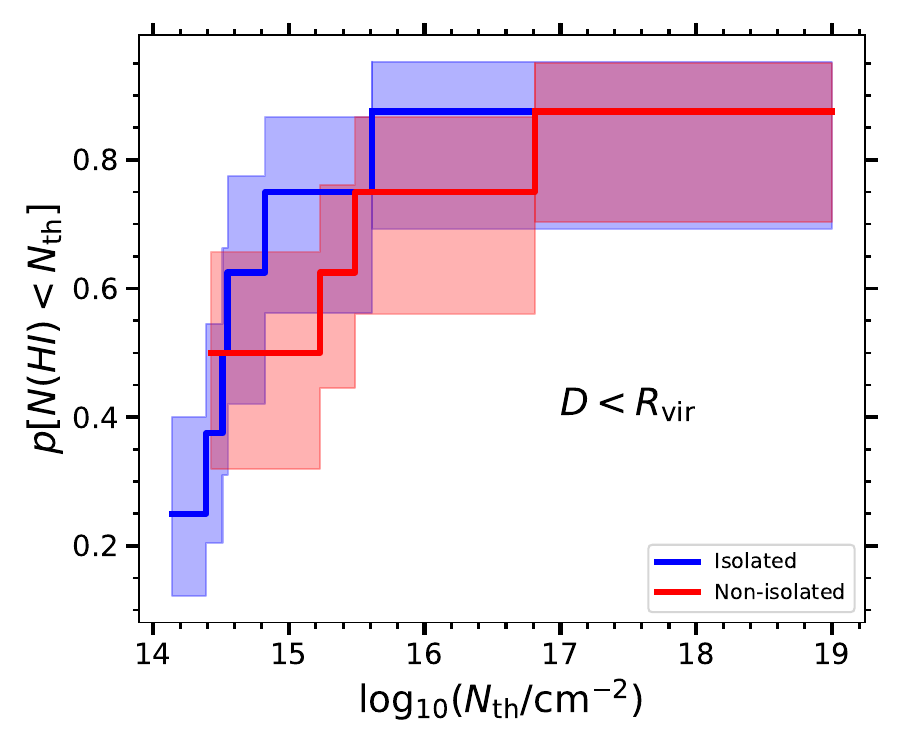}%
    \includegraphics[width=0.5\linewidth]{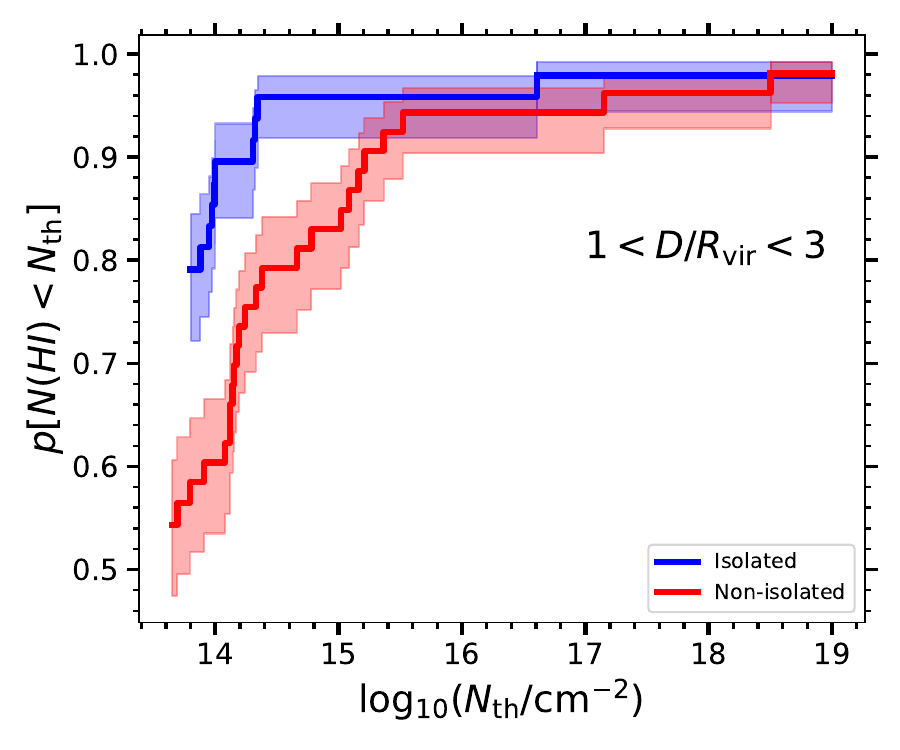}    
    \caption{{\tt Left:} Probability of detecting an absorber with $N(\HI)$ below a certain threshold plotted against the threshold $\log_{10}(N_{\rm th}/{\rm cm}^{-2})$ for galaxies with $D<R_{\rm vir}$ for the 6 sightlines with Magellan survey. The red and blue colors indicate isolated and non-isolated galaxies. The upper-limits in $N(\HI)$ are taken into consideration with the Kaplan-Meier estimates using the {\sc Python} package {\sc Survive}. {\tt Right:} Same as the left panel but for galaxies with $1\leq D/R_{\rm vir}<3$.}
    \label{fig:env_Ndist}
\end{figure*}

 In Sect.~\ref{sec:env}, we have discussed the role of environment on the $N(\HI)$ distribution inside and outside the virial radius of galaxies. The left panel of Fig.~\ref{fig:env_Ndist} shows the probability of detecting an absorber with $N(\HI)$ below a certain threshold plotted against the threshold ${\rm log}_{10}(N_{\rm th}/{\rm cm}^{-2})$ for galaxies with $D<R_{\rm vir}$ for the 6 sightlines with Magellan/IMACS observations. The blue and red colors indicate isolated and non-isolated galaxies, respectively. The upper-limits in $N(\HI)$ are taken into consideration with the Kaplan-Meier estimates using the {\sc Python} package {\sc Survive}. The right panel of Fig.~\ref{fig:env_Ndist} shows the same but for galaxies with $1 < D/R_{\rm vir}< 3$.

\section{Tabulated \HI\ measurements}

In Table.~\ref{longtable1}, we present the \HI\ measurements for the 256 MUSEQuBES galaxies used in this work.

\newpage 
\onecolumn
\begin{longtable}{ p{.04\textwidth}  p{.06\textwidth} r r  r  r  r   r   r }

\caption{The \HI\ absorption measurements and the associated galaxy properties for the MUSEQuBES sample } \label{longtable1} \\
\hline
ID & $z_{\rm gal}$ & $z_{\rm abs}$ & ${\rm log}_{10}\left(\frac{N}{{\rm cm}^{-2}}\right)$ & $\sigma \left[{\rm log}_{10}\left(\frac{N}{{\rm cm}^{-2}}\right)\right]$ & ${\rm log}_{10}\left(\frac{N_{\rm sens}}{{\rm cm}^{-2}}\right)$ & $D$ in pkpc & ${\rm log}_{10}\left(\frac{M_{\star}}{{\rm M_{\odot}}}\right)$ & ${\rm log}_{10}\left(\frac{\rm SFR}{{\rm M_{\odot}}~{\rm yr}^{-1}}\right)$  \\ 
(1) & (2) & (3) & (4) & (5) & (6) & (7) & (8) & (9)\\ 
\hline
\endfirsthead
\hline
ID & $z_{\rm gal}$ & $z_{\rm abs}$ & ${\rm log}_{10}\left(\frac{N}{{\rm cm}^{-2}}\right)$ & $\sigma\left[{\rm log}_{10}\left(\frac{N}{{\rm cm}^{-2}}\right)\right]$ & ${\rm log}_{10}\left(\frac{N_{\rm sens}}{{\rm cm}^{-2}}\right)$ & $D$ in pkpc & ${\rm log}_{10}\left(\frac{M_{\star}}{{\rm M_{\odot}}}\right)$ & ${\rm log}_{10}\left(\frac{\rm SFR}{{\rm M_{\odot}}~{\rm yr}^{-1}}\right)$  \\
(1) & (2) & (3) & (4) & (5) & (6) & (7) & (8) & (9) \\ 
\hline
\endhead
\hline
\endfoot
1 & 0.3333 & 0.3333 & 14.14 & 0.02 & 12.64 & 36.8 & 8.4 & -1.46 \\
2 & 0.3249 & 0.3236 & 16.49 & 0.02 & 12.66 & 44.2 & 8.7 & -0.84 \\
3 & 0.3284 & 0.3283 & 15.84 & 0.02 & 12.65 & 88.9 & 9.8 & -0.43 \\
4 & 0.3835 & 0.3834 & 14.80 & 0.01 & 12.61 & 50.2 & 8.4 & -1.1 \\
5 & 0.2496 & 0.2500 & 15.71 & 0.01 & 12.51 & 70.9 & 8.7 & -1.33 \\
6 & 0.5322 & 0.5333 & 14.37 & 0.03 & 13.36 & 178.4 & 8.3 & -1.35 \\
7 & 0.4021 & 0.4020 & 14.63 & 0.03 & 12.58 & 77.6 & 8.2 & -1.71 \\
8 & 0.3994 & 0.3994 & 13.66 & 0.04 & 12.58 & 155.3 & 8.1 & -1.01 \\
9 & 0.2656 & 0.2655 & 14.12 & 0.02 & 12.53 & 155.9 & 7.9 & -2.27 \\
10 & 0.4318 & 0.4315 & 14.41 & 0.02 & 12.61 & 176.5 & 10.1 & 0.38 \\
11 & 0.3227 & 0.3236 & 16.49 & 0.02 & 12.66 & 149.8 & 8.8 & -1.39 \\
12 & 0.2222 & 0.2221 & 15.15 & 0.02 & 12.31 & 96.2 & 7.2 & -1.16 \\
13 & 0.1321 & 0.1321 & 14.52 & 0.10 & 12.29 & 13.8 & 7.1 & -2.3 \\
14 & 0.2256 & 0.2259 & 16.98 & 0.30 & 12.32 & 103.3 & 7.6 & -2.29 \\
15 & 0.2909 & 0.2911 & 14.40 & 0.11 & 12.49 & 72.9 & 7.9 & -1.46 \\
16 & 0.4002 & 0.4004 & 14.61 & 0.02 & 12.51 & 67.1 & 8.7 & -0.67 \\
17 & 0.6019 & 0.6010 & 13.77 & 0.06 & 13.29 & 213.7 & 9.5 & <-1.76 \\
18 & 0.3139 & 0.3123 & 14.56 & 0.01 & 12.49 & 79.3 & 7.3 & <-2.02 \\
19 & 0.4243 & 0.4241 & 14.95 & 0.10 & 12.48 & 130.2 & 9.4 & -0.45 \\
20 & 0.3124 & 0.3123 & 14.56 & 0.01 & 12.49 & 85.9 & 8.3 & -2.09 \\
21 & 0.3089 & 0.3083 & 14.62 & 0.05 & 12.74 & 125.3 & 8.3 & -0.73 \\
22 & 0.5723 & 0.5726 & 14.98 & 0.02 & 13.52 & 27.9 & 7.6 & -1.36 \\
23 & 0.6158 & 0.6151 & 16.48 & 0.03 & 13.51 & 127.5 & 9.7 & <-1.14 \\
24 & 0.3681 & 0.3683 & 14.50 & 0.03 & 12.72 & 103.2 & 8.9 & -0.73 \\
25 & 0.6027 & 0.6019 & 15.02 & 0.02 & 13.51 & 196.6 & 10.5 & <-1.11 \\
26 & 0.6151 & 0.6151 & 16.48 & 0.03 & 13.51 & 102.4 & 9.6 & -0.12 \\
27 & 0.6153 & 0.6151 & 16.48 & 0.03 & 13.51 & 47.9 & 9.4 & 1.03 \\
28 & 0.6017 & 0.6019 & 15.02 & 0.02 & 13.51 & 126.0 & 9.3 & -0.22 \\
29 & 0.3085 & 0.3083 & 14.62 & 0.05 & 12.74 & 153.4 & 7.2 & -0.78 \\
30 & 0.3081 & 0.3083 & 14.62 & 0.05 & 12.74 & 157.2 & 8.8 & -1.12 \\
31 & 0.6028 & 0.6019 & 15.02 & 0.02 & 13.51 & 141.5 & 8.8 & -0.99 \\
32 & 0.7257 & 0.7254 & 14.60 & 0.03 & 13.52 & 202.4 & 8.2 & -1.18 \\
33 & 0.7254 & 0.7254 & 14.60 & 0.03 & 13.52 & 214.7 & 9.5 & 0.3 \\
34 & 0.3720 & 0.3715 & 14.12 & 0.03 & 12.68 & 120.2 & 9.7 & 0.02 \\
35 & 0.4166 & 0.4186 & 16.92 & 0.03 & 12.66 & 127.8 & 8.8 & -1.59 \\
36 & 0.4179 & 0.4186 & 16.92 & 0.03 & 12.65 & 122.3 & 8.3 & -1.13 \\
37 & 0.4192 & 0.4186 & 16.92 & 0.03 & 12.65 & 122.0 & 8.3 & -0.7 \\
38 & 0.5172 & 0.5170 & 14.60 & 0.12 & 13.48 & 93.0 & 9.6 & 0.24 \\
39 & 0.7175 & 0.7180 & 15.54 & 0.04 & 13.65 & 71.4 & 9.4 & -1.06 \\
40 & 0.2021 & 0.2027 & 15.24 & 0.04 & 12.61 & 100.0 & 8.6 & -1.61 \\
41 & 0.4353 & 0.4362 & 13.84 & 0.05 & 12.80 & 218.8 & 7.9 & <-0.92 \\
42 & 0.2609 & 0.2611 & 14.77 & 0.02 & 12.63 & 105.7 & 8.8 & -0.69 \\
43 & 0.7181 & 0.7180 & 15.54 & 0.04 & 13.65 & 182.6 & 9.3 & 0.04 \\
44 & 0.3992 & 0.3991 & 16.89 & 0.06 & 12.75 & 199.3 & 10.1 & <-1.46 \\
45 & 0.7177 & 0.7180 & 15.54 & 0.04 & 13.65 & 227.6 & 9.8 & 0.0 \\
46 & 0.7178 & 0.7180 & 15.54 & 0.04 & 13.65 & 154.6 & 8.9 & -0.47 \\
47 & 0.5355 & 0.5352 & 15.50 & 0.03 & 13.53 & 111.0 & 8.4 & -1.43 \\
48 & 0.5355 & 0.5358 & 14.38 & 0.02 & 13.35 & 204.8 & 10.3 & -0.06 \\
49 & 0.5717 & 0.5723 & 15.61 & 0.08 & 13.39 & 106.1 & 10.2 & <-1.08 \\
50 & 0.7293 & 0.7288 & 16.61 & 0.04 & 13.37 & 175.9 & 9.5 & 0.93 \\
51 & 0.3760 & 0.3754 & 14.29 & 0.30 & 12.52 & 71.5 & 7.9 & <-1.66 \\
52 & 0.3198 & 0.3196 & 14.55 & 0.01 & 12.60 & 49.4 & 8.5 & -1.25 \\
53 & 0.3291 & 0.3278 & 14.43 & 0.04 & 12.60 & 75.4 & 10.6 & <-1.48 \\
54 & 0.3751 & 0.3754 & 14.29 & 0.30 & 12.52 & 110.7 & 7.6 & -1.81 \\
55 & 0.3282 & 0.3278 & 14.43 & 0.04 & 12.59 & 143.6 & 8.3 & <-1.86 \\
56 & 0.1734 & 0.1736 & 14.51 & 0.04 & 12.51 & 70.9 & 9.2 & -0.44 \\
57 & 0.2961 & 0.2959 & 14.14 & 0.02 & 12.63 & 48.7 & 8.5 & -1.68 \\
58 & 0.6928 & 0.6923 & 15.02 & 0.03 & 13.40 & 154.1 & 8.3 & -1.08 \\
59 & 0.4339 & 0.4342 & 13.79 & 0.06 & 12.61 & 182.3 & 8.3 & -1.03 \\
60 & 0.6929 & 0.6923 & 15.02 & 0.03 & 13.40 & 161.6 & 9.3 & -0.4 \\
61 & 0.6926 & 0.6923 & 15.02 & 0.03 & 13.40 & 150.0 & 8.8 & -0.65 \\
62 & 0.4949 & 0.4951 & 14.24 & 0.02 & 12.95 & 109.8 & 9.5 & 0.22 \\
63 & 0.4950 & 0.4951 & 14.24 & 0.02 & 12.95 & 83.7 & 8.1 & -1.0 \\
64 & 0.2453 & 0.2455 & 13.80 & 0.03 & 12.14 & 77.4 & 7.1 & -2.84 \\
65 & 0.4832 & 0.4833 & 13.95 & 0.02 & 12.94 & 154.3 & 8.1 & -1.53 \\
66 & 0.1825 & 0.1827 & 14.83 & 0.07 & 12.14 & 50.7 & 7.9 & -1.97 \\
67 & 0.3612 & 0.3608 & 15.17 & 0.01 & 12.22 & 193.5 & 7.8 & -1.3 \\
68 & 0.3453 & 0.3451 & 16.88 & 0.01 & 12.44 & 77.2 & 9.3 & -1.59 \\
69 & 0.4252 & 0.4252 & 13.95 & 0.03 & 12.50 & 169.1 & 9.4 & -0.72 \\
70 & 0.3768 & 0.3766 & 13.93 & 0.03 & 12.45 & 96.0 & 8.7 & -0.75 \\
71 & 0.6326 & 0.6306 & 15.06 & 0.04 & 13.24 & 216.4 & 9.8 & <-1.68 \\
72 & 0.3794 & 0.3795 & 14.33 & 0.01 & 12.45 & 163.5 & 8.8 & -0.64 \\
73 & 0.3444 & 0.3451 & 16.88 & 0.01 & 12.44 & 135.4 & 10.2 & -1.23 \\
74 & 0.6330 & 0.6306 & 15.06 & 0.04 & 13.24 & 247.9 & 10.6 & <-0.8 \\
75 & 0.1487 & 0.1489 & 15.16 & 0.04 & 12.34 & 41.1 & 7.5 & -2.34 \\
76 & 0.3801 & 0.3795 & 14.33 & 0.01 & 12.45 & 134.2 & 10.4 & -1.03 \\
77 & 0.6314 & 0.6306 & 15.06 & 0.04 & 13.24 & 202.6 & 9.0 & -1.05 \\
78 & 0.6306 & 0.6306 & 15.06 & 0.04 & 13.24 & 204.6 & 8.3 & -0.87 \\
79 & 0.3439 & 0.3451 & 16.88 & 0.01 & 12.44 & 82.3 & 9.5 & <-1.82 \\
80 & 0.4468 & 0.4461 & 15.99 & 0.04 & 13.16 & 65.5 & 6.9 & <-1.78 \\
81 & 0.3799 & 0.3795 & 14.33 & 0.01 & 12.45 & 183.4 & 8.4 & -1.36 \\
82 & 0.1778 & 0.1777 & 14.80 & 0.05 & 12.60 & 42.6 & 8.7 & -1.43 \\
83 & 0.7186 & 0.7189 & 15.63 & 0.04 & 13.54 & 101.7 & 10.7 & <-0.93 \\
84 & 0.3845 & 0.3847 & 14.83 & 0.03 & 12.66 & 36.0 & 9.3 & -1.62 \\
85 & 0.5358 & 0.5362 & 15.92 & 0.02 & 13.44 & 116.9 & 9.6 & -1.09 \\
86 & 0.5361 & 0.5362 & 15.92 & 0.02 & 13.44 & 110.4 & 10.1 & 0.08 \\
87 & 0.4297 & 0.4296 & 14.40 & 0.05 & 12.70 & 77.2 & 9.1 & -0.86 \\
88 & 0.7183 & 0.7189 & 15.63 & 0.04 & 13.54 & 135.7 & 6.9 & -1.28 \\
89 & 0.5582 & 0.5576 & 14.97 & 0.12 & 13.51 & 196.4 & 8.5 & -1.05 \\
90 & 0.5578 & 0.5576 & 14.97 & 0.12 & 13.51 & 174.7 & 8.0 & -0.8 \\
91 & 0.3993 & 0.3995 & 14.43 & 0.05 & 12.67 & 176.9 & 9.2 & -0.09 \\
92 & 0.5572 & 0.5576 & 14.97 & 0.12 & 13.51 & 141.6 & 8.8 & -0.68 \\
93 & 0.3293 & 0.3286 & 18.08 & 0.07 & 12.82 & 101.9 & 8.7 & -1.55 \\
94 & 0.3989 & 0.3988 & 13.82 & 0.05 & 12.79 & 107.5 & 8.0 & -1.14 \\
95 & 0.3626 & 0.3622 & 13.96 & 0.05 & 12.79 & 92.3 & 7.2 & <-1.62 \\
96 & 0.6103 & 0.6103 & 15.42 & 0.02 & 13.58 & 113.5 & 9.4 & 0.72 \\
97 & 0.4434 & 0.4424 & 14.92 & 0.19 & 12.82 & 63.9 & 7.4 & <-1.71 \\
98 & 0.3286 & 0.3286 & 18.08 & 0.07 & 12.82 & 99.8 & 8.1 & -1.02 \\
99 & 0.5102 & 0.5102 & 14.76 & 0.09 & 13.43 & 84.0 & 6.1 & -1.38 \\
100 & 0.5097 & 0.5102 & 14.76 & 0.09 & 13.43 & 99.2 & 8.3 & -0.84 \\
101 & 0.4151 & 0.4151 & 14.29 & 0.02 & 12.77 & 142.3 & 6.2 & -0.83 \\
102 & 0.6401 & 0.6406 & 15.14 & 0.04 & 13.59 & 210.5 & 9.2 & 0.19 \\
103 & 0.3296 & 0.3286 & 18.08 & 0.07 & 12.82 & 151.6 & 9.6 & <-2.0 \\
104 & 0.2161 & 0.2161 & 13.95 & 0.02 & 12.57 & 109.9 & 8.6 & -1.01 \\
105 & 0.3299 & 0.3286 & 18.08 & 0.07 & 12.82 & 126.1 & 11.0 & <-1.41 \\
106 & 0.6407 & 0.6406 & 15.14 & 0.04 & 13.59 & 190.8 & 9.7 & -0.18 \\
107 & 0.3292 & 0.3286 & 18.08 & 0.07 & 12.82 & 107.8 & 9.9 & -1.41 \\
108 & 0.6130 & 0.6125 & 15.52 & 0.02 & 13.17 & 152.5 & 9.2 & -1.01 \\
109 & 0.3215 & 0.3219 & 14.17 & 0.01 & 12.41 & 127.3 & 9.4 & -0.2 \\
110 & 0.3214 & 0.3219 & 14.17 & 0.01 & 12.41 & 140.6 & 9.3 & -0.11 \\
111 & 0.4129 & 0.4127 & 14.08 & 0.01 & 12.40 & 101.3 & 8.8 & -1.11 \\
112 & 0.3005 & 0.2990 & 15.36 & 0.04 & 12.36 & 139.8 & 9.3 & -0.83 \\
113 & 0.3217 & 0.3219 & 14.17 & 0.01 & 12.41 & 142.6 & 8.6 & -0.43 \\
114 & 0.2502 & 0.2513 & 14.66 & 0.01 & 12.27 & 108.0 & 6.6 & -2.13 \\
115 & 0.2514 & 0.2513 & 14.66 & 0.01 & 12.27 & 74.9 & 8.5 & -0.8 \\
116 & 0.6120 & 0.6125 & 15.52 & 0.02 & 13.17 & 141.3 & 10.9 & -0.03 \\
117 & 0.6119 & 0.6125 & 15.52 & 0.02 & 13.17 & 127.4 & 9.7 & -0.34 \\
118 & 0.6119 & 0.6125 & 15.52 & 0.02 & 13.17 & 255.1 & 10.9 & <-1.04 \\
119 & 0.2451 & 0.2452 & 14.15 & 0.01 & 12.23 & 123.3 & 7.9 & -1.43 \\
120 & 0.2066 & 0.2071 & 15.21 & 0.01 & 12.23 & 108.3 & 9.1 & -0.35 \\
121 & 0.2451 & 0.2452 & 14.15 & 0.01 & 12.23 & 125.5 & 8.1 & -1.45 \\
122 & 0.2066 & 0.2071 & 15.21 & 0.01 & 12.23 & 38.2 & 8.5 & -1.03 \\
123 & 0.4242 & 0.4229 & 14.01 & 0.02 & 12.45 & 147.5 & 6.4 & <-2.32 \\
124 & 0.3847 & 0.3843 & 13.88 & 0.03 & 12.42 & 166.1 & 7.9 & -1.45 \\
125 & 0.3093 & 0.3094 & 14.32 & 0.01 & 12.42 & 114.5 & 8.4 & -0.32 \\
126 & 0.4155 & 0.4148 & 13.06 & 0.08 & 12.45 & 149.5 & 7.3 & -2.04 \\
127 & 0.3904 & 0.3905 & 19.02 & 0.02 & 12.67 & 9.5 & 6.0 & -1.86 \\
128 & 0.4164 & 0.4163 & 14.10 & 0.03 & 12.65 & 131.9 & 8.3 & -1.73 \\
129 & 0.3898 & 0.3905 & 19.02 & 0.02 & 12.67 & 104.4 & 10.0 & 0.25 \\
130 & 0.2851 & 0.2852 & 13.80 & 0.08 & 12.58 & 76.9 & 7.9 & -1.87 \\
131 & 0.3951 & 0.3953 & 13.88 & 0.03 & 12.67 & 148.7 & 7.3 & -1.67 \\
132 & 0.2111 & -1.0000 & 12.44 & -1.00 & 12.44 & 68.8 & 8.2 & -1.25 \\
133 & 0.4224 & -1.0000 & 12.59 & -1.00 & 12.59 & 207.0 & 8.3 & -1.55 \\
134 & 0.5101 & -1.0000 & 13.32 & -1.00 & 13.32 & 174.1 & 8.2 & -1.41 \\
135 & 0.5941 & -1.0000 & 13.38 & -1.00 & 13.38 & 216.3 & 9.2 & 0.5 \\
136 & 0.4789 & -1.0000 & 13.30 & -1.00 & 13.30 & 183.2 & 8.2 & -0.76 \\
137 & 0.2717 & -1.0000 & 12.38 & -1.00 & 12.38 & 35.4 & 7.1 & <-2.13 \\
138 & 0.5317 & -1.0000 & 13.24 & -1.00 & 13.24 & 170.4 & 7.6 & -1.82 \\
139 & 0.3004 & -1.0000 & 12.72 & -1.00 & 12.72 & 140.9 & 7.8 & -2.07 \\
140 & 0.4455 & -1.0000 & 12.80 & -1.00 & 12.80 & 79.7 & 10.0 & <-1.68 \\
141 & 0.4454 & -1.0000 & 12.80 & -1.00 & 12.80 & 115.4 & 9.6 & <-1.75 \\
142 & 0.4399 & -1.0000 & 12.78 & -1.00 & 12.78 & 54.9 & 9.0 & -0.74 \\
143 & 0.4408 & -1.0000 & 12.79 & -1.00 & 12.79 & 73.2 & 9.6 & 0.22 \\
144 & 0.4435 & -1.0000 & 12.80 & -1.00 & 12.80 & 82.2 & 6.9 & <-1.79 \\
145 & 0.4419 & -1.0000 & 12.79 & -1.00 & 12.79 & 139.3 & 8.9 & -1.6 \\
146 & 0.4477 & -1.0000 & 12.81 & -1.00 & 12.81 & 168.3 & 8.8 & -1.49 \\
147 & 0.4419 & -1.0000 & 12.79 & -1.00 & 12.79 & 176.2 & 8.8 & -0.96 \\
148 & 0.4384 & -1.0000 & 12.78 & -1.00 & 12.78 & 193.9 & 9.9 & 0.51 \\
149 & 0.4446 & -1.0000 & 12.80 & -1.00 & 12.80 & 169.0 & 9.4 & -0.92 \\
150 & 0.4428 & -1.0000 & 12.79 & -1.00 & 12.79 & 149.3 & 9.2 & -0.97 \\
151 & 0.5308 & -1.0000 & 13.49 & -1.00 & 13.49 & 227.0 & 10.6 & 0.07 \\
152 & 0.6811 & -1.0000 & 13.52 & -1.00 & 13.52 & 227.8 & 7.0 & <-1.06 \\
153 & 0.7461 & -1.0000 & 13.75 & -1.00 & 13.75 & 313.7 & 10.4 & <-0.38 \\
154 & 0.7023 & -1.0000 & 13.57 & -1.00 & 13.57 & 226.6 & 10.4 & <-1.0 \\
155 & 0.7438 & -1.0000 & 13.73 & -1.00 & 13.73 & 147.4 & 9.5 & -0.62 \\
156 & 0.7024 & -1.0000 & 13.57 & -1.00 & 13.57 & 206.1 & 9.1 & <-1.28 \\
157 & 0.5159 & -1.0000 & 13.43 & -1.00 & 13.43 & 75.2 & 8.9 & 0.78 \\
158 & 0.6489 & -1.0000 & 13.50 & -1.00 & 13.50 & 164.9 & 9.2 & 0.52 \\
159 & 0.5827 & -1.0000 & 13.51 & -1.00 & 13.51 & 254.6 & 7.4 & <-0.32 \\
160 & 0.5080 & -1.0000 & 13.41 & -1.00 & 13.41 & 225.4 & 7.5 & <-1.9 \\
161 & 0.5308 & -1.0000 & 13.49 & -1.00 & 13.49 & 185.6 & 8.1 & -1.36 \\
162 & 0.4758 & -1.0000 & 12.81 & -1.00 & 12.81 & 158.2 & 10.6 & <-0.54 \\
163 & 0.4394 & -1.0000 & 12.64 & -1.00 & 12.64 & 133.1 & 8.1 & <-1.17 \\
164 & 0.5101 & -1.0000 & 13.39 & -1.00 & 13.39 & 211.6 & 9.5 & -0.56 \\
165 & 0.7378 & -1.0000 & 13.56 & -1.00 & 13.56 & 66.2 & 8.5 & -1.12 \\
166 & 0.7104 & -1.0000 & 13.44 & -1.00 & 13.44 & 161.6 & 8.1 & -0.46 \\
167 & 0.6607 & -1.0000 & 13.48 & -1.00 & 13.48 & 138.8 & 8.0 & <-1.39 \\
168 & 0.6606 & -1.0000 & 13.48 & -1.00 & 13.48 & 142.1 & 9.4 & -0.12 \\
169 & 0.4490 & -1.0000 & 12.85 & -1.00 & 12.85 & 182.5 & 10.6 & 0.87 \\
170 & 0.4490 & -1.0000 & 12.85 & -1.00 & 12.85 & 188.1 & 10.7 & 0.78 \\
171 & 0.7369 & -1.0000 & 13.74 & -1.00 & 13.74 & 255.4 & 10.1 & <-0.46 \\
172 & 0.4784 & -1.0000 & 12.68 & -1.00 & 12.68 & 197.3 & 8.9 & -0.65 \\
173 & 0.4581 & -1.0000 & 12.58 & -1.00 & 12.58 & 171.1 & 9.0 & 1.01 \\
174 & 0.1392 & -1.0000 & 12.42 & -1.00 & 12.42 & 83.9 & 7.9 & -2.25 \\
175 & 0.1203 & -1.0000 & 12.39 & -1.00 & 12.39 & 20.2 & 9.3 & -1.61 \\
176 & 0.1201 & -1.0000 & 12.39 & -1.00 & 12.39 & 11.3 & 10.4 & <-2.37 \\
177 & 0.2790 & -1.0000 & 12.51 & -1.00 & 12.51 & 105.2 & 7.3 & <-2.18 \\
178 & 0.4395 & -1.0000 & 12.53 & -1.00 & 12.53 & 212.2 & 8.9 & -1.33 \\
179 & 0.5256 & -1.0000 & 13.34 & -1.00 & 13.34 & 227.4 & 6.6 & <-1.61 \\
180 & 0.6476 & -1.0000 & 13.32 & -1.00 & 13.32 & 155.3 & 8.7 & <-1.05 \\
181 & 0.6469 & -1.0000 & 13.32 & -1.00 & 13.32 & 156.3 & 8.3 & -1.11 \\
182 & 0.6531 & -1.0000 & 13.32 & -1.00 & 13.32 & 124.9 & 8.7 & -0.59 \\
183 & 0.5205 & -1.0000 & 13.32 & -1.00 & 13.32 & 171.3 & 9.6 & 0.09 \\
184 & 0.6173 & -1.0000 & 13.31 & -1.00 & 13.31 & 116.3 & 8.1 & -1.14 \\
185 & 0.6178 & -1.0000 & 13.31 & -1.00 & 13.31 & 161.5 & 8.9 & -0.39 \\
186 & 0.6183 & -1.0000 & 13.31 & -1.00 & 13.31 & 189.6 & 8.1 & 0.24 \\
187 & 0.5112 & -1.0000 & 13.35 & -1.00 & 13.35 & 164.5 & 8.4 & -1.59 \\
188 & 0.7318 & -1.0000 & 13.54 & -1.00 & 13.54 & 125.7 & 9.3 & -1.17 \\
189 & 0.6153 & -1.0000 & 13.37 & -1.00 & 13.37 & 193.0 & 8.1 & -1.32 \\
190 & 0.5553 & -1.0000 & 13.48 & -1.00 & 13.48 & 167.1 & 8.6 & 0.35 \\
191 & 0.4506 & -1.0000 & 12.35 & -1.00 & 12.35 & 98.7 & 7.0 & <-1.78 \\
192 & 0.3452 & -1.0000 & 12.24 & -1.00 & 12.24 & 56.3 & 6.0 & <-1.93 \\
193 & 0.5394 & -1.0000 & 13.03 & -1.00 & 13.03 & 234.9 & 7.7 & <-1.87 \\
194 & 0.5385 & -1.0000 & 13.03 & -1.00 & 13.03 & 245.4 & 8.0 & -1.61 \\
195 & 0.5623 & -1.0000 & 13.07 & -1.00 & 13.07 & 183.8 & 6.5 & <-1.72 \\
196 & 0.5394 & -1.0000 & 13.03 & -1.00 & 13.03 & 189.7 & 8.4 & -0.99 \\
197 & 0.3437 & -1.0000 & 12.44 & -1.00 & 12.44 & 138.5 & 9.6 & -0.46 \\
198 & 0.1953 & -1.0000 & 12.34 & -1.00 & 12.34 & 75.0 & 7.7 & -2.26 \\
199 & 0.5151 & -1.0000 & 13.44 & -1.00 & 13.44 & 143.5 & 8.6 & -0.76 \\
200 & 0.6335 & -1.0000 & 13.24 & -1.00 & 13.24 & 220.2 & 8.9 & -1.09 \\
201 & 0.5099 & -1.0000 & 13.23 & -1.00 & 13.23 & 140.5 & 11.4 & <-1.12 \\
202 & 0.5121 & -1.0000 & 13.23 & -1.00 & 13.23 & 118.5 & 10.4 & -1.06 \\
203 & 0.5138 & -1.0000 & 13.45 & -1.00 & 13.45 & 69.1 & 9.9 & <-1.46 \\
204 & 0.5151 & -1.0000 & 13.44 & -1.00 & 13.44 & 99.9 & 10.5 & <-1.3 \\
205 & 0.5150 & -1.0000 & 13.44 & -1.00 & 13.44 & 119.8 & 10.3 & <-1.32 \\
206 & 0.5116 & -1.0000 & 13.23 & -1.00 & 13.23 & 158.8 & 9.8 & <-1.33 \\
207 & 0.5125 & -1.0000 & 13.23 & -1.00 & 13.23 & 147.3 & 7.9 & -1.44 \\
208 & 0.5102 & -1.0000 & 13.23 & -1.00 & 13.23 & 171.6 & 10.2 & -1.17 \\
209 & 0.5113 & -1.0000 & 13.23 & -1.00 & 13.23 & 56.4 & 9.9 & <-1.56 \\
210 & 0.4879 & -1.0000 & 13.40 & -1.00 & 13.40 & 137.9 & 9.1 & -0.76 \\
211 & 0.7480 & -1.0000 & 13.65 & -1.00 & 13.65 & 131.5 & 8.5 & -0.91 \\
212 & 0.6589 & -1.0000 & 13.47 & -1.00 & 13.47 & 113.1 & 7.4 & <-1.11 \\
213 & 0.4779 & -1.0000 & 13.13 & -1.00 & 13.13 & 255.4 & 9.7 & -0.79 \\
214 & 0.3304 & -1.0000 & 12.82 & -1.00 & 12.82 & 130.2 & 9.5 & <-2.12 \\
215 & 0.3568 & -1.0000 & 12.79 & -1.00 & 12.79 & 115.1 & 8.9 & -0.22 \\
216 & 0.3870 & -1.0000 & 12.79 & -1.00 & 12.79 & 163.8 & 6.8 & -1.57 \\
217 & 0.4950 & -1.0000 & 13.42 & -1.00 & 13.42 & 236.7 & 8.4 & <1.45 \\
218 & 0.6513 & -1.0000 & 13.59 & -1.00 & 13.59 & 201.5 & 10.5 & -0.06 \\
219 & 0.6504 & -1.0000 & 13.59 & -1.00 & 13.59 & 129.7 & 8.9 & -1.18 \\
220 & 0.6314 & -1.0000 & 13.60 & -1.00 & 13.60 & 167.8 & 8.7 & -1.44 \\
221 & 0.5010 & -1.0000 & 13.42 & -1.00 & 13.42 & 196.0 & 7.4 & <-1.65 \\
222 & 0.3073 & -1.0000 & 12.40 & -1.00 & 12.40 & 153.0 & 7.2 & <-1.95 \\
223 & 0.3625 & -1.0000 & 12.37 & -1.00 & 12.37 & 130.1 & 6.8 & <-1.67 \\
224 & 0.5399 & -1.0000 & 13.15 & -1.00 & 13.15 & 287.0 & 6.8 & -0.8 \\
225 & 0.6597 & -1.0000 & 13.19 & -1.00 & 13.19 & 238.9 & 9.4 & <-0.73 \\
226 & 0.7150 & -1.0000 & 13.21 & -1.00 & 13.21 & 250.3 & 8.8 & -0.51 \\
227 & 0.5450 & -1.0000 & 13.17 & -1.00 & 13.17 & 200.3 & 10.0 & <-1.1 \\
228 & 0.5469 & -1.0000 & 13.18 & -1.00 & 13.18 & 181.2 & 8.1 & <-1.15 \\
229 & 0.6100 & -1.0000 & 13.16 & -1.00 & 13.16 & 165.5 & 6.0 & <-1.41 \\
230 & 0.5091 & -1.0000 & 13.08 & -1.00 & 13.08 & 159.1 & 7.7 & -0.61 \\
231 & 0.5307 & -1.0000 & 13.14 & -1.00 & 13.14 & 141.6 & 9.4 & <-1.25 \\
232 & 0.7398 & -1.0000 & 13.29 & -1.00 & 13.29 & 163.9 & 9.0 & <-1.27 \\
233 & 0.2682 & -1.0000 & 12.25 & -1.00 & 12.25 & 151.6 & 8.7 & -0.96 \\
234 & 0.3159 & -1.0000 & 12.43 & -1.00 & 12.43 & 128.2 & 7.6 & <-2.35 \\
235 & 0.2654 & -1.0000 & 12.24 & -1.00 & 12.24 & 58.9 & 7.8 & -2.3 \\
236 & 0.2674 & -1.0000 & 12.25 & -1.00 & 12.25 & 70.5 & 10.2 & -0.7 \\
237 & 0.4056 & -1.0000 & 12.44 & -1.00 & 12.44 & 98.0 & 7.1 & <-1.94 \\
238 & 0.4196 & -1.0000 & 12.45 & -1.00 & 12.45 & 158.3 & 8.7 & -0.73 \\
239 & 0.3449 & -1.0000 & 12.68 & -1.00 & 12.68 & 123.8 & 6.2 & <-1.69 \\
240 & 0.4231 & -1.0000 & 12.66 & -1.00 & 12.66 & 119.5 & 6.8 & -1.88 \\
241 & 0.3451 & -1.0000 & 12.67 & -1.00 & 12.67 & 98.2 & 6.2 & -1.38 \\
242 & 0.6385 & -1.0000 & 13.47 & -1.00 & 13.47 & 151.8 & 10.0 & <-1.15 \\
243 & 0.5232 & -1.0000 & 13.37 & -1.00 & 13.37 & 112.3 & 9.3 & -0.51 \\
244 & 0.7241 & -1.0000 & 13.51 & -1.00 & 13.51 & 218.4 & 6.8 & <-1.68 \\
245 & 0.5227 & -1.0000 & 13.37 & -1.00 & 13.37 & 158.4 & 9.1 & -0.26 \\
246 & 0.5230 & -1.0000 & 13.37 & -1.00 & 13.37 & 129.1 & 8.2 & -0.39 \\
247 & 0.6390 & -1.0000 & 13.47 & -1.00 & 13.47 & 97.4 & 7.2 & <-2.45 \\
248 & 0.5593 & -1.0000 & 13.45 & -1.00 & 13.45 & 139.3 & 9.5 & -1.59 \\
249 & 0.7016 & -1.0000 & 13.45 & -1.00 & 13.45 & 89.2 & 8.3 & -1.67 \\
250 & 0.7275 & -1.0000 & 13.53 & -1.00 & 13.53 & 243.4 & 10.1 & -0.61 \\
251 & 0.5741 & -1.0000 & 13.48 & -1.00 & 13.48 & 191.4 & 8.4 & -0.69 \\
252 & 0.6273 & -1.0000 & 13.46 & -1.00 & 13.46 & 161.0 & 7.1 & <-1.63 \\
253 & 0.6630 & -1.0000 & 13.46 & -1.00 & 13.46 & 216.2 & 8.8 & -0.34 \\
254 & 0.7155 & -1.0000 & 13.48 & -1.00 & 13.48 & 160.7 & 8.0 & <-1.26 \\
255 & 0.6831 & -1.0000 & 13.45 & -1.00 & 13.45 & 187.1 & 7.9 & <-1.73 \\
256 & 0.6628 & -1.0000 & 13.46 & -1.00 & 13.46 & 194.3 & 8.2 & -1.18 \\

\bottomrule
\end{longtable}
\noindent\textbf{Notes:} (1) Running ID of the galaxy (2) Galaxy redshift (3) Corresponding \HI\ column density-weighted absorption redshift (4) Total \HI\ column density (5) Error in the total \HI\ column density (6) 3$\sigma$ upper limit on column density obtained from the line-free region of the spectra (7) Impact parameter (8) Stellar mass of the galaxy (9) Star formation rate of the galaxy (`$<$' indicates 3$\sigma$ upper limit on SFR). The flag value of -1 in columns for $z_{\rm abs}$ and $\sigma \left[{\rm log}_{10}\left(\frac{N}{{\rm cm}^{-2}}\right)\right]$ represents non-detections of \HI\ absorption. In these cases, the ${\rm log}_{10}\left(\frac{N}{{\rm cm}^{-2}}\right)$ column corresponds to the  3$\sigma$ upper limit.  
\twocolumn

\label{LastPage}
\end{document}